\documentclass[%
 superscriptaddress,
 amsmath,amssymb,
 aps,
 prb,
 twocolumn,
 longbibliography %, linenumbers
]{revtex4-2}
\usepackage[]{lineno}
\usepackage{graphicx}% Include figure files
\usepackage{dcolumn}% Align table columns on decimal point
\usepackage{bm}% bold math
\usepackage{xcolor}
\usepackage{tikz}
\usepackage{xfrac}
\usepackage{blindtext}
\usepackage{times}
\usepackage{multirow}

\usepackage{color}
\usepackage[colorlinks=true, urlcolor=blue, linkcolor=blue, citecolor=blue, pdftex]{hyperref}

\usepackage[normalem]{ulem}

\definecolor{darkblue}{HTML}{004D6B}
\definecolor{darkred}{HTML}{8c1515}
\definecolor{darkgreen}{HTML}{006400}
\usepackage{hyperref}
\hypersetup{
	colorlinks=true,
	urlcolor=darkred,
	citecolor=darkblue,
	linkcolor=darkred,
	breaklinks
}

\usepackage[caption=false]{subfig}

\makeatletter
\@ifundefined{thepage}{}{}
\@ifundefined{thelinenumber}{}{}

\makeatletter
\newcommand*\patchAmsMathEnvironmentForLineno[1]{%
  \expandafter\let\csname old#1\expandafter\endcsname\csname #1\endcsname
  \expandafter\let\csname oldend#1\expandafter\endcsname\csname end#1\endcsname
  \renewenvironment{#1}%
     {\linenomath\csname old#1\endcsname}%
     {\csname oldend#1\endcsname\endlinenomath}}%
\newcommand*\patchBothAmsMathEnvironmentsForLineno[1]{%
  \patchAmsMathEnvironmentForLineno{#1}%
  \patchAmsMathEnvironmentForLineno{#1*}}%
\AtBeginDocument{%
  \patchBothAmsMathEnvironmentsForLineno{equation}%
  \patchBothAmsMathEnvironmentsForLineno{align}%
  \patchBothAmsMathEnvironmentsForLineno{flalign}%
  \patchBothAmsMathEnvironmentsForLineno{alignat}%
  \patchBothAmsMathEnvironmentsForLineno{gather}%
  \patchBothAmsMathEnvironmentsForLineno{multline}%
}
\makeatother
\graphicspath{{}}

\renewcommand{\Im}{\operatorname{Im}} % 

\usepackage{array}
\newcolumntype{C}[1]{>{\centering\arraybackslash}p{#1}}

\usetikzlibrary{decorations.markings,arrows,decorations.pathreplacing,calc,intersections,through,backgrounds}

\newcommand{\ie}{{\it i.e.},\ }
\newcommand{\eg}{{\it e.g.},\ }

\begin{document}
%\linenumbers
%\switchlinenumbers

\title{Supersymmetry on the lattice: Geometry, Topology, and Flat Bands}

\author{Krishanu Roychowdhury}
\affiliation{Theory Division, Saha Institute of Nuclear Physics, 1/AF Bidhannagar, Kolkata 700064, India}
\affiliation{Homi Bhabha National Institute, Training School Complex, Anushaktinagar, Mumbai
400094, India}
\affiliation{Max-Planck-Institut f\"{u}r Physik komplexer Systeme,
N\"{o}thnitzer Strasse 38, 01187 Dresden, Germany}
\author{Jan Attig}
\affiliation{Institute for Theoretical Physics, University of Cologne, 50937 Cologne, Germany}
\author{Simon Trebst}
\affiliation{Institute for Theoretical Physics, University of Cologne, 50937 Cologne, Germany}
\author{Michael J. Lawler}
\affiliation{Laboratory of Atomic And Solid State Physics, Cornell University, Ithaca, NY 14853, USA}
\affiliation{Department of Physics, Binghamton University, Binghamton, NY, 13902, USA}

\begin{abstract}
In quantum mechanics, supersymmetry (SUSY) posits an equivalence between two elementary degrees of freedom, bosons, and fermions defined by local rules. Here we apply it to find connections between bosonic and fermionic lattice models in the realm of condensed matter physics and uncover a novel 5-fold way topology it demands in these systems. At the single-particle level, our connections pair a bosonic and fermionic lattice model, either describing the hopping of number-conserving particles or local couplings between fermion parity conserving particles. The pair are isospectral except for zero modes, such as flat bands, quadratic band touchings, and nexus points, whose existence is undergirded by the Witten index of the SUSY theory. We develop a unifying framework to formulate these SUSY connections in terms of general lattice graph correspondences. Notably, in this framework, the supercharge operator that generates SUSY is Hermitian and can itself be interpreted as a hopping Hamiltonian on a bipartite lattice, a feature that enables the discovery of materials or model lattices hosting the SUSY partners. To illustrate the power of SUSY, we present 16 use cases of SUSY, that span topics including frustrated magnets, Kitaev spin liquids, and topological superconductors, the majority of which turn out to provide insights into the discovery and design of flat bands and topological materials. 
\end{abstract}
\date{\today}

\maketitle

%\tableofcontents

%%%%%%%%%%%%%%%%%%%%%%%%%%%%%%%%%%%%%%%%%%%%%%%%%%%%%%
% INTRODUCTION
%%%%%%%%%%%%%%%%%%%%%%%%%%%%%%%%%%%%%%%%%%%%%%%%%%%%%%

\section{Introduction}
Supersymmetry (SUSY) has set foot into condensed matter physics in several isolated areas, beginning with disorder \cite{Efetov1999}, then in the study of strongly interacting theories \cite{Fendley_2003a, Fendley_2003b, Huijse_2011, Bauer2013, Fendley2018}, and recently with the advent of topological mechanics \cite{KaneLubensky2013, Nash2015gyromaterials, Paulose2015, Huber2016, Rocklin2016, Chen2016topomech, Meeussen2016, socolar2017mechanical, Attig2019, Wakao2020}. Some of this work parallels high energy physics, which also aims for insights into strongly interacting field theories \cite{Golfand1971, Ramond1971, Neveu1971}, and to produce fermionic theories from bosonic ones. There, together with a certain level of naturalness, SUSY has gained prominence by going beyond being a mere trick to providing a leading theory of physics beyond the standard model \cite{StandardModel}. No analogous vision in condensed matter exists, but topological mechanics suggests one: SUSY enables us to add {\sl locality} to the classification of condensed matter by (conventional) symmetry and topology \cite{Chiu2016classification} and thereby \emph{produces simple rules underlying their design}.

Topological mechanics arose from recognizing that the dynamical equations governing balls-and-springs models admit a Dirac-like `square rooting' connection to fermionic systems \cite{KaneLubensky2013}. The common practice of Maxwell counting in these mechanical systems, the difference between the number of degrees of freedom and the number of constraints, then turns out \cite{KaneLubensky2013, Lawler2016} to be a determination of the Witten index \cite{Witten1982} pointing to an underlying SUSY connection \cite{Attig2019}. Practitioners immediately adopted this discovery, working out many linear theories with free fermion partners \cite{Nash2015gyromaterials, Huber2016, huber_HOTI_phonons, Suesstrunk2015, Rocklin2016, Chen2016topomech, Meeussen2016, susstrunk2016classification, Suesstrunk2017, bahl_HOTI_microwave, Paulose2015, Peri2019, thomale_HOTI_electrical, Attig2019}, complete with topological invariants protecting zero modes in bosonic systems. They get around the absence of topologically protected zero modes in free bosonic systems \cite{gurarie2003bosonic} by using {\sl local constraints}. Topology is then the preservation of zero modes provided the number of constraints does not change. Adding more constraints removes zero modes. They have even shown that this topology protects the zero modes at the non-linear level \cite{lo2021}. So we now have mechanical systems with topological zero modes protected by local constraints.

Practitioners of topological mechanics also envision most of their examples as engineered metamaterials, but locality is a property of physical systems that arises naturally. Frustrated magnets offer a striking example, where residual entropy arises in underconstrained systems. This has been seminally established by Maxwell constraint counting in geometrically frustrated systems such as the classical kagome or pyrochlore antiferromagnets \cite{Moessner1998, Moessner1998b}. While such residual degeneracies in frustrated magnets are commonly referred to as accidental, as there is no apparent symmetry protection, the similarity to concepts in topological mechanics has led some of us to explore their stability in the presence of distortions or disorder \cite{Roychowdhury2018}. What was found is that the robustness of accidental degeneracies can intimately be linked to the preservation of locality; certain types of distortions and disorder do not lift the frustration if the number of local constraints is unaltered. Thus in real kagome antiferromagnets like Cs$_2$ZrCu$_3$F$_{12}$ or Cs$_2$CeCu$_3$F$_{12}$, it is the exponential fall-off of exchange constants away from nearest neighbors that seems to produce topologically protected low energy modes~\cite{Roychowdhury2018}. Stepping back one might be tempted to think of the formation of these accidental degeneracies in frustrated magnets, in analogy to topological mechanics, as the consequence of a hidden SUSY -- a perspective that we will explore in this paper.

If we view topological mechanics as a vision for 'symmetry+topology+locality', what we know so far is that the classification of locally constrained systems is also much richer than the classification {\sl without} locality. The classes in the ten-fold way of electronic band theory \cite{Altland1997, Chiu2016classification} obey a periodicity in the dimension of the system: If a topological invariant exists in dimension $d$, it either exists in dimension $d+8$ for some classes or $d+2$ for others \cite{Schnyder2008classification, Kitaev2009periodic, ryu2010topological, ludwig2015topological}. These invariants are either ${\mathbb Z}$ or ${\mathbb Z}_2$ valued. Such a classification can be extended to {\sl finite frequency} topological modes of bosonic systems~\cite{susstrunk2016classification, Xu2020, zhou2020topological, Wakao2020}, but not to zero-frequency bosonic modes. Classifying the rigidity matrices in topological mechanics has also led to a table of invariants, but these depend on both spatial dimension $d$ {\sl and} the Maxwell counting index $\nu$. While only a three-fold way was discovered, \cite{roychowdhury2018classification, Xu2020}, the periodicity and the invariants of the ten-fold way are observed for $\nu=0$ but not for other values of $\nu$. No periodicity arises in the $\nu\neq 0$ regime, and it includes new invariants such as ${\mathbb Z}_2 \times {\mathbb Z}_2$, ${\mathbb Z}_{12}$, ${\mathbb Z}_{15}$, ${\mathbb Z}_{24}$, and ${\mathbb Z}_{24}\times {\mathbb Z}_{3}$. This broader classification can be understood by realizing the existence of an underlying SUSY where the rigidity matrices act as supercharges -- as we explain in this paper. So, including locality in the structure of topological phases appears to open the door to the discovery and control of new unexpected low energy modes in condensed matter.

In this paper, we develop a unifying SUSY framework that uses locality as the principal ingredient to bridge concepts from topological mechanics to frustrated magnetism to band theory. This framework evolves around a mapping between lattice models of free fermions and bosons -- the most elementary description of condensed matter systems (Section~\ref{sec:SUSY_lattice_models}). The mapping is based on a general graph construction that provides both a visual and algebraic understanding of the underlying SUSY connection. It allows us to explore the relationship between SUSY, topology, and locality for many examples, which include, (i) fundamental connections between some of the most widely studied lattice geometries such as the kagome and honeycomb lattices (discussed in the next section) or the pyrochlore and diamond lattices (Section~\ref{sec:complex_SUSY}), (ii) the construction of mechanical analogs of Kitaev spin liquids (Section~\ref{sec:topo_mechanics}), or (iii) a correspondence between degenerate coplanar spin spiral states and Fermi surfaces (Section~\ref{sec:spin_spirals}). On a conceptual level, we also find the framework allows for a five-fold way classification of topological invariants for non-interacting bosonic systems via their fermionic SUSY partners (Section~\ref{sec:classification} and \ref{subsec:real_supersymmetry}) and through these invariants, points a way to discovery, control, and design of unexpected low energy modes, nexus points, and flat bands in solid state physics.

%%%%%%%%%%%%%%%%%%%%%%%%%%%%%%%%%%%%%%%%%%%%%%%%%%%%%%
% Supersymmetric lattice models
%%%%%%%%%%%%%%%%%%%%%%%%%%%%%%%%%%%%%%%%%%%%%%%%%%%%%%

\section{Supersymmetric lattice models}
\label{sec:SUSY_lattice_models}

We can begin our way towards SUSY by discussing a basic property of block matrices. 
By squaring a Hermitian matrix of the form 
\begin{align}
\label{H1}
{\bf H} = \begin{pmatrix}
      & {\bf R} \\
      {\bf R}^\dagger &
    \end{pmatrix},
\end{align}
with a generic matrix ${\bf R}$ of arbitrary dimensions, one obtains a block-diagonal matrix with two diagonal blocks
\begin{align}
\label{H21}
{\bf H}^2 = \begin{pmatrix}
      {\bf RR^\dagger} & \\
      & {\bf R^\dagger R} 
    \end{pmatrix} \,,
\end{align}
in which the two blocks ${\bf RR^\dagger}$ and ${\bf R^\dagger R}$ are {\it isospectral} except for zero modes which result from a potential dimension mismatch between the kernel of ${\bf R}$  and ${\bf R}^\dagger$ if ${\bf R}$ is not a square matrix. It will be this simple matrix relation upon which we will build our supersymmetric lattice construction in the following.

\begin{figure}[b]
    \centering
    \includegraphics[width=\columnwidth]{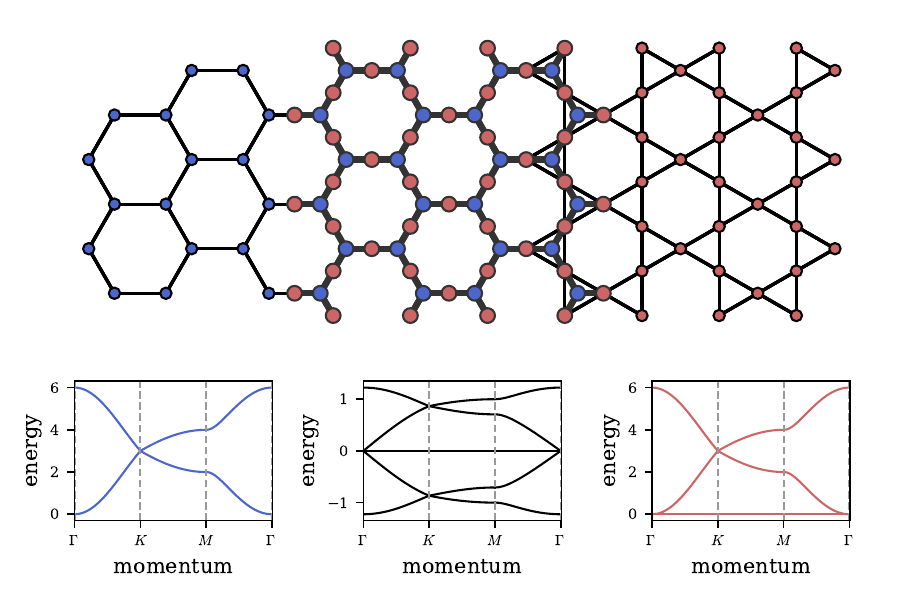}
    \caption{\textbf{SUSY lattice model correspondence.} Shown are three lattice geometries, the honeycomb (left, blue), honeycomb-X (middle, black), and kagome (right, red) lattices and their respective tight-binding band structures in the lower row. The honeycomb and kagome spectra are isospectral up to an additional flat band in the kagome model. The band structure of the honeycomb-X model (middle panel) can be related to the other two via squaring (or, vice versa, taking a square root). As such, the three spectra can be connected via the matrix correspondence \eqref{H21}, where one identifies the honeycomb and kagome lattices as supersymmetric (SUSY) partners and the honeycomb-X lattice with the supercharge. For the topological classification according to Table \ref{table:classificationBDI}, we find, noting that the Witten index here is $\nu = 1$, that the nexus point in the supercharge spectrum has a non-trivial topological invariant given by $\pi_1 = +1$, see also the illustration in Fig.~\ref{fig:homotopy} of the Appendix.}
    \label{fig:susy-kagome-honeycomb}
\end{figure}

To do so, let us consider the instructive example visualized in Fig.~\ref{fig:susy-kagome-honeycomb}, which makes a connection between the familiar honeycomb lattice (on the left, blue) and the kagome lattice (on the right, red). If one calculates the band structures of their respective tight-binding Hamiltonians, \eg by diagonalizing their nearest-neighbor hopping or graph adjacency matrices, one ends up with the two spectra plotted in the left and right panels of the lower row. These spectra are identical -- up to a flat band in the kagome band structure. That is, we find an isospectrality akin to what we have seen for the matrix correspondence \eqref{H21}, which leads us to identify the two blocks ${\bf RR^\dagger}$ and ${\bf R^\dagger R}$ in \eqref{H21} with the tight-binding Hamiltonians of the two lattices at hand. The additional flat band, or zero mode, in the kagome spectrum, can then be traced back to the difference in the dimensions of the two blocks, which is simply the difference in the number of sites (three) in the kagome lattice unit cell versus the two sites of the honeycomb lattice unit cell. That is, we could infer the entire spectrum of the kagome tight-binding model from the one of the honeycomb lattice without any actual calculation.

From a more abstract point of view, which we will lay out in the following, the matrix correspondence~\eqref{H21} allows one to identify a SUSY theory with a pair of Hamiltonians which are mandated to be isospectral and where the additional zero modes arise from a non-trivial Witten index \cite{Witten1982}. In this sense, we have just connected the honeycomb and kagome lattices as supersymmetric partners. One could, for instance, adorn the honeycomb lattice with non-interacting fermions -- the textbook example of the graphene band structure with its Dirac cone, while placing non-interacting bosonic modes on the kagome lattice -- as one routinely considers in the context of studying the frustrated Heisenberg antiferromagnet on this lattice \cite{harris1992possible, zhitomirsky2004exact, zhitomirsky2005high, xing2020selective}, thereby pointing towards a SUSY connection between the two seemingly different worlds of Dirac semi-metals and ground-state manifolds of frustrated magnets.

In the following, we provide a more systematic understanding of the matrix correspondence~\eqref{H21} in terms of SUSY. When this SUSY matrix correspondence is applied to pairs of fermionic and bosonic lattice models, this perspective naturally leads one to discuss the consequences of band structure topology on the fermionic side  (routinely considered, e.g., for electronic spectra) for the bosonic counterpart. On the level of the associated Bloch wavefunctions, we show that this allows one to define, for bosonic systems, a supersymmetric extension of the conventional Berry connection and curvature (or more generally, the quantum metric). Once this formal framework is established, we will recast the matrix squaring of \eqref{H21} and its underlying SUSY relation of lattice models in the general terms of graph theory. This ultimately enables us to make statements on the nature of SUSY that are strikingly pictorial in that they are simple graph substitution rules. 

Our SUSY framework allows us to explicate other prescriptions in the literature of squaring and square-rooting fermionic band structures \cite{Schomerus2017, Xu2020, Ezawa_2020, Mizoguchi2020, Mizoguchi2021, Marques_2021, Marques20212n, dias2021matryoshka}. One important result is that the generator of the SUSY itself can be interpreted as a square-root Hamiltonian derived from the adjacency matrices of certain types of lattice graphs (and as such has a graph representation itself), while the supersymmetric partner Hamiltonians are the squared systems. In our introductory example, it is the honeycomb-X lattice~\footnote{The honeycomb-X lattice is also referred to as decorated honeycomb or heavy-hexagon lattice in some communities.} in the middle of Fig.~\ref{fig:susy-kagome-honeycomb} that corresponds to this SUSY generator. Its spectrum exhibits not only a flat band in the middle of its particle-hole symmetric spectrum (inherited from square-rooting the flat-band kagome spectrum) but also a Dirac cone right at this particle-hole symmetric point (which it inherits from the quadratic band minima of the lowest dispersive bands in the honeycomb/kagome band structures). It is for the observation of such remarkable features, that such square-root band structures have attracted interest in the construction of lattice models for `square-root semimetals' \cite{Mizoguchi2021} or `square-root topological insulators' \cite{Schomerus2017, Xu2020, Ezawa_2020, Mizoguchi2020, Marques_2021}.

%%%%%%%%%%%%%%%%%%%%%%%%%%%%%%%%%%%%%%%%%%%%%%%%%%%%%%
% SUSY 
%%%%%%%%%%%%%%%%%%%%%%%%%%%%%%%%%%%%%%%%%%%%%%%%%%%%%%

\subsection{Supersymmetry}
\label{subsec:supersymmetry}

To set the stage, let us provide a more formal introduction to how SUSY can be used to connect elementary fermionic and bosonic degrees of freedom as well as non-interacting systems of many such degrees of freedom. With an eye towards the topological classification of such non-interacting systems, we then discuss how certain antiunitary symmetries, particularly relevant to the classification of free-fermion systems, transform under SUSY. This allows us to inspect topological invariants and their generalizations in supersymmetric settings.

Let us consider a system consisting of both complex fermionic and bosonic degrees of freedom. The central object that enables a supersymmetric identification between such fermions and bosons is a fermion-odd {\it supercharge} operator
 \begin{equation}
    \mathcal{Q} = c^\dagger{\bf R} b \,,
    \label{eq:ComplexSUSYCharge}
 \end{equation}
where $c^\dagger$ denotes a fermionic creation operator, $b$ a bosonic annihilation operator, and $Q$ satisfies $Q^2=0$. 
With this supercharge operator at hand, one can now generate a SUSY Hamiltonian via
\begin{align}
  \label{eq:complexsusy_H_SUSY}
  \mathcal{H}_{\text{SUSY}} = \{\mathcal{Q} , \mathcal{Q}^\dagger \}
  &= c^\dagger {\bf R R^\dagger} c + b^\dagger {\bf R^\dagger R} b \nonumber \\
 &\equiv 
 \begin{pmatrix} c^\dagger & b^\dagger \end{pmatrix}
 \begin{pmatrix}
      {\cal H}_F & \\
      & {\cal H}_B
 \end{pmatrix} 
 \begin{pmatrix} c \\ b
 \end{pmatrix}\,.
 \end{align}
From this, the two matrices, ${\bf R R^\dagger}$ and ${\bf R^\dagger R}$, which constitute the two diagonal blocks of ${\bf H}^2$ in \eqref{H21}, can be readily identified as a free fermionic and a free bosonic Hamiltonian matrix, respectively.

For a square matrix ${\bf R}$, ${\cal H}_F$ and ${\cal H}_B$ are entirely isospectral (including potential zero modes if any). For a rectangular matrix ${\bf R}$, however, there will always be a mismatch in the number of zero eigenvalues which we can characterize by the index 
\begin{align}\label{eq:witten1}
\nu &= {\rm dim}({\rm kernel}[{\bf R}])-{\rm dim}({\rm kernel}[{\bf R}^\dagger]) \nonumber \\
&= {\rm col}[{\bf R}]-{\rm row}[{\bf R}] \,,
\end{align}
called the Maxwell-Calladine index in topological mechanics \cite{KaneLubensky2013}. In the many-body problem, it is the one-body sector of the Witten index \cite{Witten1982} $\mathrm{Tr}\, (-1)^F\equiv \mathrm{Tr}\,e^{i\pi c^\dagger c}$,  
and so is a topological invariant of a free SUSY theory \footnote{Sutherland might have been the first to identify the mismatch of sublattice sites in a bipartite graph with such an index \cite{sutherland1986localization} (with Shastry being credited for pointing out its relation to the Atiya-Singer index theorem \cite{atiyah1963index}) and later, Lieb associated it with the nullity of the corresponding hopping matrix \cite{Lieb1989}. Due to a mapping between chiral Hamiltonians and supercharges, this Shastry-Sutherland-Lieb index is the single particle Witten index $\nu$ of a corresponding non-interacting SUSY problem (see Fig.~\ref{fig:susymapping}).}. So long as $\nu \neq 0$, this implies SUSY can exist in the ground state. When $\nu=0$, no zero modes can exist on either side. Further, the situation for $\nu \neq 0$ is a definitive indication of flat bands to appear in the band structure of either ${\cal H}_F$ ($\nu<0$) and ${\cal H}_B$ ($\nu>0$). For example, the Witten index is $\nu=1$ for the honeycomb-kagome correspondence highlighted in the introduction which therefore has to give rise to a flat band in the kagome band structure as discussed before. 

%%%%%%%%%%%%%%%%%%%%%%%%%%%%%%%%%%%%%%%%%%%%%%%%%%%%%%
% SUSY from graphs
%%%%%%%%%%%%%%%%%%%%%%%%%%%%%%%%%%%%%%%%%%%%%%%%%%%%%%

\subsection{Lattice graphs}
\label{sec:lattice_graphs}
In more abstract terms, one can identify the hopping matrices of Eq.~\eqref{eq:complexsusy_H_SUSY} with a (weighted) {\sl adjacency matrix} of an underlying graph structure.
For some lattice graph with vertices $\{v_i\}$ an adjacency matrix is defined as 
\begin{equation}
    \mathbf{A}_{ij} = \begin{cases}
        1 \quad v_i \text{ connected to } v_j \\
        0 \quad \text{otherwise} \,.
    \end{cases}
\end{equation}
The {\sl weighted adjacency} matrix extends this definition by including a weight $w_{ij}$ for every non-zero element of $\mathbf{A}_{ij}$.

%%%%%%%%%%%%%%%%%%%%%%%%%%%%%%%%%%%%%%%%%%%%%%%%%%%%%%
\subsubsection*{Squaring graphs}

What happens when we square an adjacency matrix $\mathbf{A}$? To find out, let us compute the elements of the squared matrix 
\begin{equation}
    (\mathbf{A}^2)_{ik} = \sum_j \mathbf{A}_{ij} \mathbf{A}_{jk} \,,
    \label{eq:graph_edge_weights_squaring}
\end{equation}
which we can readily interpret as elements of another adjacency matrix, but with a different set of connections. Where $\mathbf{A}$ connected vertices $v_i$ and $v_j$ with weight $A_{ij}$, $\mathbf{A}^2$ chains two of those connections together to connect next-nearest neighboring vertices $v_i$ and $v_k$. So on a pictorial level, squaring an adjacency matrix is equivalent to singling out next-nearest neighbors of the original graph in a new graph.

\begin{figure}
    \centering
    \includegraphics[width=\columnwidth]{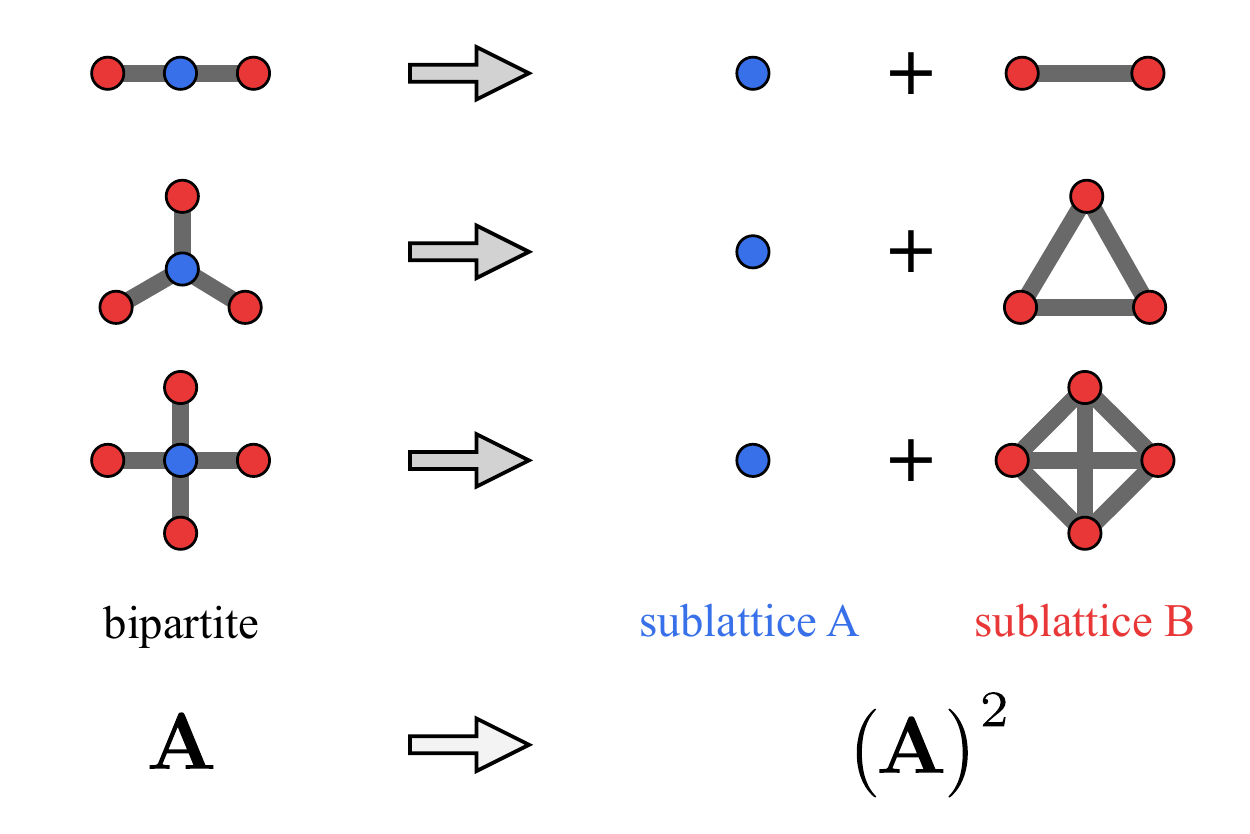}
    \caption{\textbf{Squaring graphs.} By squaring the (weighted) adjacency matrix $\mathbf{A}$ of a given graph, one can arrive at a squared graph by interpreting $\mathbf{A}^2$ as its adjacency matrix. For a given bipartite graph, this squaring gives an off-diagonal block matrix [as in \eqref{H1}], which in its graph theoretical representation is equivalent to a decomposition in its two subgraphs. This process can be reversed to a `graph square rooting' procedure by going from right to left in the figure, which in the graph theoretical representation, amounts to a substitution of graph cliques of size $z$ (\ie fully connected plaquettes with $z$ sites) by $z$-connected sites.}
    \label{fig:lattice-construction-plaquettes}
\end{figure}

These statements can be further refined if $\mathbf{A}$ describes the adjacency of a {\sl bipartite} graph. In this case, the matrix itself can be brought into a two-block structure upon sorting the vertices $\{v_i\}$ of the original bipartite graph into two distinct sets $\{v_i^{\rm I}\}$ and $\{v_i^{\rm II}\}$, encompassing vertices from subgraphs ${\rm I}$ and ${\rm II}$, respectively. The bipartite-ness of the graph then dictates that these two blocks are in fact off-diagonal blocks since vertices in one set (subgraph) are then connected only to vertices in the other set (subgraph) but not to the ones in their own set (subgraph). The adjacency matrix of a bipartite graph thus takes the form
\begin{align}
\label{eq:A_bipartitte}
     {\bf A} = \begin{pmatrix}
      &{\bf A}_{\rm I-II}  \\
      {\bf A}_{\rm II-I}& 
    \end{pmatrix} \,,
\end{align}
where ${\bf A}_{\rm I-II}$ and ${\bf A}_{\rm II-I}$ are the off-diagonal blocks that describe the connections between subgraph ${\rm I}$ and ${\rm II}$ and vice versa. Upon squaring, taking two steps in the graph at a time, the next-nearest neighbors of vertices in one subgraph are then necessarily also in the same subgraph, bringing the squared adjacency matrix into block-diagonal form
\begin{align}
\label{eq:A2}
	{\bf A}^2 = \begin{pmatrix}
      {\bf A}_{\rm I} & \\
      & {\bf A}_{\rm II}
    \end{pmatrix} \,,
\end{align}
where each block describes the coupling within one of the two subgraphs. Therefore, the original bipartite graph {\sl decomposes} upon squaring its adjacency matrix into two distinct subgraphs that are not connected anymore -- a procedure which we have visualized in Fig.~\ref{fig:lattice-construction-plaquettes}.

\begin{table*}[t]
  \begin{tabular}{l||c|c|c|c}
  \hline\hline
  lattice & \multirow{2}{*}{graph} &  \multirow{2}{*}{graph} &  \multirow{2}{*}{graph} &  \multirow{2}{*}{reference} \\
   correspondence& & & &  \\
  \hline \hline
  \multirow{2}{*}{SUSY}  & fermion lattice  & supercharge & boson lattice  & \multirow{2}{*}{this work}\\
  				     & (square)  & (root) & (square)  & \\ \hline
  line graphs & root & split graph & line graph &  Refs.~\cite{Mielke1991,Mielke1992,Miyahara2005}  \\
  Lieb lattices & sublattice & Lieb lattice & sublattice &  Ref.~\cite{Lieb1989} \\
  graph theory & premedial & & medial & Ref.~\cite{tait1877} \\
  \hline \hline
   \multirow{5}{*}{examples} & honeycomb & honeycomb-X & kagome & Fig.~\ref{fig:susy-kagome-honeycomb}\phantom{0} \\
  	 				& square-octagon & square-octagon-X & squagome & Fig.~\ref{fig:susy-squareoctagon-squagome}\phantom{0} \\
  					& fcc & diamond & fcc & Fig.~\ref{fig:fcc-diamond-correspondence} \\
					& diamond & diamond-X & pyrochlore & Fig.~\ref{fig:susy-diamond-pyro} \\
					& hyperoctagon & hyperoctagon-X & hyperkagome & Fig.~\ref{fig:susy-hyperoctagon-hyperkagome} \\
  \hline\hline
  \end{tabular}
  \caption{{\bf Graph correspondences.} 
   		Comparison of our SUSY correspondences and alternate lattice correspondences, including the line graph construction often used in connection with flat-band lattice models, generalizations of the Lieb lattice beyond the square lattice, and the graph theoretical notion of medial and premedial lattices. The lower half of the table lists some representative examples in two and three spatial dimensions with a reference to the triptych-like figures illustrating their respective tight-binding band structures.}
  \label{table:graph_correspondences}
\end{table*}

%%%%%%%%%%%%%%%%%%%%%%%%%%%%%%%%%%%%%%%%%%%%%%%%%%%%%%
\subsubsection*{Graph square roots}

Let's ask whether we can also invert this graph operation -- can we define the square root of an adjacency matrix so that we end up with another adjacency matrix? That is, is there a meaningful way to construct the square root of a given graph?

The algebraic perspective taken above, might not be of immediate help here: If we are given the adjacency matrix $\mathbf{M}$ of some graph, we do not want to simply identify a matrix $\mathbf{A}$ such that $\mathbf{A}^2 = \mathbf{M}$ (or alternatively $\mathbf{A} = \sqrt{\mathbf{M}}$), since this would lead us, in most cases, to a highly-connected graph, which would neither be bipartite nor a typical lattice graph. Instead, the crux is that the matrix $\mathbf{A}$ actually has {\sl enlarged} dimensions with regard to $\mathbf{M}$, which upon squaring takes on an off-diagonal block structure with one of the two blocks becoming equivalent to $\mathbf{M}$.

But the above graph interpretation of the squaring operation points to a way to answer these questions: If squaring a bipartite graph leads to a decomposition into its two subgraphs, one can invert this operation by taking the two subgraphs and declare them to be the two constituent subgraphs of a combined bipartite lattice -- which would then be the `square-root graph' of the two. But if one is given only a single graph how does one find its counterpart graph so that the two can be joined together into a bipartite graph?

This subgraph matching can be facilitated by an algorithm that inverts the graph-theoretical interpretation of graph squaring (see Appendix \ref{app:square_rooting_algorithm} for a detailed description). Using the illustration of Fig.~\ref{fig:lattice-construction-plaquettes}, we see the effect of graph squaring by going from left to right in this figure: Any $z$-coordinated site within a given bipartite graph will result, upon squaring, in a {\sl fully connected plaquette} with $z$ vertices, which in the language of graph theory is also called a {\sl clique}. To do the inverse, \ie to find the bipartite square-root graph for a given graph by constructing its matching subgraph, the algorithm now works in the opposite direction (from right to left): It takes a $z$-clique and replaces it with a $z$-coordinated vertex that connects to all constituents of the prior clique. Performing such replacements in an iterative manner, where one starts with the largest clique and proceeds to smaller cliques in subsequent steps, one is eventually left with the desired matching subgraph and the entire bipartite graph -- the legitimate square-root graph we are looking for. 

%%%%%%%%%%%%%%%%%%%%%%%%%%%%%%%%%%%%%%%%%%%%%%%%%%%%%%
\subsection{Connections to other lattice correspondences}\label{sec:otherlatticecorrespondences}
%%%%%%%%%%%%%%%%%%%%%%%%%%%%%%%%%%%%%%%%%%%%%%%%%%%%%%

SUSY correspondences imply lattice correspondences. They have been used to connect their respective band structures and predict flat bands.  But SUSY does not require any specific rules relating to lattice sites, or that the lattice models are even non-interacting. It only requires fermion degrees of freedom on one sublattice, bosonic on the other, and a parity odd operator defining the supercharge.

\begin{figure}[b]
    \centering
    \includegraphics[width=0.8\columnwidth]{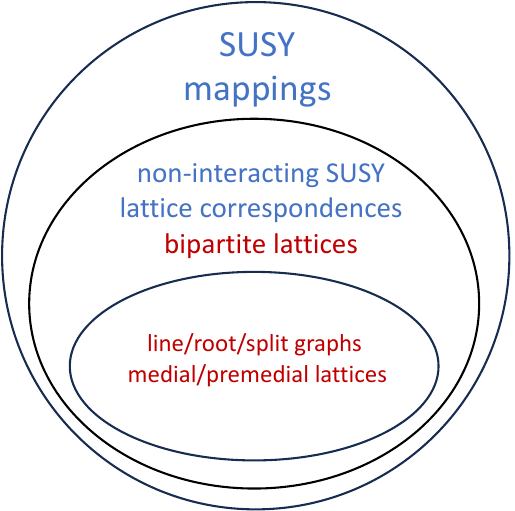}
    \caption{\textbf{Hierarchy of lattice correspondences.} Supercharges define a general relation between bosonic and fermionic systems. Non-interacting supercharges are a subset and define lattice correspondences. They are the same lattice correspondences defined by bipartite lattices \cite{cualuguaru2022general} via the mapping presented in Fig. \ref{fig:susymapping}.  The flat bands found, for example, in Ref. \cite{cualuguaru2022general}, obtained by a generalization of the line graph construction to all bipartite lattices, correspond to the same flat bands found in the supercharge spectra presented in all the triptych figures in this manuscript. A subset of non-interacting SUSY/ bipartite lattice models is the root graphs to line graphs via split graph correspondences, also known as the premedial lattice to medial lattice pairings.
    }
    \label{fig:lattice-correspondances}
\end{figure}

We will argue below that SUSY is more general than the line graph correspondences that connect a root graph to a line graph via a split graph and illustrate this with an example. At the non-interacting limit, we will also argue the bipartite lattice correspondences recently introduced in Ref. \cite{cualuguaru2022general} are equal to SUSY correspondences. Adding to these interacting SUSY lattice models, SUSY, therefore, transcends a hierarchy of lattice correspondences, as shown in Table \ref{table:graph_correspondences} and Fig.~\ref{fig:lattice-correspondances}. 

%%%%%%%%%%%%%%%%%%%%%%%%%%%%%%%%%%%%%%%%%%%%%%%%%%%%%%
\subsubsection{Lieb lattices, split graphs, and bipartite lattices}
%%%%%%%%%%%%%%%%%%%%%%%%%%%%%%%%%%%%%%%%%%%%%%%%%%%%%%

The Lieb lattice, split graphs, and general bipartite lattices are SUSY graph correspondences. The {\sl Lieb lattice} \cite{Lieb1989} is the lattice we have denoted as square-X lattice -- a square lattice where one adds a site to each bond. This bipartite lattice decomposes into two sublattices, the original square lattice and the checkerboard lattice. In the notation of graph theory, the Lieb lattice is considered as a {\sl split graph} \cite{ma2020spin} and it can readily appear as the supercharge $Q = c^\dagger{\bf R} b$ with fermions on the X sites and bosons on the other sites or vice versa, see Fig.~\ref{fig:susy_trio_square_checkerboard}. But we can think of all X lattices considered in the triptych-like figures of Section~\ref{sec:complex_SUSY}, such as the honeycomb-X lattice and hyperhoneycomb-X lattice, as \emph{generalized Lieb lattices}. This generalization identifies Lieb lattices with split graphs.

Recently, Ref. \onlinecite{cualuguaru2022general} generalized split graphs to all bipartite lattices. But a general bipartite lattice also defines a supercharge, for we can place fermions on the A sublattice and bosons on the B sublattice. This mapping between supercharges and bipartite lattices, discussed in more detail in Fig. \ref{fig:susymapping}, is used throughout this manuscript to construct spectra for the supercharge. Hence, bipartite lattices define the same lattice correspondences as non-interacting SUSY models.

%%%%%%%%%%%%%%%%%%%%%%%%%%%%%%%%%%%%%%%%%%%%%%%%%%%%%%
\subsubsection{Our SUSY algorithm and line graph/root graph, medial-lattice/pre-medial-lattice correspondences}
%%%%%%%%%%%%%%%%%%%%%%%%%%%%%%%%%%%%%%%%%%%%%%%%%%%%%%

The SUSY algorithm presented in appendix \ref{app:square_rooting_algorithm}, maps a bosonic lattice to a fermionic lattice and vice-versa and appears similar to the line graph construction. Our SUSY examples and the line graph construction both have tight binding spectra hosting flat bands \cite{chiu2020fragile}. A {\sl line graph} of a given lattice, called the {\sl root graph}, is obtained from connecting the vertices on the center of all bonds. line graphs and root graphs also go by the name of {\sl medial} and {\sl pre-medial} lattices, respectively\cite{tait1877}. Examples include the kagome lattice, which is the line graph of the honeycomb lattice (see Fig.~\ref{fig:susy-kagome-honeycomb}), or the 3D Shastry-Sutherland lattice, which is the line graph of the hyperoctagon lattice (see Fig.~\ref{fig:susy-hyperoctagon-hyperkagome}). In these examples, there is exactly one type of clique in the root graph, a two-site bond, and the graph square-rooting algorithm derives the line graph. Hence, for any graph with this property, the SUSY graph correspondence of section \ref{sec:lattice_graphs} is the line graph correspondence. 

%%%%%%%%%%%%%%%%%%%%%%%%%%%%%%%%%%%%%%%%%%%%%%%%%%%%%%
\subsubsection{SUSY lattice correspondences beyond line graphs}
%%%%%%%%%%%%%%%%%%%%%%%%%%%%%%%%%%%%%%%%%%%%%%%%%%%%%%

\begin{figure}[b]
    \centering
    \includegraphics[width=\columnwidth]{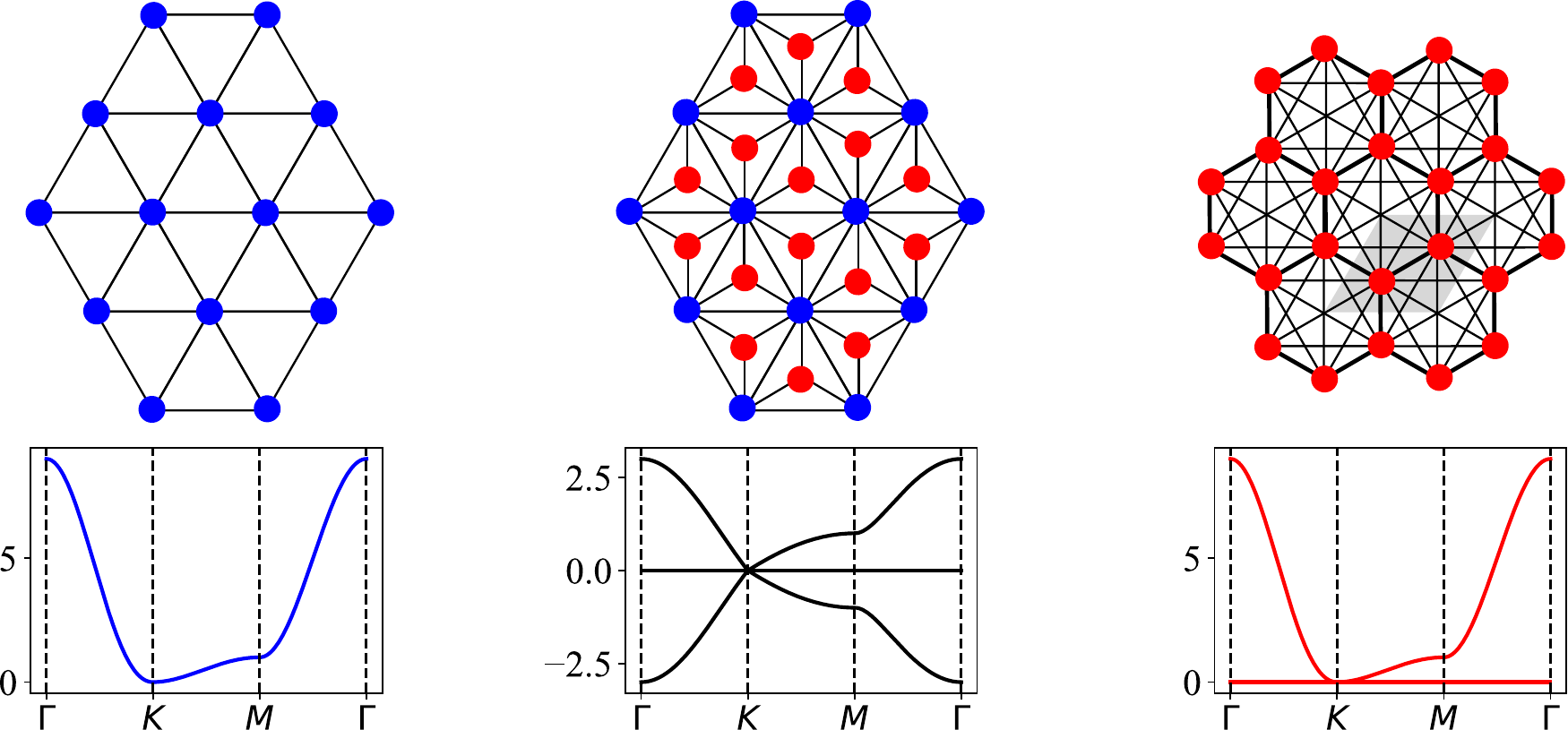}
    \caption{\textbf{SUSY beyond the line graph construction.} Shown is the SUSY correspondence between complex fermions on a triangular lattice with nearest neighbor hopping (blue, left) and complex bosons on a honeycomb lattice with further neighbor hoppings (red, right) that falls beyond the line graph lattice construction. For the topological classification according to Table \ref{table:classificationBDI}, we find, noting that the Witten index here is $\nu = 1$, that the nexus point in the supercharge spectrum has a non-trivial topological invariant of $\pi_1 = +1$.}
    \label{fig:new-susy-beyond-lattice-construct}
\end{figure} 
 
Let us consider examples beyond the line graph construction. All ``isostatic'' cases with an equal number of fermions and bosons, i.e. a Witten index of $\nu=0$, are beyond the line graph construction. An example is presented in Fig. \ref{fig:triangular-honeycomb-correspondence}. These do not demand flat bands but imply other degeneracies such as a Dirac point. 

An example beyond the line graph construction that does imply flat bands is presented in Fig.~\ref{fig:new-susy-beyond-lattice-construct}. The supercharge lattice in the middle, which is a bipartite dice lattice, harbors fermions on the triangular sublattice coupled with bosons at the centers of the triangular plaquettes (of that triangular sublattice). Squaring this supercharge decouples the fermions from the bosons with the former hopping on the triangular lattice and the latter on a honeycomb lattice. However, though the tight-binding Hamiltonian describes the fermionic model with nearest neighbor hoppings only, the bosonic Hamiltonian includes first, second, and third neighbor hoppings on the honeycomb lattice. The appearance of a single flat band on the bosonic side follows from the Witten index of the supercharge being $\nu=1$, or alternatively, by the honeycomb lattice having one more atom in its unit cell compared to the triangular lattice. This example goes beyond the line graph construction because one graph is not the line graph of the other. But, just like the spectra in our line graph examples, it exhibits flat bands and nexus points. More examples of this type are shown in Appendix~\ref{app:MoreSUSYExamples}. 

%%%%%%%%%%%%%%%%%%%%%%%%%%%%%%%%%%%%%%%%%%%%%%%%%%%%%%
\subsubsection{Summary}
%%%%%%%%%%%%%%%%%%%%%%%%%%%%%%%%%%%%%%%%%%%%%%%%%%%%%%

In summary, our SUSY graph correspondence and the graph square-rooting algorithm capture all of the lattice correspondences discussed in this section (section \ref{sec:otherlatticecorrespondences}). The supercharge defines a bipartite lattice and is analogous to the Lieb lattice and split graphs. The mapping between the fermion lattice and boson lattice is analogous to the mapping between the line graph and its root graph (or medial and premedial lattices). 

While the other lattice correspondences have not been explicitly discussed as SUSY correspondences, some authors have used the line graph construction \cite{Mielke1991, Mielke1992, Miyahara2005} and Lieb lattices \cite{Lieb1989} to construct flat-band models. For a pedagogical overview see Refs.~\onlinecite{Flach2018} and \onlinecite{Liu2014}. These examples are all SUSY lattice correspondences.

But SUSY is more general than these mappings, more general even than lattice correspondences. We have presented an example above to emphasize that it goes beyond the line graph construction and can help us search for new flat-band materials. In Appendix \ref{app:MoreSUSYExamples}, we present more examples, aiming to show the breadth of the SUSY lattice correspondence. But the interacting spin model presented in \ref{sec:GeoMagnets} is not merely a lattice correspondence. It has flat bands in the large-$S$ limit but remains supersymmetric for any $S$. Hence, the SUSY discussed in this manuscript is a general framework capable of capturing classes of lattice models hosting robust flat bands or other gapless modes beyond those considered to date in the literature.

%%%%%%%%%%%%%%%%%%%%%%%%%%%%%%%%%%%%%%%%%%%%%%%%%%%%%%
% SUSY as graph correspondence
%%%%%%%%%%%%%%%%%%%%%%%%%%%%%%%%%%%%%%%%%%%%%%%%%%%%%%

\subsection{SUSY as graph correspondence}
\label{sec:susyasgraph}

The two previous sections have presented two ways of arriving at a block-diagonal  Hamiltonian of the form of \eqref{H21} -- first, by squaring the supersymmetric charge operator to arrive at the block-diagonal Hamiltonian $\mathcal{H}_{\text{SUSY}}$ of \eqref{eq:complexsusy_H_SUSY} and, second, by squaring the adjacency matrix of a bipartite lattice, ${\bf A}^2$, of \eqref{eq:A2}.
Equating them all 
\begin{align}
\label{eq:SUSY_square}
\begin{pmatrix}
      {\bf RR^\dagger} & \\
      & {\bf R^\dagger R} 
    \end{pmatrix}
  = 
  \begin{pmatrix}
      {\cal H}_F & \\
      & {\cal H}_B
    \end{pmatrix}
  = 
  \begin{pmatrix}
      {\bf A}_{\rm I} & \\
      & {\bf A}_{\rm II}
    \end{pmatrix} 
    \,,
\end{align}
brings us to the essence of the framework that we develop in this manuscript: 
We can identify the two sublattices of a bipartite graph (on the right) as fermionic and bosonic partners in a SUSY theory (middle), whose tight-binding models must be isospectral, up to zero modes (on the left).  
Taking the square root of \eqref{eq:SUSY_square} gives
\begin{align}
\label{eq:SUSY_squareroot}
   \begin{pmatrix}
      & {\bf R} \\
      {\bf R}^\dagger &
    \end{pmatrix}  
  = 
   \begin{pmatrix}
      &\mathcal{Q} \\
      \mathcal{Q}^\dagger &
    \end{pmatrix}  
  = 
 \begin{pmatrix}
      &{\bf A}_{\rm I-II}  \\
      {\bf A}_{\rm II-I}& 
    \end{pmatrix}
    \,,
\end{align}
which lets us identify the bipartite lattice geometry, given by the adjacency matrix {\bf A} (right), with the supercharge (middle),
and whose tight-binding spectrum must be the square root of the tight-binding spectra of its two subgraphs, up to zero modes (left).

To illustrate these statements, we can go back to our initial example of the SUSY correspondence of Fig.~\ref{fig:susy-kagome-honeycomb}:  The honeycomb and kagome lattices are, in this framework, supersymmetric partners which are connected via the honeycomb-X lattice -- their parent bipartite lattice or, equivalently, the supercharge. The energy spectra of the three lattices are indeed connected to one another as described above; the honeycomb and kagome tight-binding Hamiltonians are isospectral up to a zero mode on the kagome side, while the spectrum of the honeycomb-X lattice is indeed the square root spectrum of the other two. It has symmetric positive and negative energy branches, or particle-hole symmetry in the parlance of condensed matter physics, corresponding to the positive/negative square roots. The zero mode of the kagome lattice survives as a mid-spectrum flat band in the honeycomb-X spectrum. The quadratic band minima at the $\Gamma$-point of the honeycomb and kagome energy spectra become linear modes forming a Dirac cone at the $\Gamma$-point in the honeycomb-X spectrum. 
In fact the two latter observations -- particle-hole symmetry and (higher-order) linear band crossings -- are generic features of the energy spectra of bipartite lattices, whose origin can be naturally explained with our SUSY framework. 
  
\begin{figure}[t]
    \centering
    \includegraphics[width=0.8\columnwidth]{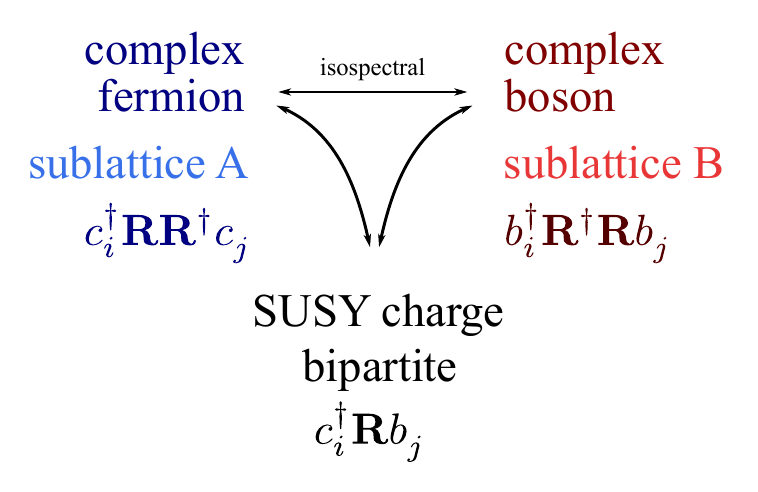}
    \caption{\textbf{SUSY matching complex bosons and fermions} -  Schematic relation between complex bosons and fermions which can be connected by a supercharge defined on a bipartite lattice. In this case, both models are residing on its respective A and B sublattices. }
    \label{fig:susy-techniques-complex}
\end{figure}

We can also put our SUSY graph correspondence to work in a constructive manner. One natural way would be to start from a bipartite lattice, consider it to be a supercharge, and then identify two isospectral tight-binding models by decomposing it into its two sublattices. In the conceptual summary of our SUSY graph correspondence in Fig.~\ref{fig:susy-techniques-complex} this corresponds to a start at the bottom and then working our way up.
However, in practical settings, one might be more interested in producing the supersymmetric partner
for a given tight-binding model, \eg by starting on the top left of Fig.~\ref{fig:susy-techniques-complex} with a bosonic tight-binding model and asking whether it has a fermionic counterpart (or vice versa starting with a fermionic model and asking whether it has a bosonic counterpart).
To construct such a SUSY partner, our graph square-root algorithm 
(which inverts the graph squaring of Fig.~\ref{fig:lattice-construction-plaquettes} as detailed in Appendix~\ref{app:square_rooting_algorithm}) 
comes into play -- it allows one to simultaneously construct both the supersymmetric lattice partners as well as the square-root bipartite lattice that decomposes into the two sublattices. If, for instance, one starts with the kagome lattice, one would readily identify the honeycomb lattice as its SUSY partner. Multiple other examples will follow in the next section making connections, e.g., between the square-octagon and squagome lattices (Fig.~\ref{fig:susy-squareoctagon-squagome}) 
or, in three spatial dimensions, the diamond and pyrochlore lattices (Fig.~\ref{fig:susy-diamond-pyro}) 
or the hyperoctagon and hyperkagome lattices (Fig.~\ref{fig:susy-hyperoctagon-hyperkagome}).

All in all, the consequences of identifying SUSY with a graph correspondence seem quite substantial. The following parts of this manuscript are devoted to corroborating this by numerous examples in which the graph language greatly benefits the analysis and contextualization of various bosonic and fermionic lattice models. 

%%%%%%%%%%%%%%%%%%%%%%%%%%%%%%%%%%%%%%%%%%%%%%%%%%%%%%

\subsection{Symmetry, Supersymmetry, and Topology}
\label{sec:classification}

To complete our general discussion of SUSY-related bosonic and fermionic lattice models, we want to expand the underlying SUSY formalism to also reflect on Hamiltonian symmetries, band structure topology, and general classification of the connected models.

To start this discussion, it is important to revisit the supercharge operator in \eqref{eq:ComplexSUSYCharge}. It not only generates a pair of isospectral fermionic and bosonic Hamiltonians, it is itself associated with a third Hermitian operator \footnote{{We could also construct another Hermitian charge $-i(Q-Q^\dagger)$ if we need to study the full ${\cal N}=2$ SUSY algebra.}}
\begin{align}
  \label{eq:complexsusy_charge}
  \mathcal{Q}_{\rm H} = \mathcal{Q} + \mathcal{Q}^\dagger
  &= c^\dagger {\bf R} b + b^\dagger {\bf R}^\dagger c \nonumber \\
 &\equiv 
 \begin{pmatrix} c^\dagger & b^\dagger \end{pmatrix}
 \begin{pmatrix}
       & {\bf R}\\
      {\bf R}^\dagger & 
 \end{pmatrix} 
 \begin{pmatrix} c \\ b
 \end{pmatrix}\,.
 \end{align}
This operator is an arbitrary Hermitian quadratic form that anticommutes with fermion parity $e^{i\pi c^\dagger c}$.  Its eigenspectrum is presented in the middle panel of the triptych-like figures, such as Fig.~\ref{fig:susy-kagome-honeycomb}. For more details, see also the eigenstate mapping in Fig.~\ref{fig:susymapping} of the Appendix.
Any deformation of the model parameters that preserves SUSY will map ${\cal Q}_{\rm H}$ to some other ${\cal Q}_{\rm H}'$ and so we can define these SUSY preserving deformations as simply a deformation of ${\cal Q}_{\rm H}$ itself. The fermionic and bosonic Hamiltonians, on the other hand, are not generally deformable under SUSY preserving deformations for they must maintain {\sl nonnegative} eigenvalues. Hence, $\mathcal{Q}_{\rm H}$ defines the topological classification of quadratic SUSY problems. 

To classify ${\cal Q}_{\rm H}$, we will begin by identifying the symmetries of ${\cal H}_F$ and ${\cal H}_B$ that are important for the topological classification of fermionic problems. Then we will see how these symmetries map under SUSY so that knowing a symmetry of ${\cal H}_F$ aids in determining its form for ${\cal H}_B$ or vice versa. Finally, this understanding of symmetry will lead us to the classification of supersymmetric systems via $\mathcal{Q}_{\rm H}$.

It has become widely appreciated that one can use electronic band structure calculations to readily deduce topological properties \cite{bhbook}. Doing so rests on the fact that for {\sl fermionic} systems, topological invariants are intimately connected to certain unitary and antiunitary symmetries of the single particle Hamiltonian matrix \cite{Chiu2016classification}.

In the case of a fermionic Hamiltonian, the action of the following set of symmetries is of pivotal importance: time-reversal symmetry ($\cal T$),  particle-hole symmetry ($\cal P$), and chiral/sublattice symmetry ($\cal C={\cal T}\cdot{\cal P}$). $\cal T$ and $\cal P$ are anti-unitary symmetries that commute ($\cal T$) or anticommute ($\cal P$) with the single particle Hamiltonian matrix, while $\cal C$ is an anticommuting unitary symmetry. They square to the possible values of ${\cal T}^2=\pm1, {\cal P}^2=\pm1, {\cal C}^2=0,1$. Combining these symmetries leads to ten topologically distinct classes of Hamiltonians describing non-interacting fermionic systems \cite{Altland1997}. 

\begin{figure}[t]
	\centering
	\includegraphics[width=1.0\columnwidth]{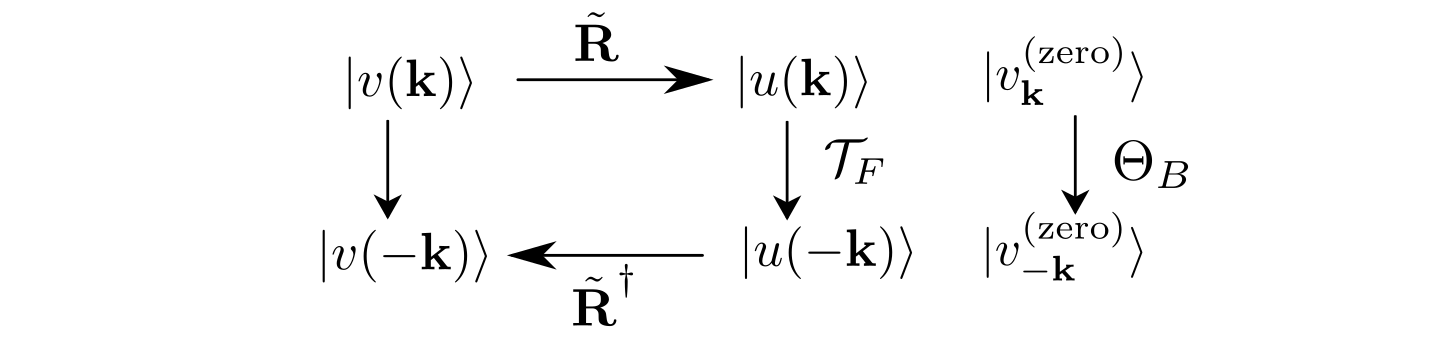}
	\caption{{\bf Time reversal symmetry under SUSY map.} The category to define the bosonic time-reversal within the space of the finite-energy bosonic states $|v({\bf k})\rangle$ by combining the SUSY mapping to their fermionic partner states $|u({\bf k})\rangle$ and the time-reversal in the space of the fermionic states, ${\cal T}_F$. Another time-reversal operator $\Theta_B$ acts within the manifold of the zero modes $|v^{({\rm zero})}_{\bf k}\rangle$ separated by the projector ${\cal P}_{\rm fb}$. Together, they define the full bosonic time-reversal ${\cal T}_B$ in \eqref{eq:timerevbos1} .}
	\label{fig:susyT}
\end{figure}

Consider first the time-reversal symmetry for fermions and its mapping to bosons. Working in Fourier space, the fermionic and bosonic eigenstates of ${\cal H}_F$ and ${\cal H}_B$, the Bloch wavefunctions $|u({\bf k})\rangle$ and $|v({\bf k})\rangle$, are related as
\begin{align}
	\label{eq:Berry3}
 	|u({\bf k})\rangle =
 	\frac{{\bf R}({\bf k})}{\sqrt{\omega({\bf k})}}|v({\bf k})\rangle \quad {\rm and} \quad
 	|v({\bf k})\rangle =
 	\frac{{\bf R}^\dagger({\bf k})}{\sqrt{\omega({\bf k})}}|u({\bf k})\rangle \,.
\end{align}
This mapping is then associated with the operator
\begin{align}\label{eq:normpreservingmap}
    \tilde{\bf R}({\bf k})\equiv \frac{{\bf R}({\bf k})}{\sqrt{\omega({\bf k})}}\,, 
\end{align}
which defines a norm-preserving map between the finite energy eigenstates of the concerned Hilbert spaces. But by construction, this excludes the zero modes of the energy spectra $\omega({\bf k})$ of the two isospectral Hamiltonians ${\cal H}_F$ and ${\cal H}_B$. If the flat bands arise when $\nu> 0$, {\ie} when ${\bf R}$ is a rectangular matrix and annihilates the flat band states, then 
\begin{align}\label{eq:susyflatband}
 {\bf R}({\bf k})~|v_{\bf k}^{\rm (zero)}\rangle=0 \,,
\end{align}
when $|v_{\bf k}^{\rm (zero)}\rangle$ is a state in the flat band manifold. Let us assume ${\cal P}_{\rm fb}$ is a projector onto this manifold. The full time-reversal operator spanning the entire bosonic Hilbert space (including the flat bands) then is related to the fermionic time reversal operator via (see Fig.~\ref{fig:susyT})
\begin{align}\label{eq:timerevbos1}
 {\cal T}_B({\bf k}) &= [1-{\cal P}_{\rm fb}({\bf k})]{\bf R}^\dagger(-{\bf k}){\cal T}_F({\bf k}){\bf R}({\bf k})[1-{\cal P}_{\rm fb}({\bf k})]  \nonumber \\ 
 &~~~~~~~~~~~~~~~~~~~~~~~~~~~~~~~+ {\cal P}_{\rm fb}(-{\bf k})\Theta_B({\bf k}){\cal P}_{\rm fb}({\bf k}) \,,
\end{align} 
where $\Theta_B$ is a bosonic time-reversal operator that operates only within the flat band manifold ({\eg} it can be $\Theta_B=U{\cal K}$ where $U$ is a unitary operator and ${\cal K}$ is the complex conjugation, both restricted to the flat band manifold). Therefore, the matrix form of ${\cal T}_B$ is block-diagonal
\begin{align}\label{eq:timerevbos2}
 {\cal T}_B =
 \left[\begin{array}{c|c} 
   {\bf R}^\dagger{\cal T}_F{\bf R} &  \\ 
   \hline 
    & \Theta_B
\end{array}\right],
\end{align}
where the upper block, corresponding to all finite energy states, is separated from the lower one, which consists of the zero modes (flat bands), by the projector ${\cal P}_{\rm fb}$. The time-reversal operator in \eqref{eq:timerevbos1} satisfies 
${\cal T}_B(-{\bf k}){\cal T}_B({\bf k})=1$. 
If $\nu\leq0$, the flat bands either do not arise or they arise on the fermionic side. Then the mapping is simpler
\begin{align}\label{eq:timerevbos3}
 {\cal T}_B({\bf k}) &= {\bf R}^\dagger(-{\bf k}){\cal T}_F({\bf k}){\bf R}({\bf k})\,.
\end{align} 
As a result, we generally expect the time-reversal symmetry to map between the fermions and bosons. 

Such a mapping cannot be constructed for the bosonic particle-hole operator ${\cal P}_B$ when $\nu>0$. This is because, when restricted to the flat band manifold which maps onto itself under particle-hole conjugation, it reduces to the time-reversal operator ${\cal T}_B$ only. As a result, such identification in this case also fails to apply for the bosonic chiral symmetry operator ${\cal C}_B={\cal T}_B\cdot{\cal P}_B$ which is trivial in the presence of the flat bands. These situations are easily established when the SUSY for $\nu\neq 0$ identifies a bipartite fermionic system (\eg on the honeycomb lattice) with a non-bipartite bosonic system (\eg on the kagome lattice). The absence of a mapping between the fermions and bosons for ${\cal P}$ and ${\cal C}$ is then directly associated with the loss of the bipartite property of the lattice. 

With both the symmetries and the Hermitian operator $\mathcal{Q}_{\rm H}$ identified, we are now in a position to classify our SUSY models. First, consider the case with no symmetry, just supersymmetry. We see that fermion parity $e^{i\pi c^\dagger c}$ is an anticommuting unitary operator which acts at the single particle level as the anticommuting unitary matrix 
\begin{align}
    {\cal C}_{\rm SUSY} = \begin{pmatrix} 
    {\bf I} & \\ & -{\bf I}
    \end{pmatrix}.
\end{align}
So all SUSY problems are chiral. Then adding the additional symmetries $\mathcal{T}$, $\mathcal{P}$ and $\mathcal{C}$ that act on both fermions and bosons as discussed above we see that we cannot additionally add a $\mathcal{C}$ since a chiral symmetry is already present. And, if we add $\mathcal{T}$, we automatically obtain a $\mathcal{P}$ for $\mathcal{C}=\mathcal{T}\cdot \mathcal{P}$. As a result, we avoid the difficulty with deriving ${\cal P}_B$ from ${\cal P}_F$ pointed out above. 
The bipartite-ness of $\mathcal{Q}_{\rm H}$ naturally enables a $\mathcal{P}$ symmetry. Hence, in classifying the SUSY problems, either we do not have $\cal{T}$ and $\cal{P}$ or we have them both with one derivable from the other. 

With that, we arrive at a {\sl five-fold way} classification of SUSY models characterized by the absence of $\mathcal{T}_{\rm SUSY}$/$\mathcal{P}_{\rm SUSY}$ (class AIII) and the four classes with $\mathcal{T}_{\rm SUSY}$, $\mathcal{P}_{\rm SUSY}$ having $\mathcal{T}_{\rm SUSY}^2=\pm1$, $\mathcal{P}_{\rm SUSY}^2=\pm1$ (classes BDI, CI, CII, DIII). Previously, two of us classified the related problem of rigidity matrices \cite{roychowdhury2018classification, Xu2020} and found a three-fold classification (classes AIII, BDI, and CII). So by classifying ${\cal Q}_{\rm H}$, we have now found two previously unknown classes of SUSY Hamiltonians.

To construct a classification table, we need to identify the topology of the classifying space of $\mathcal{Q}_{\rm H}$ in each of the five symmetry classes. Following the classification of rigidity matrices \cite{roychowdhury2018classification}, we can arrive at this by carrying out a singular value decomposition of a generic rectangular ${\bf R}$ matrix of dimension $M\times N$ as ${\bf R} = {\bf U}{\boldsymbol \Sigma}{\bf V}^\dagger$ ($\bf U$ and $\bf V$ are unitary matrices of dimension $M\times M$ and $N\times N$ respectively) and smoothly flatten the singular values ${\boldsymbol\Sigma}\to{\bf I}_{M\times N}$. The resulting flattened matrix then lives in a potentially non-trivial topological space defined by the gapping condition that no singular values vanish. The effect of this transformation on $\mathcal{Q}_{\rm H}$ is to place it in the form
\begin{equation}\label{flattenedsusy1}
    \mathcal{Q}_{\rm H}\to 
    \begin{pmatrix} {\bf U} &\\ & {\bf V}\end{pmatrix}
    \begin{pmatrix} & {\bf I}_{M\times N}\\ {\bf I}_{N\times M} & \end{pmatrix}
    \begin{pmatrix} {\bf U}^\dagger & \\ & {\bf V}^\dagger\end{pmatrix}.
\end{equation}
Inserting $e^{i(\pi/4)\boldsymbol\sigma_y}$, we can rotate this expression into an eigenvalue decomposition
\begin{equation}
\mathcal{Q}_{\rm H}\to 
    {\bf P}
    \begin{pmatrix} {\bf I}_{M\times M} & & \\ & -{\bf I}_{M\times M} & \\
    & & {\bf 0}_{(N-M)\times (N-M)}
    \end{pmatrix}
    {\bf P}^\dagger,
\end{equation}
with
\begin{align}
    {\bf P}=
    \begin{pmatrix} {\bf U} &\\ & {\bf V}\end{pmatrix}e^{i(\pi/4)\boldsymbol\sigma_y},
\end{align}
where, without loss of generality, we have considered $M<N$. For the case $\nu=0$, where ${\bf R}$ is a square matrix with $M=N$, the singular value gapping condition is identical to the eigenvalue gapping condition for all five classes AIII, BDI, CI, CII, and DIII. For $\nu\neq 0$, however, we arrive at a space of Hermitian matrices where some eigenvalues are forced to vanish by SUSY. These eigenvalues appear as protected flat bands in SUSY band structure calculations. The gapping condition now corresponds to a pair of (positive and negative) eigenvalues vanishing to create additional zero eigenvalues. An example in band structures is a {\sl nexus point} \cite{Heikkil2015, Bradlyn2016, chang2017nexus, das2020topological}: a point where multiple energy bands merge in a three- or higher-fold degenerate fashion (including, in particular, a possible combination with flat bands, which will be the case for most of our examples). Hence, by flattening the eigenvalues of  $\mathcal{Q}_{\rm H}$, we map it onto certain spaces of matrices that can have non-trivial topology.

\begin{table}[b]
  \begin{tabular}{c|ccc||ccc}
  \hline\hline
  \multicolumn{4}{c||}{BDI} & \multicolumn{3}{c}{Figures with examples} \\
  $\quad \nu \quad$ & $\quad \pi_1 \quad$& $\quad \pi_2 \quad$ & $\quad \pi_3 \quad$
    & $\quad \pi_1 \quad$& $\quad \pi_2 \quad$ & $\quad \pi_3 \quad$ \\
  \hline
  1 & $\mathbb{Z}_2$ 	& 0 				& $\mathbb{Z}$ & (\ref{fig:susy-kagome-honeycomb},\ref{fig:susy_trio_square_checkerboard}) & -- & -- \\
  2 & 0 				& $\mathbb{Z}$ 	& $\mathbb{Z}$ & -- & (\ref{fig:susy-hyperhoneycomb},\ref{fig:susy-diamond-pyro},\ref{fig:susy-hyperoctagon-hyperkagome}) & (\ref{fig:susy-hyperhoneycomb},\ref{fig:susy-diamond-pyro},\ref{fig:susy-hyperoctagon-hyperkagome}) \\
  3 & 0 				& 0 				& $\mathbb{Z}$ & -- & -- & -- \\
  4 & 0 				& 0 				& 0  			 & -- & -- & -- \\
  \hline\hline
  \end{tabular}
  \caption{{\bf Topological classification of SUSY Hamiltonians with finite Witten index.} The table (on the left) indicates topological invariants ($\mathbb{Z}_2$, $\mathbb{Z}$) as a function of Witten index $\nu$. It is organized not by the spatial dimension used in the ten-fold way table \cite{Schnyder2008classification, Kitaev2009periodic, ryu2010topological, ludwig2015topological} but by \emph{homotopy groups} $\pi_n$. Mathematically, the latter corresponds to maps from $n$-dimensional sphere ${\cal S}_n$ to the flattened supercharge in \eqref{flattenedsusy1}. In physical terms, $\pi_1$ is also known as the Berry phase and $\pi_2$ as the Chern number, which we here associate with features (such as nexus points) of the zero-energy band of the supercharge band structure (depicted in the middle panel of the triptych-like figures of Section~\ref{sec:complex_SUSY}). Shown are results for symmetry class BDI \cite{roychowdhury2018classification}; four more symmetry classes are discussed in Appendix~\ref{app:classification} leading to a {\sl five-fold way} classification scheme, fully tabulated in Table \ref{table:classification33} of the Appendix. On the right, we tabulate example systems (illustrated in the respectively linked figures) with non-trivial topological index, including the supercharge on the following lattice geometries: honeycomb-X (Fig.~\ref{fig:susy-kagome-honeycomb}), square-X/Lieb lattice (Fig.~\ref{fig:susy_trio_square_checkerboard}), hyperhoneycomb-X (Fig.~\ref{fig:susy-hyperhoneycomb}), diamond-X (Fig.~\ref{fig:susy-diamond-pyro}), and hyperoctagon-X (Fig.~\ref{fig:susy-hyperoctagon-hyperkagome}). The non-trivial homotopy groups associated with certain features in their respective band structure are illustrated in Fig.~\ref{fig:homotopy} of the Appendix.
  }
  \label{table:classificationBDI}
\end{table}

The final steps are to identify the topology of the classifying spaces and to compute topological invariants to identify protected zero modes. We carry these steps out in Appendix \ref{app:classification} where we present complete example calculations along with tables of homotopy groups associated with each class, dimension, and Witten index $\nu$. It turns out that all of the examples in the next section, Section~\ref{sec:complex_SUSY}, are in the BDI symmetry class. To highlight their potential topological zero modes, we therefore present in Table \ref{table:classificationBDI} the BDI table produced in Appendix \ref{app:classification} and the figures that constitute the associated examples.

In summary, the SUSY band structures that fit into the formalism of this paper fall into the five-fold classification of chiral Hamiltonians. For the case $\nu=0$, we can resort to the ten-fold way to classify $\mathcal{Q}_{\rm H}$ while for $\nu\neq0$ we can rely on Ref.~\cite{roychowdhury2018classification} for the class AIII (unitary), BDI (orthogonal) and CII (symplectic). Building on these references, Appendix \ref{app:classification} presents the five tables classifying the SUSY band structures. 

%%%%%%%%%%%%%%%%%%%%%%%%%%%%%%%%%%%%%%%%%%%%%%%%%%%%%%
% Examples
%%%%%%%%%%%%%%%%%%%%%%%%%%%%%%%%%%%%%%%%%%%%%%%%%%%%%%

\section{Frustrated magnets}
\label{sec:complex_SUSY}

In putting our SUSY correspondence to work, let us turn to the phenomenology of frustrated magnets as one realm to highlight the conceptual insights one might quickly derive in connecting them to SUSY-related free-fermion systems. One such insight relates to the Maxwell counting for geometrically frustrated magnets, which we will discuss in the language
of our SUSY correspondence in Section~\ref{sec:GeoMagnets}. Another insight is that SUSY allows for a classification of extensive ground-state manifolds in classical spin models, which we will subsequently turn to in subsection \ref{sec:spin_spirals}. Our SUSY correspondence can also be employed to predict magnon dispersions for certain frustrated magnets in large magnetic fields, which we will discuss in subsection \ref{sec:magnon_dispersions}, and parton dispersions for certain quantum spin liquids, in subsection \ref{sec:parton_dispersions}.

%%%%%%%%%%%%%%%%%%%%%%%%%%%%%%%%%%%%%%%%%%%%%%%%%%%%%%
% Geometrically Frustrated Magnets
%%%%%%%%%%%%%%%%%%%%%%%%%%%%%%%%%%%%%%%%%%%%%%%%%%%%%%
\subsection{Maxwell counting for geometrically frustrated magnets}
\label{sec:GeoMagnets}     

The first case study of our SUSY formalism in the context of frustrated magnetism concerns the special class of geometrically frustrated magnets that can satisfy a total spin constraint on each {\sl simplex} (or fully connected plaquette in the parlance of the current manuscript) of the lattice \cite{Moessner1998, Moessner1998b}. Common examples include kagome and pyrochlore Heisenberg antiferromagnets where these simplices correspond to triangles and tetrahedra, respectively. But less common examples are also possible such as distorted kagome antiferromagnets  \cite{Roychowdhury2018} and even the square lattice Neél antiferromagnet since a two-site bond can also be viewed as a simplex \cite{Lawler2016} or a fully connected plaquette, as indicated in Fig.~\ref{fig:lattice-construction-plaquettes}.  To see the presence of these spin constraints, we need to write the spin model Hamiltonian as a {\sl perfect square}
\begin{equation}
	\label{eq:geospinmodel}
    H = \sum_{\langle ij\rangle}J_{ij}{\bf S}_i\cdot{\bf S_j} + \text{const.} = \frac{1}{2}\sum_\Delta\bigg[\sum_{i\in\Delta} a_i{\bf S}_i\bigg]^2 \,,
\end{equation}
where $\Delta$ denotes the simplex of the lattice, we keep spin rotation invariance for simplicity and generalize the total spin constraint to include the $a_i$ factors.  Note the similarity of this perfect square formulation to a balls-and-springs model with potential energy $\frac{1}{2}k e_m^2$, $e_m$ the extension of spring $m$, such as those discussed below in section \ref{sec:topo_mechanics}. This special form of the Hamiltonian enables a SUSY correspondence \cite{Lawler2016}, which we will cast in our general framework in the following.

We can identify both a quantum and a classical SUSY model from the perfect square Hamiltonian \eqref{eq:geospinmodel}. 
In the quantum case, the individual terms of the Hamiltonian do not generally commute with one another and therefore 
cannot be simultaneously satisfied in the ground state. In the classical case, however, each term in the Hamiltonian can be simultaneously satisfied, thereby defining a set of {\em Maxwell constraints}. Let us, in the following, first visit the quantum case and the behavior of magnon excitations of an ordered state and then turn to the classical case to illustrate the role of SUSY and the relation between the Witten index and Maxwell counting in geometrically frustrated magnets.

\subsubsection*{Quantum antiferromagnets}
In the quantum case, we can define a supersymmetric charge from the total spin on a simplex by associating a fermion with each component of this spin in the large-$S$ limit \cite{Lawler2016}. For simplicity, we will study a pure XY quantum model, similar to those studied for their connections to gauge theories \cite{Balents2002} or deconfined quantum criticality \cite{Dang2011}. For such models, we can construct the supercharge
\begin{equation}
	\label{eq:susychargegeospin}
    {\cal Q} = \sqrt{\frac{J}{2}}c^\dagger_{\Delta} S^-_{\Delta} \,,
\end{equation}
where $S^-_{\Delta} = \sum_{j\in \Delta} ( S_{jx}+iS_{jy})$. It has a U(1) symmetry in which spin rotations around the $z$-axis, $S_\Delta^-\to e^{i\theta}S_\Delta^-$, are absorbed into a phase change of the fermions $c_\Delta^\dagger \to e^{-i\theta}c_\Delta^\dagger$, and leads to the SUSY Hamiltonian $\frac{1}{2}\{{\cal Q},{\cal Q}^\dagger\}$
\begin{equation}
    H_{\rm SUSY}\! =\! \frac{J}{2}\!\sum_{\Delta}\!\left({\bf S}_\Delta^2\! +\! (1\!-\!2c^\dagger_\Delta c^{ }_\Delta)S^z_{\Delta}\right) - J\!\!\!\sum_{\langle\Delta\Delta'\rangle}c^\dagger_\Delta S^z_{\Delta\Delta'}c_{\Delta'} \,,
    \label{eq:InteractingSUSY}
\end{equation}
where the first term is the perfect square Hamiltonian of \eqref{eq:geospinmodel} with $a_i=1$, ${\bf S}_\Delta$ the total $xy$-spin of simplex $\Delta$, $S^z_\Delta$ the total z-component of simplex $\Delta$, and $S^z_{\Delta\Delta'}$ the $z$-component of the spin on the site shared by neighboring simplices $\Delta$ and $\Delta'$. In this way, we arrive at an {\sl interacting SUSY problem} where fermions and bosons know about each other's existence. Since the spin model Hamiltonian of \eqref{eq:geospinmodel} does not involve spins interacting with fermions, this new model is not directly related to the original one. But we will see that in the large-$S$ limit, the fermions and bosons {\em decouple} and, at the quadratic level, the magnons of a geometrically frustrated magnet of the kind we are discussing here will have a fermionic SUSY partner.  

As the avid reader might have already noticed, the formulation in terms of an effective total spin on a simplex bears some similarity to the lattice construction algorithm of Fig.~\ref{fig:lattice-construction-plaquettes} (and outlined in Appendix~\ref{app:square_rooting_algorithm}) as it groups edges of the lattice in terms of fully connected plaquettes. Building a supercharge by combining such a fully connected plaquette of the original lattice with a new particle in its center, a fermion in this case, is graphically equivalent to introducing a new vertex in the center of a clique and connecting it with all existing vertices of this clique. For instance, in the case of an XY model on the kagome lattice, the $c$ and $c^\dagger$ are placed on the honeycomb lattice, formed by the center of the triangles of the kagome lattice -- or, in the parlance of this manuscript, the SUSY partner of the kagome lattice.

To decouple the fermions and bosons, in a subsequent step, we expand around a ground state of the magnetic system. We do so by expressing the on-site spin operators $S^{\pm}_j=S^x_j {\pm} i S^y_j$ and $S^z_j$ in terms of Holstein-Primakoff \cite{holstein1940field} bosonic (magnon) annihilation and creation operators $b_j$ and $b_j^\dagger$ as 
\begin{align}\label{eq:holsteinpremakoff}
   S^+_j=\sqrt{2S-n_j}~b_j~;~S^-_j=b^\dagger_j\sqrt{2S-n_j}~;~S^z_j=S-n_j, 
\end{align}
where $n_j=b_j^\dagger b_j^{\phantom \dagger}$ measures the on-site magnon occupancy. In general, we do so differently on each site, choosing the $z$-direction to point along the local magnetic ordering vector. 

We can attempt to expand the supercharge to order $\sqrt{S}$ in a large-$S$ expansion. Doing so, we obtain a supercharge similar in form to \eqref{eq:ComplexSUSYCharge}
\begin{equation}\label{eq:SUSYChargeR1R2}
    {\cal Q} = c^\dagger {\bf R}_1 b + c^\dagger{\bf R}_2 b^\dagger \,,
\end{equation}
but here there are {\em two} matrices ${\bf R}_1$ and ${\bf R}_2$ unlike in \eqref{eq:ComplexSUSYCharge}. We can also wonder if the approximation preserves the SUSY algebra that demands ${\cal Q}^2 = 0$. We find
\begin{equation}
    {\cal Q}^2 = c^\dagger {\bf R}_1{\bf R}_2^T c^\dagger\,.
\end{equation}
This would be a pairing term in the fermion Hamiltonian. For all the cases we discuss below, we find fermion pairing vanishes and ${\cal Q}^2 = 0$, the ${\cal N}=2$ SUSY algebra is preserved by the large-$S$ approximation.

Proceeding to derive the SUSY Hamiltonian, via $H_{\rm SUSY} = \frac{1}{2} \{Q,Q^\dagger\}$, yields
\begin{multline}
    H_{\rm SUSY} = \frac{1}{2} c^\dagger({\bf R}_1{\bf R}_1^\dagger - {\bf R}_2{\bf R}_2^\dagger)c  \\+ 
    \frac{1}{2} b^\dagger({\bf R}_1^\dagger {\bf R}_1 + {\bf R}_2^T{\bf R}_2^*)b +
    \left(b\,{\bf R}_2^\dagger {\bf R}_1b + {\rm h.c.}\right)\,.
\end{multline}
Thus, expanding around a ground state of the XY model (spontaneously) violates the U(1) symmetry of magnons {\em but not of the fermions} -- the superfluid of magnons are partnered with metallic fermions. 

Now, in this XY model, there are many large-$S$ ground states, each with its own magnon dispersions. These ground states define the frustration: the spins struggle to choose the best from among all the ground state options. One way to understand this frustration is to take a walk along a path in the ground state manifold and, stopping at points along the walk, study the magnon band structure. Let us do so in a kagome lattice example.

One ``ferrimagnetic" walk in the kagome lattice XY model defined above, is to start from the antiferromagnetic ``$q=0$ state'' defined by three spin ordering vectors ${\bf A}$, ${\bf B}$, ${\bf C}$ lying in the $xy$-plane with ${\bf A}+{\bf B}+{\bf C}=0$; placing one on each of the three sites in the unit cell.  Then add a $z$-component uniformly to all three ordering vectors, capturing this change by the polar angle $\theta$ that is $\pi/2$ in the $q=0$ state and zero in the fully tilted simple ferromagnetic state. This walk then takes us from a classic antiferromagnetic state to the fully polarized ferromagnetic state.

\begin{figure}
    \centering
    \includegraphics[width=1.0\columnwidth]{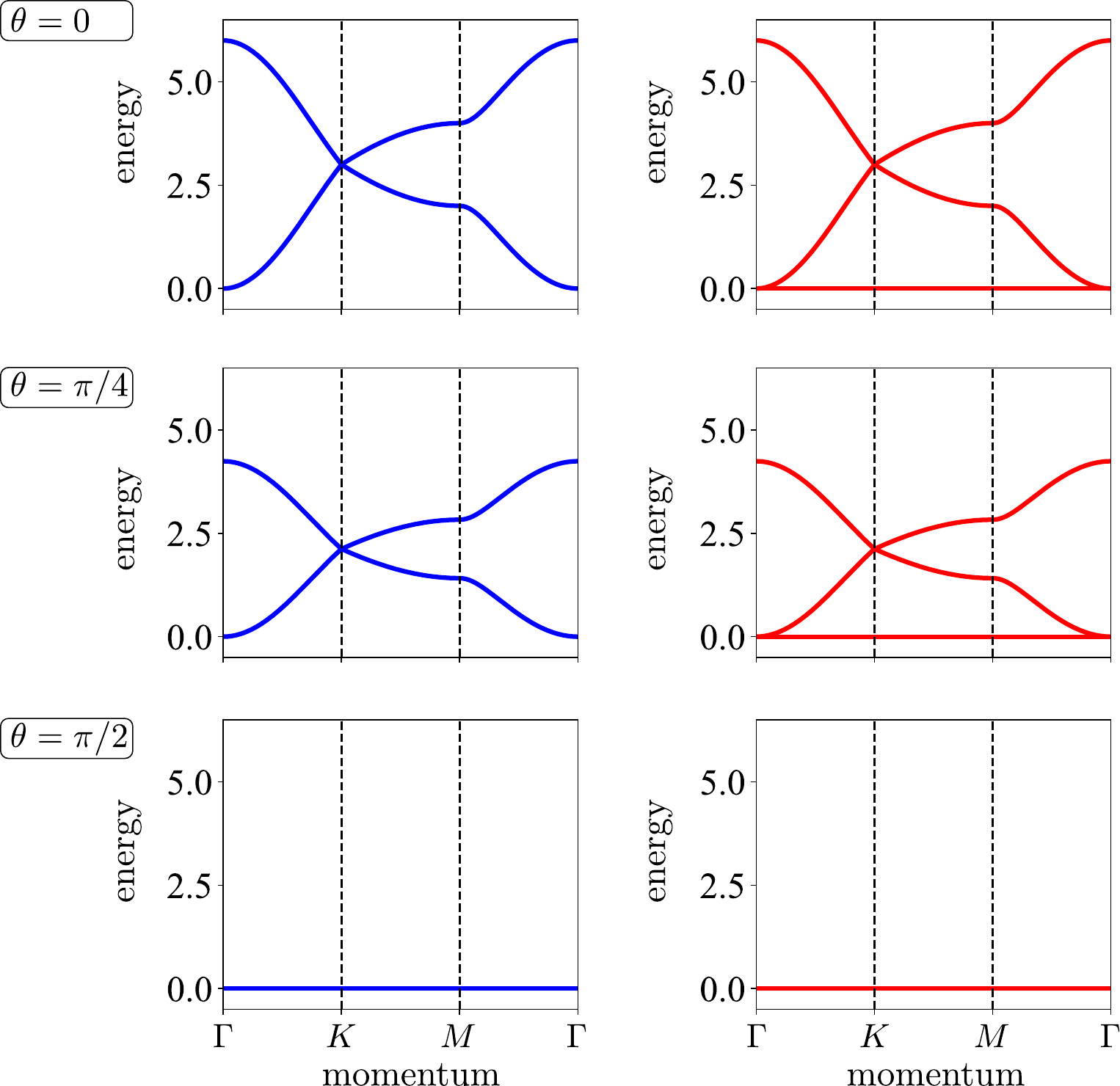}
    \caption{{\bf Kagome lattice magnons (right) and their SUSY partner (left)}. Spectra are shown along a ``ferrimagnetic" path in the XY model ground state manifold: at the $z$-polarized ferromagnetic point $\theta=0$ (top),  at $\theta=\pi/4$ in the ferrimagnetic region (middle),  at the antiferromagnetic $q=0$ point $\theta=\pi/2$ (bottom). These band structures show how ferromagnetic magnons have a quadratic band-touching semimetal as a SUSY partner.
     Similar to the band structure in Fig.~\ref{fig:susy-kagome-honeycomb}, we note that the Witten index is $\nu = 1$ for all three panels with the quadratic band touching (in the two upper panels) exhibiting a non-trivial $\pi_1$-topology. 
    }
    \label{fig:xymodel}
\end{figure}

In Fig.~\ref{fig:xymodel}, we present the evolution of the magnon band structure and their SUSY partner along the above walk. It shows at all points except $\theta=\pi/2$, that the ferrimagnetic states reproduce Fig.~\ref{fig:susy-kagome-honeycomb} -- the magnons have a flat band with a quadratic band touching at the $\Gamma$-point (with a non-trivial $\pi_1$-topology) and a semimetal as a SUSY partner. These states preserve the XY spin rotational symmetry of the spins and the $U(1)$ symmetry of a metal. But the bandwidth depends on the point along the walk and vanishes as the purely antiferromagnetic point $\theta\to\pi/2$ is reached. At this special point, flat bands emerge in {\sl both} the magnon and partner fermion models.

We can take a second walk in the infinite-dimensional space of ground states of the large-$S$ spin model, this time following a purely antiferromagnetic trajectory. Here we begin with the $q=0$ state and rotate it out of the $xy$-plane, keeping $S_\Delta^z=0$. This can be achieved by rotating about the ${\bf B}$ direction by an amount $\alpha$, a rotation that lifts the ${\bf A}$ spins above the plane and ${\bf C}$ spins below the plane. Surprisingly, this rotation has no effect on the band structure: all of these antiferromagnetic states have a vanishing magnon bandwidth.

These results suggest thermal order-by-disorder would be different from quantum order-by-disorder. Thermal order-by-disorder is an entropic selection of ground states that results from warming up the system from its $T=0$ ground state. For the kagome XY model, we would expect a much larger entropy for the antiferromagnetic states than the ferromagnetic states and so an antiferromagnetic state would be selected at large-$S$ but finite $T$. On the other hand, the pure ferromagnetic state at $\theta=0$ is an eigenstate of the full interacting Hamiltonian. This state has the least quantum fluctuations and so is the most stable ground state. It should be selected at finite $S$, $T\to0$. So the two limits do not commute and we expect the ordering tendencies will be different between thermal and quantum fluctuations, much like the $q=0$ is selected by quantum fluctuations while the higher entropy $\sqrt{3}\times\sqrt{3}$ state (not discussed here) is selected by thermal fluctuations in the XXZ kagome antiferromagnet \cite{chernyshev2015order}.

Before concluding, we must comment on the full interacting theory. The Hamiltonian of \eqref{eq:InteractingSUSY}, preserves magnon number and has many known eigenstates \cite{honecker2020loop}. Hence, the ferromagnetic (zero-magnon) ground state, the one-magnon states, the two-magnon states, etc. are all mapped to themselves by  $H_{\rm SUSY}$ and captured by small matrices. Specifically, we find the ``all-down" spin state with zero fermions per site and the  ``all-up" spin state with one fermion per site is exactly at zero energy. In addition to these two zero-magnon states, we find the one-magnon states, produced by the linear combinations of the all-down states raised by one unit of angular momentum, $S_i^+|-S,-S,\ldots\rangle$, or lowering the all-up states by one unit of angular momentum, $S_i^-|S,S,\ldots\rangle$, have exactly the band structure of the bosonic dispersions plotted in Fig. \ref{fig:xymodel} with the flat band at zero energy and a bandwidth of $6JS$. Hence, there are an infinite number of zero modes among the one-magnon states. Similarly, we find two zero energy states among the one-fermion states, the states corresponding all-down spins with one fermion occupying the ${\bf k}=0$ state and the all-up state with one hole occupying the ${\bf k}=0$ hole-state. There are clearly more zero-energy states than these. In total, we find an infinite number of exact {\sl ferromagnetic} eigenstates in the SUSY model of \eqref{eq:InteractingSUSY}. We have not identified any antiferromagnetic eigenstates and do not expect to do so. We conjecture these all are lifted to finite energy by quantum fluctuations and are only at zero energy in the classical $S\to\infty$ limit. Hence, we still expect ferromagnetism to be selected by quantum fluctuations.

%%%%%%%%%%%%%%%%%%%%%%%%%%%%%%%%%%%%%%%%%%%%%%%%%%%%%%

\subsubsection*{Classical Maxwell Counting}
In the previous discussion, we computed the band structure of a large-$S$ kagome XY model and found a flat band of magnons partnered with metallic fermions. Let us turn our attention to the existence of this flat band, for in these models, it is the fundamental cause of their frustration effects. 

The supercharge of \eqref{eq:susychargegeospin}, introduces one complex fermion on each simplex and, at the classical level, two total spin constraints imposed on the ground state by the ${\bf S}_\Delta^2$-term. Expressing the complex fermion as two real fermions $c^\dagger_\Delta = \gamma_{\Delta x}-i\gamma_{\Delta y}$ suggests that we have {\sl fermionized the constraints}: $\gamma_{\Delta x}$ corresponds to the $S_{\Delta x}=0$ constraint and $\gamma_{\Delta y}$ to the $S_{\Delta y}=0$ constraint. Similarly, one complex boson $b_i$ suggests that we have two real ``degrees of freedom'' per site. Hence, from a real-fermion, real-boson perspective, the single particle Witten index, which counts the difference between the number of bosons and fermions, corresponds in the classical limit to Maxwell counting: The number of ``degrees of freedom'' minus the number of constraints is twice the number of complex boson operators $b_i$ minus twice the number of complex fermion operators $c_\Delta$. In this way, we can exactly reproduce Moessner and Chalker's Maxwell counting \cite{Moessner1998, Moessner1998b} in a supersymmetric theory. 

For the specific case of the kagome XY model discussed above, the Maxwell counting works out to four constraints per unit cell (which has two triangles) and 6 real degrees of freedom (three spins). Hence, the Maxwell counting tells us there is a minimum of $\nu = 6-4 = 2$ real degrees of freedom, i.e.\ {\sl one} complex degree of freedom, per unit cell. This therefore demands the existence of one flat band in the magnon-number preserving band structure as presented in Fig. \ref{fig:xymodel}.

Maxwell counting involves more than identifying flat bands, it also enables topology through topological mechanics \cite{KaneLubensky2013}. In this vein, topological properties of magnons in distorted kagome antiferromagnets were studied in Ref.~\onlinecite{Roychowdhury2018} by placing them in the form of \eqref{eq:geospinmodel}. This study found two classes of problems associated with a triangulated surface in spin space called {\sl spin origami} \cite{shender1993kagome,chandra1993anisotropic,ritchey1993spin}. Flattenable spin origami with a flat band of zero energy magnons, and non-flattenable spin origami with Fermi-surface-like degeneracy of magnons. These results are due to SUSY \cite{Lawler2016}, but SUSY was not employed directly in obtaining them. Nevertheless, the topological property of spin waves discussed in this paper is precisely that expected by the SUSY encoded in \eqref{eq:susychargegeospin} upgraded to Heisenberg spins, an upgrade that is possible \cite{Lawler2016} with the real formalism discussed in Section \ref{sec:topo_mechanics}.

In summary, we have demonstrated the use of SUSY and our lattice construction to find highly frustrated magnets, materials whose magnons exhibit a flat band at different points on their large-$S$ ground state manifold. Using an XY model as an example, a model with a global $U(1)$ symmetry, and following Ref. \onlinecite{Lawler2016}, we wrote down a supercharge assigning a fermion creation operator to the $xy$-plane component of the total spin constraint on a simplex. This approach reproduced the Moessner-Chalker-Maxwell counting \cite{Moessner1998, Moessner1998b} identifying highly frustrated magnets as underconstrained systems, recognizing that such counting formally is associated with a SUSY system. This formal connection revealed a hidden U(1) symmetry of the magnons in the example studied -- their fermionic partner is a semimetal instead of a superfluid. In addition, our identification of such systems is broader than Moessner and Chalker, capturing lattices that go beyond corner-sharing simplices. 

%%%%%%%%%%%%%%%%%%%%%%%%%%%%%%%%%%%%%%%%%%%%%%%%%%%%%%
% Ground State Manifolds
%%%%%%%%%%%%%%%%%%%%%%%%%%%%%%%%%%%%%%%%%%%%%%%%%%%%%%

\subsection{Ground-state manifolds}
\label{sec:spin_spirals}

As a second case study for our SUSY framework, we will exclusively turn to {\sl classical} spin models, which are often considered the first step in the search for unconventional forms of magnetism. Conceptual advances such as the discussion of residual entropies \cite{Wannier1950, Ramirez1999} as a defining signature of frustration \cite{Moessner2001, Moessner2006} or the identification of cooperative paramagnetism \cite{Ramirez1994}, emergent Coulomb phases \cite{Henley2010}, or order-by-disorder phenomena \cite{Villain1980} have been formulated in the context of such classical models alongside the establishment of classical spin liquids \cite{Balents2010} and spin ice \cite{Bramwell2001, Castelnovo2008}. Many of these phenomena are evolving around extensive ground-state manifolds of geometrically frustrated Heisenberg antiferromagnets.

Here we will apply our SUSY framework to accomplish two conceptual goals in this context. First, we will explicate how the SUSY lattice correspondence leads one to quickly identify lattice geometries for which Heisenberg antiferromagnets are likely to exhibit extensive ground-state manifolds.  One prime example is the (classical) kagome antiferromagnet which we connected to the honeycomb-kagome case study in our introduction (Fig.~\ref{fig:susy-kagome-honeycomb}). We will also see many other instances in two and three spatial dimensions in this section, which are all exemplified by similar triptych-like figures, such as Fig.~\ref{fig:susy-diamond-pyro} which makes the case for the pyrochlore antiferromagnet. Second, we will use our framework to recast a spin-fermion correspondence in terms of our SUSY framework, which had been formulated by some of us \cite{Attig2017} to provide a link between the spin spiral ground-state manifolds of frustrated spin models and Fermi surfaces of electronic tight-binding models.

%%%%%%%%%%%%%%%%%%%%%%%%%%%%%%%%%%%%%%%%%%%%%%%%%%%%%%
\subsubsection*{Luttinger-Tisza method}

The starting point for our discussion is a classical Heisenberg antiferromagnet whose Hamiltonian we write as
\begin{equation}
	\mathcal{H}_\text{Heisenberg} = \sum_{ij} M_{ij} \, {\bf S}_i \cdot {\bf S}_j \,,
	\label{eq:susy-ham-classical-heisenberg}
\end{equation}
where the three-component vectors ${\bf S} = (S^x, S^y, S^z)$ denote $O(3)$ spins and $M_{ij}$ describes the (antiferromagnetic) coupling between two spins at real space coordinates $i$ and $j$. For a given lattice structure, the individual $M_{ij}$'s reflect 
the connectivity of this geometry and constitute the entries of a (weighted) adjacency matrix $\bf M$.

\begin{figure}[b]
	\centering
	\includegraphics[width=.9\columnwidth]{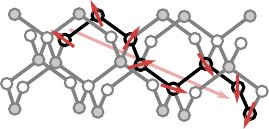}
	\caption{{\bf Coplanar spin spirals.} Schematic drawing of a coplanar spin spiral on the diamond lattice. The Heisenberg spins on every site enclose a fixed angle when traversing through the lattice along the spiral wavevector (as indicated by the faint background arrow). Such phases are relevant as ground states of many classical frustrated antiferromagnets.}
	\label{fig:coplanar_spiral}
\end{figure}

We are interested in the ground states minimizing the energy of this Hamiltonian, which, under certain circumstances, one can identify analytically using the Luttinger Tisza (LT) method \cite{Luttinger1951, Luttinger1946}. This method is based on the observation that any ground state minimizing \eqref{eq:susy-ham-classical-heisenberg} is also a ground state of the unconstrained problem where $|{\bf S}_i|\neq1$. Hence, solving the unconstrained problem first using linear algebra can enable the solution of the constrained problem. For certain lattices, this attempt at a solution always succeeds and leads to {\sl coplanar spin spirals} (see Fig.~\ref{fig:coplanar_spiral}) with the same unit cell as the underlying lattice. Not all classical spin ground states can be characterized in this way, but Heisenberg models have a tendency to do so \cite{Chen2021}.

Specifically, the LT approach proceeds by Fourier transforming the interaction matrix of \eqref{eq:susy-ham-classical-heisenberg} to its momentum space representation $\mathbf{M}({\bf k})$ and diagonalizing it for a given momentum $\bf k$ -- a step that is strongly reminiscent of tight-binding calculations and which we will build upon in the following. Before doing so, let us point out a key distinction here in that one still has to reconstruct the real-space coplanar spin spiral once one has identified the momentum ${\bf k}$ of the minimal eigenvalue. The wavevector of this spin spiral is simply given by ${\bf k}$ and phases of individual spins within the real-space basis can be read off the eigenstate itself as long as the equal-length {\sl hard-spin constraint} is fulfilled -- a constraint which we have effectively relaxed when simply diagonalizing $\mathbf{M}({\bf k})$. For Bravais lattices, however, this constraint must be generically fulfilled \cite{Lyons1960}, 
while for non-Bravais lattices this must not be the case and, by enforcing the constraint, one might end up selecting a subset of states found by the minimization. 

%%%%%%%%%%%%%%%%%%%%%%%%%%%%%%%%%%%%%%%%%%%%%%%%%%%%%%
\subsubsection*{Extensive degeneracies and flat bands}

One approach of equating the LT approach to a band structure calculation is to identify, on the level of matrix equivalences, 
the spin interaction matrix with the (bosonic) right-hand side of Fig.~\ref{fig:susy-techniques-complex}. We do this using the algorithm presented in Appendix \ref{app:square_rooting_algorithm}, which allows us to express ${\bf M} = {\bf R^\dagger R}$.
In doing so, the lattice correspondence of our SUSY framework identifies our LT calculation on some lattice (such as the kagome) with a free-fermion calculation on some other lattice (such as the honeycomb) as isospectral up to zero modes. But it is exactly the possible formation of such zero-energy flat bands that we are after when asking whether the Heisenberg antiferromagnet on some given lattice possibly has an extensive ground-state manifold. The kagome-honeycomb SUSY identification alluded to above is precisely of this sort and lets us conclude that the kagome antiferromagnet has an extensive ground-state degeneracy. 

\begin{figure}[t]
	\centering
	\includegraphics[width=\columnwidth]{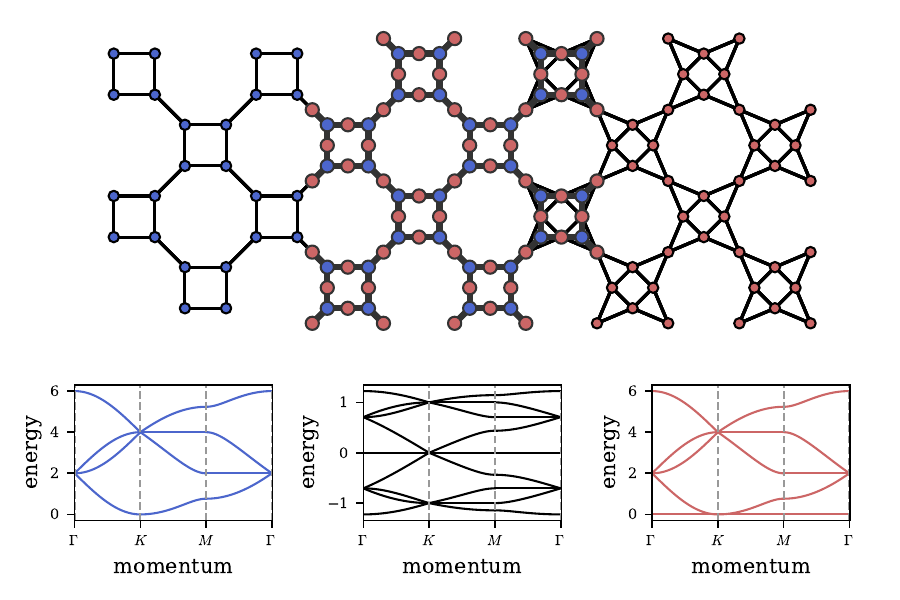}
	\caption{\textbf{SUSY correspondence of square-octagon and squagome lattices.} 
	Complex fermions (blue, left) on the square-octagon lattice are supersymmetrically linked to 
	complex bosons (red, right) on the squagome lattice. 
	The mapping can be established with a supercharge which can be interpreted as the adjacency matrix of a square-octagon-X lattice 
	(center plot), \ie a square-octagon lattice with additional sites on every bond. 
	For the topological classification according to Table \ref{table:classificationBDI},
	we find, noting that the Witten index here is $\nu = 2$, that the nexus point in the supercharge spectrum 
	has a trivial topological invariant of $\pi_2 = 0$, see also the illustration in Fig.~\ref{fig:homotopy} of the Appendix.}
	\label{fig:susy-squareoctagon-squagome}
\end{figure}

This idea can be readily generalized to other lattice geometries. In two spatial dimensions, one might consider the squagome antiferromagnet (Fig.~\ref{fig:susy-squareoctagon-squagome}), which has recently drawn some attention for the possible experimental realization of a spin liquid state in KCu$_6$AlBiO$_4$(SO$_4$)$_5$Cl \cite{Fujihala2020}. The squagome is SUSY-related via our lattice correspondence to the square-octagon lattice (hence its name), whose smaller unit cell has two sites less than the one of the squagome lattice; this gives rise to a Witten index of $\nu = 2$ and two flat bands in the squagome spectrum pointing to an extensive ground-state degeneracy and the formation of a classical spin liquid ground state \cite{Rousochatzakis2013,Morita2018} similar to the case of the kagome antiferromagnet.

\begin{figure}[t]
	\centering
	\includegraphics[width=\columnwidth]{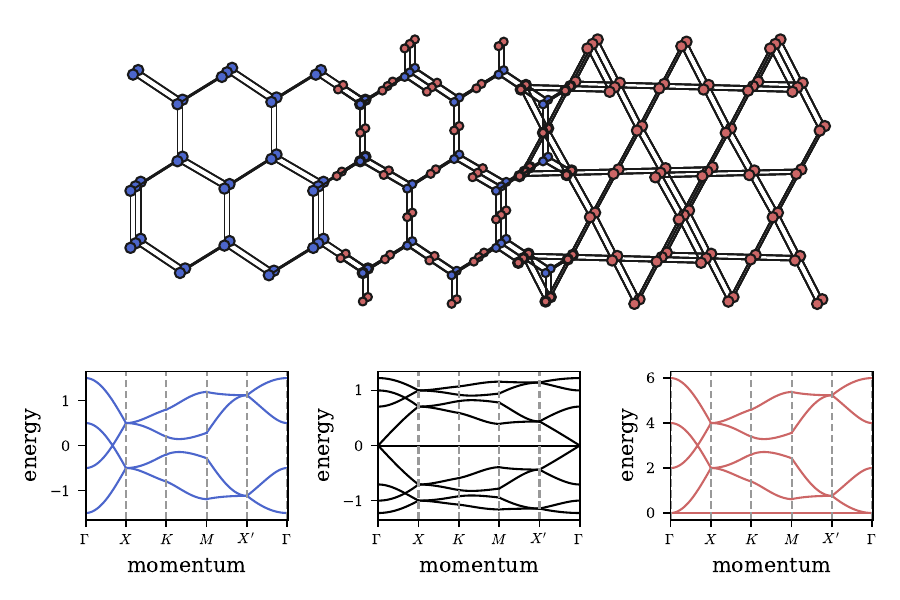}
	\caption{{\bf The hyperhoneycomb lattice and its SUSY partner.}
			The hyperhoneycomb (10,3)b lattice (left) is connected to a three-dimensional lattice of corner-sharing triangles
			which bears some similarity to the two-dimensional kagome lattice but should be distinguished from the 
			hyperkagome lattice of Fig.~\ref{fig:susy-hyperoctagon-hyperkagome}. 
			For the topological classification according to Table \ref{table:classificationBDI},
			we find, noting that the Witten index here is $\nu = 2$, that the nexus point in the supercharge spectrum 
			has a non-trivial topological invariant of $\pi_2 = +1$, see also the illustration in Fig.~\ref{fig:homotopy} of the Appendix.}
	\label{fig:susy-hyperhoneycomb}
\end{figure}

In three spatial dimensions, a well-known spin model with an extensive ground-state manifold is the classical pyrochlore antiferromagnet (Fig.~\ref{fig:susy-diamond-pyro}), which can also be captured by our SUSY-framework as the pyrochlore lattice is SUSY-connected via our lattice correspondence to the diamond lattice -- the two additional sites in the pyrochlore unit cell (with regard to the one of the diamond lattice) indicate a Witten index of $\nu = 2$ and the formation of two flat bands in the pyrochlore spectrum, the harbingers of an extensive ground-state degeneracy, spin liquid, and spin ice physics \cite{Moessner1998,Moessner1998b}. 

As a final example, we can also construct a possibly interesting three-dimensional lattice geometry, which has not received much attention so far, by applying our SUSY graph correspondence to the tri-coordinated hyperhoneycomb lattice (Fig.~\ref{fig:susy-hyperhoneycomb}), which has been investigated in the context of the Kitaev material $\beta$-Li$_2$IrO$_3$ \cite{Takayama2015}. To do so, we employ our lattice construction (Fig.~\ref{fig:lattice-construction-plaquettes}) to arrive at a three-dimensional structure of corner-sharing triangles, depicted on the right-hand side of Fig.~\ref{fig:susy-hyperhoneycomb}.
The latter might have deserved to be named hyperkagome, which however has been taken by the distinct lattice geometry of Fig.~\ref{fig:susy-hyperoctagon-hyperkagome} that is SUSY-related to the hyperoctagon lattice (both of which share a screw symmetry that is absent in the lattice geometries at hand). The point here is that our SUSY correspondence allows us to readily infer that the Heisenberg antiferromagnet on this lattice of corner-sharing triangles will exhibit an extensive ground-state degeneracy (\ie a flat band in its LT spectrum) and as such likely a spin liquid ground state. This observation is expected to generally hold for the SUSY-partners of the entire family of tri-coordinated lattices in three spatial dimensions \cite{Wells1977,OBrien2016,Eschmann2020}.

%%%%%%%%%%%%%%%%%%%%%%%%%%%%%%%%%%%%%%%%%%%%%%%%%%%%%%
\subsubsection*{Spin spirals and Fermi surfaces}

\begin{figure}[b]
	\centering
	\includegraphics[width=\columnwidth]{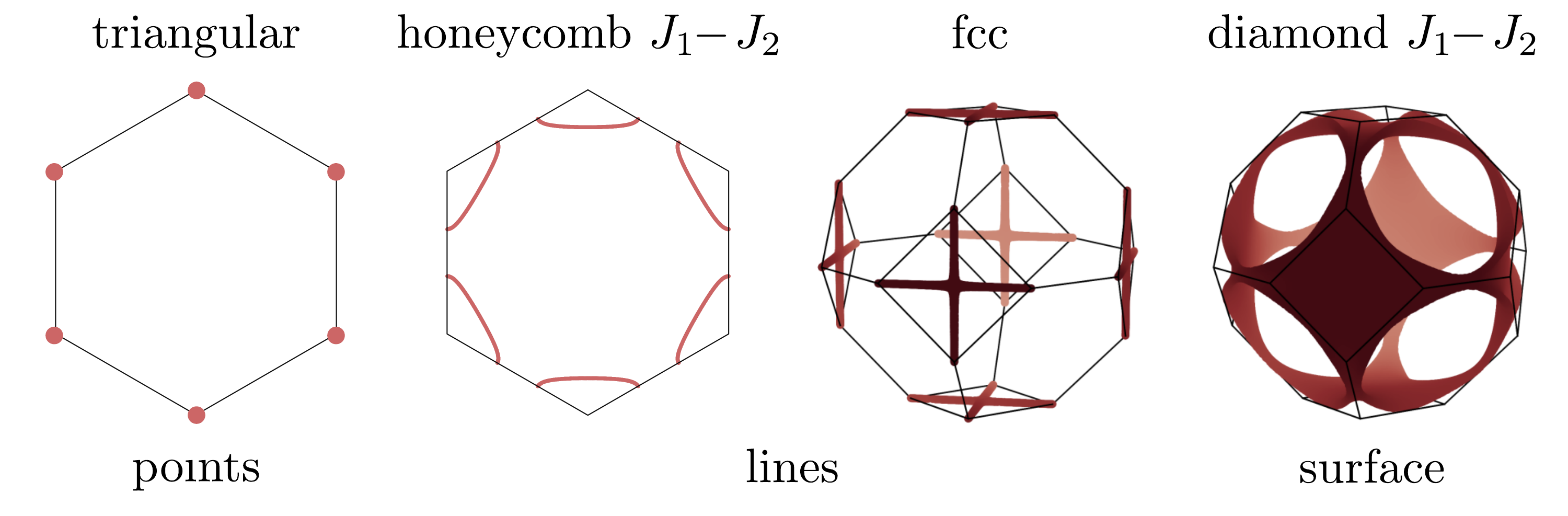}
	\caption{{\bf Spin spiral manifolds.} 
			Shown are the ground-state manifolds of coplanar spin spirals, exemplified by their respective $\bf k$ wavevectors
			in momentum space. Heisenberg antiferromagnets on different lattice geometries exhibit manifolds of varying
			dimensionality -- points for the triangular lattice, lines for the fcc lattice and $J_1-J_2$ honeycomb model, as well
			as entire surfaces for the  $J_1-J_2$ diamond lattice.  }
	\label{fig:SpinSpiralManifolds}
\end{figure}

\begin{figure}[t]
	\centering
	\includegraphics[width=\columnwidth]{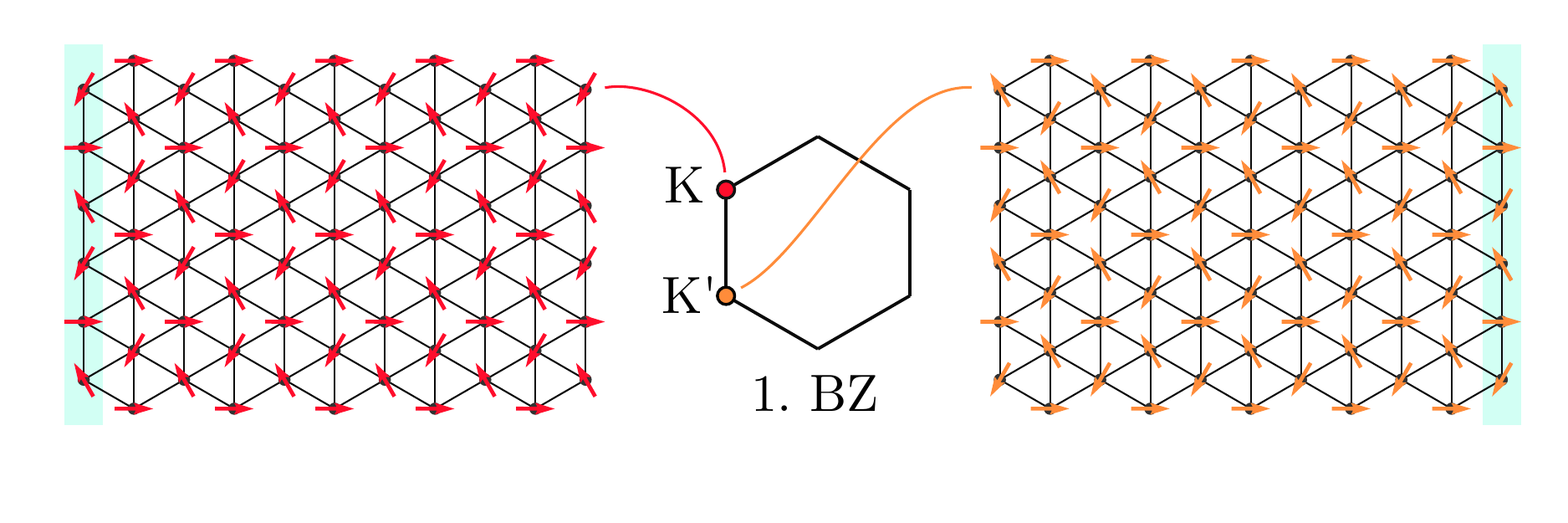}
	\caption{{\bf 120-degree order} in the classical Heisenberg antiferromagnet on the triangular lattice. The two possible ordering patterns (left and right) correspond to wavevectors $\bf K$ and $\bf K'$ at the corners of the Brillouin zone (BZ).}
	\label{fig:120degree}
\end{figure}

Another  approach of relating the LT approach for classical Heisenberg models to electronic band structure calculations 
is motivated by the observation that, for many geometrically frustrated antiferromagnets, the manifold of coplanar spin spiral 
ground states resembles a Fermi surface \cite{Attig2017}. Probably the most striking example is the $J_1-J_2$ Heisenberg
antiferromagnet on the diamond lattice \cite{Bergman2007}, for which the manifold of spin spiral states evolves as a function
of $J_2/J_1$ from a spherical geometry (for small $J_2/J_1$) to an open topology as depicted in Fig.~\ref{fig:SpinSpiralManifolds} 
(for intermediate $J_2/J_1$), and collapses into one-dimensional lines in the limit of $J_2/J_1 \to \infty$, which corresponds to 
two decoupled face-centered cubic (fcc) lattices, also illustrated in  Fig.~\ref{fig:SpinSpiralManifolds}. 
While the above cases relate to classical spin liquid ground states and systems with (sub)extensive ground state manifolds, 
one can also make such a connection between coplanar spin
spirals and nodal electronic states for ordered classical states. Take, for instance, the triangular lattice Heisenberg antiferromagnet with its well-known 120-degree ordered ground states (Fig.~\ref{fig:120degree}); in momentum space, the two possible ordering patterns of this 120-degree order correspond to momentum vectors $\bf K$ and $\bf K'$ at the corners of the Brillouin zone -- the well-known location of the Dirac cones of the honeycomb tight-binding model for free fermions. That one can indeed relate these two situations in a one-to-one SUSY correspondence in a similar way as one can connect the line-like or surface-like spin spiral manifolds above to the nodal lines or Fermi surfaces of SUSY-related electronic models will be the second SUSY correspondence for LT calculations that we will discuss here.

\begin{figure}[t]
	\centering
	\includegraphics[width=\columnwidth]{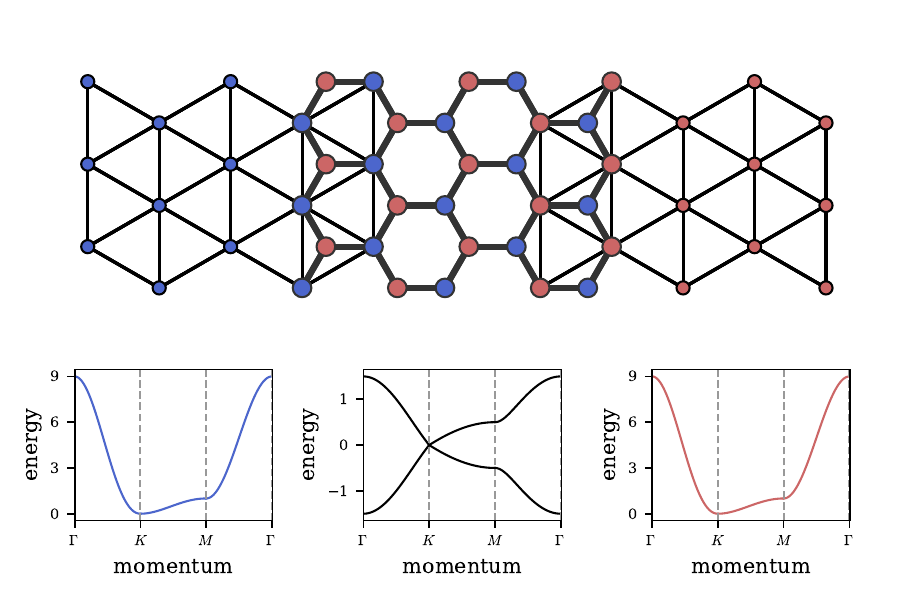}
	\caption{\textbf{SUSY corresponding triangular lattices.}  Complex fermions (blue, left) on a triangular lattice are supersymmetrically linked to complex bosons (red, right) on the same triangular lattice. 
	The mapping can be established with a supercharge which can be interpreted as the adjacency matrix of a honeycomb lattice 
	whose two sublattices are the two triangular lattices, respectively. 
	For the topological classification according to Table \ref{table:classification34},
	we find, noting that the Witten index here is $\nu = 0$, that the Dirac point at the $K$-point 
	in the supercharge spectrum has a non-trivial topological invariant of $\pi_1 = +1$.}
	\label{fig:triangular-honeycomb-correspondence}
\end{figure}

The conceptual difference to the first scenario above is that we are now aiming to connect a {\sl ground-state} property of the classical spin model, captured by the minimal energies in the LT spectrum, to what is typically a {\sl mid-spectrum} feature -- the Fermi surface of an electronic tight-binding model. But in the language of our SUSY correspondences, this immediately brings to mind that we might want to connect to the spectrum of the SUSY {\sl charge} and the tight-binding spectrum of its lattice model. Let us exemplify this for the 120-degree order of the triangular lattice antiferromagnet. Our SUSY lattice correspondence (Fig.~\ref{fig:lattice-construction-plaquettes}) connects the triangular lattice via a square root to the honeycomb lattice whose two triangular sublattices are SUSY partners as established before and illustrated in our triptych-like form in Fig.~\ref{fig:triangular-honeycomb-correspondence}. But when drawing our attention to the supercharge itself and its tight-binding spectrum in the middle panel of Fig.~\ref{fig:triangular-honeycomb-correspondence}, we indeed find the correspondence that we have been looking for -- the minima of the LT spectrum [left and right in Fig. \ref{fig:triangular-honeycomb-correspondence} at $\bf K$ (and $\bf K'$, not shown)] get mapped to the Dirac points in the middle of the well-known electronic tight-binding spectrum of the honeycomb lattice and the square-rooting along the way turns the quadratic band minima of the LT calculation into the quintessential linear dispersions of the Dirac cones.

\begin{figure}[b]
	\centering
	\includegraphics[width=0.7\columnwidth]{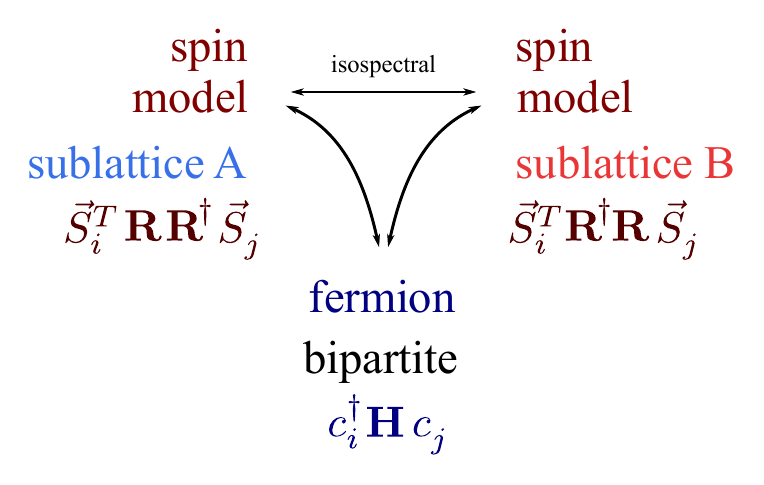}
	\caption{\textbf{SUSY matching spin-spiral groundstates and free fermion Fermi surfaces.} 
			Correspondence between a bipartite fermion model and two distinct spin models 
			on its two sublattices is shown in which the Fermi surface resembles the respective 
			ground state manifolds of coplanar spin spirals.}
	\label{fig:susy-techniques-spin}
\end{figure}

Let us formulate this SUSY perspective on the spin-fermion correspondence \cite{Attig2017} in more mathematical terms using our
previously established framework. To this end, we use the correspondence between a supercharge and a free {\sl chiral} fermion lattice model on the lattice geometry of the supercharge.
Such a free chiral fermion lattice model of the form
\begin{equation}
    \mathcal{H} = \begin{pmatrix} c_A^\dagger & c_B^\dagger \end{pmatrix}
 \begin{pmatrix}
       & {\bf R}\\
      {\bf R}^\dagger & 
 \end{pmatrix} 
 \begin{pmatrix} c_A \\ c_B
 \end{pmatrix}
\end{equation}
maps onto a supercharge via $c_A\to c$ and $c_B\to b$. Hence, the Dirac points in the spectrum of the supercharge actually correspond to Dirac points in this associated fermionic Hamiltonian. Going from the supercharge to one of the SUSY partnering sublattices one has to square this matrix $\bf H$ yielding a block-diagonal form as in \eqref{eq:SUSY_square}. In the spin-fermion correspondence, we now identify one of the two blocks ${\bf R}{\bf R}^\dagger$ or ${\bf R}^\dagger{\bf R}$ with the spin interaction matrix $\bf M$ that we diagonalize in the LT approach, as demonstrated in Fig.~\ref{fig:susy-techniques-spin}.
Of course, this squaring of the original matrix has the effect that the two positive/negative energy branches of the original fermionic tight-binding model get mapped onto one another and that the Fermi energy features in the middle of the particle-hole symmetric spectrum get mapped onto the minimal energy features of the squared Hamiltonian.
We have summarized this SUSY correspondence between a classical spin model on some lattice geometry with a fermionic tight-binding model on its square-root lattice in the illustration of Fig.~\ref{fig:susy-techniques-spin}. 

\begin{figure}[b]
	\centering
	\includegraphics[width=\columnwidth]{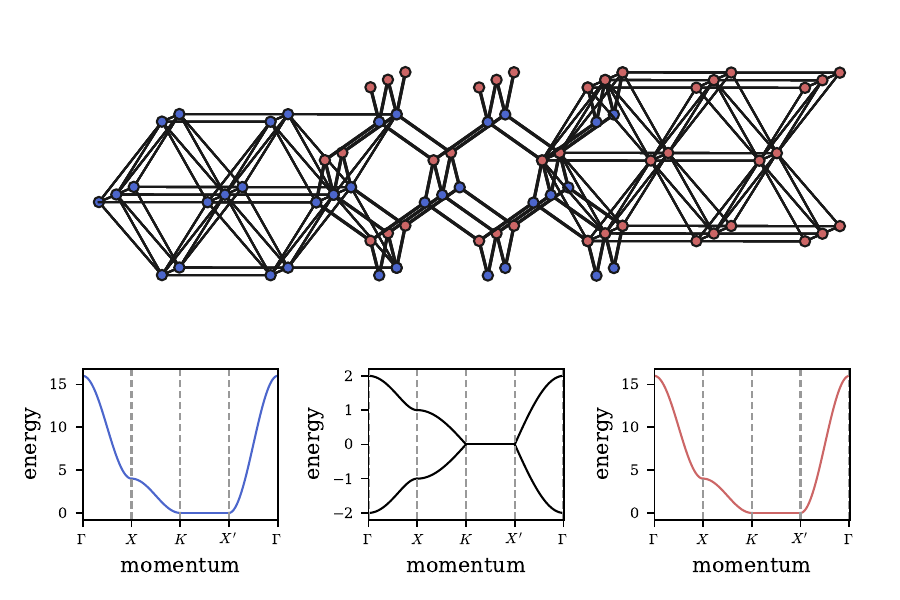}
	\caption{\textbf{Spin-fermion SUSY correspondence for the fcc antiferromagnet.}
	Complex fermions (blue, left) on the fcc lattice are supersymmetrically linked to complex bosons (red, right) on the same fcc lattice. 	The mapping can be established with a supercharge which can be interpreted as the adjacency matrix of a diamond lattice 	whose two sublattices are the two fcc lattices, respectively. For the topological classification according to Table \ref{table:classification34}, we find, noting that the Witten index here is $\nu = 0$, that the line of Dirac points between the $K$ and $K'$ points in the supercharge spectrum has a non-trivial topological invariant of $\pi_1 = +1$ (a Berry phase of $\pi$ around the line node).
	}
	\label{fig:fcc-diamond-correspondence}
\end{figure}

Having established this SUSY framework for the spin-fermion correspondence we can return to the cases of spin models with multiple spin spiral ground states whose (sub)extensive ground-state degeneracies can be captured by some non-trivial manifold in momentum space. The first example here might be the one of face-centered cubic (fcc) lattice Heisenberg antiferromagnet, which exhibits a subextensive ground-state manifold of spin spiral states whose wavevectors constitute a {\sl line} in momentum space \cite{Smart1966}, see Fig.~\ref{fig:SpinSpiralManifolds}. Putting our SUSY correspondence to work, we can connect this manifold to the Fermi surface of the fermionic tight-binding model on its square-root graph -- the diamond lattice (with its two fcc sublattices), which indeed exhibits a line-like Fermi surface (Fig.~\ref{fig:fcc-diamond-correspondence}).   

\begin{figure}[t]
	\centering
	\includegraphics[width=\columnwidth]{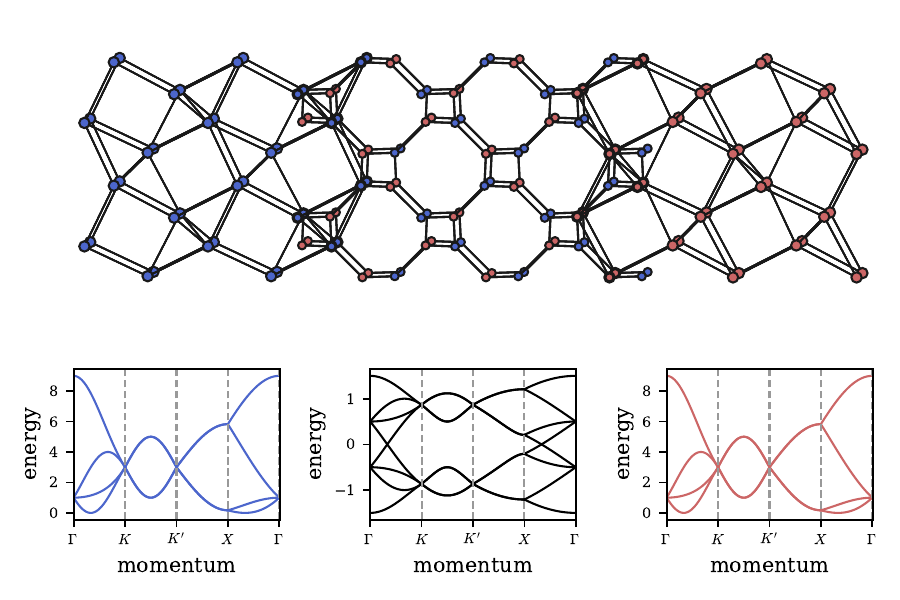}
	\caption{\textbf{Spin-fermion SUSY correspondence for the three-dimensional Shastry-Sutherland antiferromagnet.}
	Complex fermions (blue, left) on the hyperoctagon lattice (middle) are supersymmetrically linked to complex bosons (red, right) 
	on a three-dimensional generalization of the Shastry-Sutherland lattice, see Fig.~\ref{fig:susy_trio_shastry_sutherland} in the Appendix. 
	The ground-state manifold of spin-spiral states for this 3D Shastry-Sutherland model is illustrated in Fig.~\ref{fig:3DShastrySutherland-groundstate-manifold}. For the topological classification according to Table \ref{table:classification34}, we find, noting that the Witten index here is $\nu = 0$, that the Dirac point between the $\Gamma$ and the $K$ point and that between the $X$ and $\Gamma$ in the supercharge spectrum are topologically trivial (yielding a trivial $\pi_1$).
	}
	\label{fig:3DShastrySutherland-hyperoctagon-correspondence}
\end{figure}

\begin{figure}[t]
	\centering
	\includegraphics[width=\columnwidth]{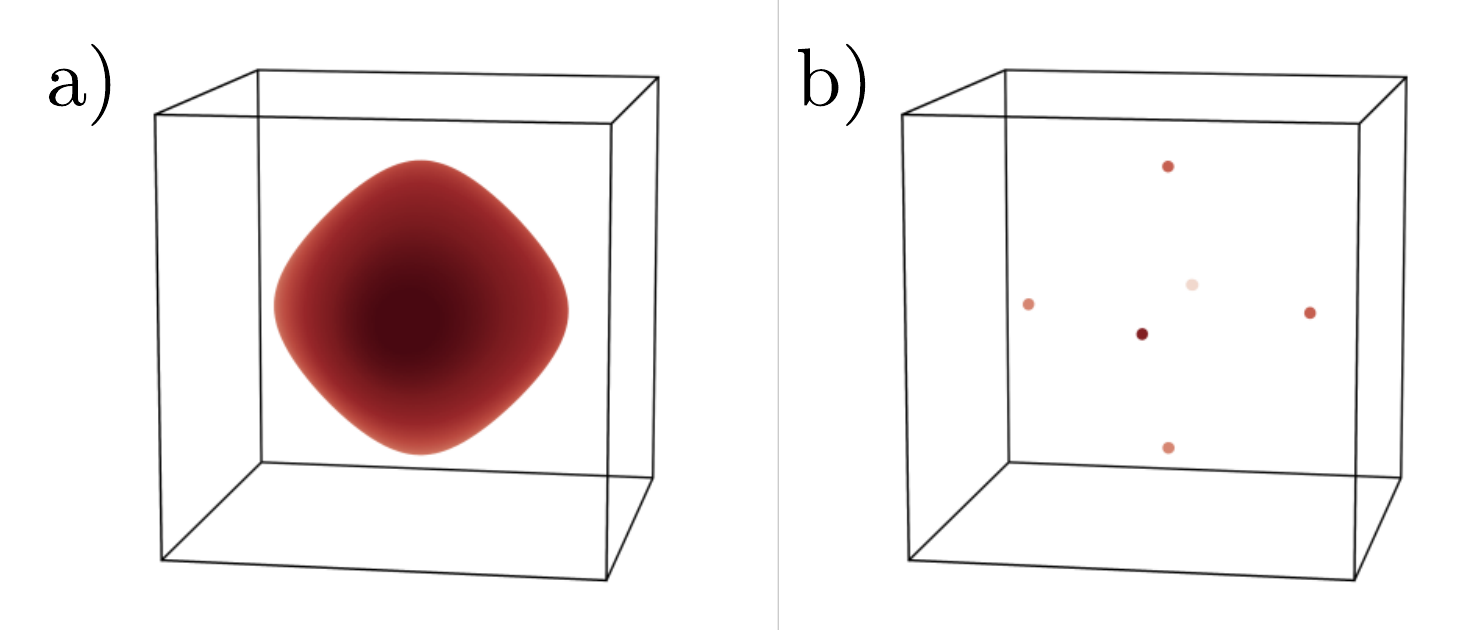}
	\caption{\textbf{Spin-spiral ground-state manifold of the three-dimensional Shastry-Sutherland antiferromagnet.}
	While the unconstrained diagonalization of the spin interaction matrix leads to a two-dimensional manifold of spin spiral states (panel a), 
	enforcing the Luttinger-Tisza constraint reduces the actual ground states to a set of individual ${\bf k}$-points (panel b).
	}
	\label{fig:3DShastrySutherland-groundstate-manifold}
\end{figure}

One might also be able to go one step further and construct, via our SUSY correspondence, a non-trivial spin model with a spin spiral {\sl surface} describing its ground states that has hitherto not been studied. One attempt in doing so is to start from the hyperoctagon lattice geometry as supercharge and take its lattice square (see Fig.~\ref{fig:3DShastrySutherland-hyperoctagon-correspondence}) to arrive at a three-dimensional variant of the Shastry-Sutherland model (akin to a similar SUSY connection in two spatial dimensions between the square-octagon and Shastry-Sutherland lattices exemplified in Fig.~\ref{fig:susy_trio_shastry_sutherland} of the Appendix). Since the fermionic tight-binding band structure on the hyperoctagon lattice exhibits a regular Fermi surface \cite{Trebsthyperoctagon2014}, this brings us, at first sight, to a full spin spiral surface (of codimension one) in the case of the three-dimensional Shastry-Sutherland model 
\footnote{
This might be an interesting spin model for future exploration, as it might, for instance, exhibit a spin-Peierls instability akin to its fermionic counterpart \cite{Hermanns2015} which might relax the spin spiral surface to a spin spiral line upon inclusion of phonon modes.}.
Unfortunately, however, this spin spiral surface is not stable upon enforcing the Luttinger Tisza constraint, which is not fulfilled by the vast majority of {\bf k}-points constituting the spin spiral surface, but only a finite set of individual \emph{points}, see Fig.~\ref{fig:3DShastrySutherland-groundstate-manifold}. 
We will leave it as an open challenge for future work to identify a nearest-neighbor Heisenberg antiferromagnet that indeed exhibits a full spin spiral surface, preferably via the SUSY correspondence at hand.

Finally, there is the class of  $J_1$-$J_2$ Heisenberg models \cite{Chen2021} for which the spin-fermion SUSY correspondence is particularly interesting as it seems to generically map spin systems to fermionic systems with a full Fermi surface, \ie a nodal manifold of co-dimension one. This is the case for the aforementioned $J_1$-$J_2$ model on the diamond lattice \cite{Bergman2007}, a restricted $J_1$-$J_2^*$ model on the body-centered cubic (bcc) lattice \cite{Shender1982,Attig2017} or the $J_1$-$J_2$ model on the honeycomb lattice \cite{Mulder2010} where the Fermi surface (or line in the two-dimensional model) exists in various shapes and topologies for a wide range of parameters $J_2/J_1$. For the $J_1$-$J_2$ model on the fcc lattice \cite{Henley1987,Revelli2019} one also finds a spin-spiral surface, albeit only for a single coupling parameter $J_2 = J_1/2$.
Conceptually, these $J_1$-$J_2$ models differ from what we have discussed so far in that they allow for couplings {\sl within} the same sublattice of a bipartite lattice or are defined for a non-bipartite lattice in the first place. As such, our SUSY lattice construction which is supposed to start from a clean bipartite graph does not immediately apply. However, one can still create a proper `graph squaring' in this case \cite{Attig2017}, by retaining the {\sl same} lattice geometry but doubling the degrees of freedom on every site
\footnote{The  $J_1$-$J_2$ fcc model is somewhat special here, as its underlying lattice geometry is non-bipartite. However, at the singular coupling of $J_2 = J_1 / 2$, one can replace checkerboard plaquettes spanned by $J_1$ and $J_2$ in the underlying fcc lattice by newly added 4-coordinated sites to form a fermion lattice, thereby allowing for a spin-fermion correspondence.}. We refer the interested reader to Ref.~\onlinecite{Attig2017} for further details of this SUSY mapping (though the language of SUSY was not yet adopted in that article).
    
%%%%%%%%%%%%%%%%%%%%%%%%%%%%%%%%%%%%%%%%%%%%%%%%%%%%%%

\subsection{Magnon dispersions}
\label{sec:magnon_dispersions}
In the third case study of our SUSY framework, we will switch our attention from frustrated magnets at zero magnetic field to those in large magnetic fields. These exhibit number-conserving magnons---bosonic excitations that arise  
near saturation fields~\cite{Schnack2002,zhitomirsky2004exact, schnack2006exact, hagymasi2021magnetization} -- and again the SUSY correspondence will reveal the existence of frustration. 

%%% Michael Made it to here %%%%%%

Let us start with an instructive example: for the kagome antiferromagnet in a magnetic field, it has been observed 
that {\sl below} the saturation field, localized one-magnon states populate the hexagonal motifs of the kagome lattice in the densest possible packing -- a triangular {\sl magnon crystal}
\cite{Schnack2020,zhitomirsky2004exact}. Such magnon localization is intimately related to and a precursor of the existence of the flat band in the nearby polarized state \cite{schnack2006exact}. Here we explain the existence of the flat band using our SUSY framework: the magnons have a SUSY partner on the honeycomb lattice. We will, in the first step, construct the fermionic analog of such a (topological) magnon spectrum. Reverting this procedure in a second step, 
we demonstrate how one can then predict non-trivial magnon phenomenology from their fermionic SUSY partners.

To accomplish the first step, going from bosonic magnon dispersions to its SUSY fermionic counterpart, we again return to our principal example of the honeycomb-kagome correspondence from the introduction 
(Fig.~\ref{fig:susy-kagome-honeycomb}). 
Here we start on the kagome side and consider a spin-$S$ kagome Heisenberg antiferromagnet subject to a uniform magnetic field $h>0$
\begin{align}
    {\cal H}=J\sum_{\langle i,j\rangle} {\bf S}_i\cdot {\bf S}_j - h\sum_i S_i^z\,.
\end{align}
At high fields,  beyond a saturation value of $h=6JS$, the ground state is a fully polarized state \cite{Schnack2002,zhitomirsky2004exact,zhitomirsky2005high}. To obtain the excitation spectrum of magnons in this phase we express the on-site spin operators $S^{\pm}_j=S^x_j {\pm} i S^y_j$ and $S^z_j$ in terms of bosonic (magnon) annihilation and creation operators $b_j$ and $b_j^\dagger$ following the Holstein-Primakoff expansion of \eqref{eq:holsteinpremakoff}.
Keeping up to terms quadratic in the bosonic operators in the large-$S$ limit (assuming $h\propto S$), we arrive at a bosonic tight-binding  Hamiltonian on the kagome lattice
\begin{align}
    {\cal H}^{\text{sw}}_{ij} 
    = JS\sum_{\langle i,j\rangle} \left( b_i^\dagger b_j^{\phantom\dagger} + {\rm h.c.} \right)
    + (h-4JS)\sum_i n_i \,,
\end{align}
with a hopping of strength $JS$ and chemical potential $(h-4JS)$. Translating to  Fourier space, the corresponding Bloch Hamiltonian reads  
\begin{multline}\label{eq:magnonham1}
H^B_{ij}({\bf k})= h-4JS +\\
S\sum_{\bf L} J_{\text{kagome}}({\bf d}_i-{\bf d}_j-{\bf L})e^{i{\bf k}\cdot({\bf d}_i-{\bf d}_j-{\bf L})}
\end{multline}
where $J_{\text{kagome}}(\delta{\bf r}) = J$ if $\delta{\bf r}$ represents the nearest neighbor separation on the kagome lattice with basis vectors ${\bf d}_i$ and Bravais lattice vectors ${\bf L}$ and zero otherwise.
The spectrum consists of a flat branch of magnons at $\omega_{\bf k}=h-6JS$ and two dispersive branches at $\omega_{\bf k}=h-3JS\pm JS\sqrt{3+2\sum_i\cos({{\bf k}\cdot{\bf d}_i})}$ identical to Fig.~\ref{fig:susy-kagome-honeycomb} on the kagome side shifted by a constant $(h-6JS)$ (to obtain exactly the same spectrum, set the field to exactly the point of saturation $h = 6JS$). This high-field limit of a frustrated kagome Heisenberg antiferromagnet, therefore, provides us with a natural setup to realize the simplest tight-binding model of bosons on the kagome lattice. 
  
%%%%%%%%%%%%%%%%%%%%%%%%%%%%%%%%%%%%%%%%%%%%%%%%%%%%%%

Let us now proceed to explicitly construct its fermionic partner, which via our SUSY lattice correspondence we expect to live on the honeycomb lattice and be exactly isospectral to the bosonic kagome model (up to flat bands). 
Our starting point is the magnon Hamiltonian $H_B({\bf k})$ in \eqref{eq:magnonham1}, where for simplicity we set the magnetic field to the point of saturation $h=6JS$ to bring the lowest eigenvalue of $H_B({\bf k})$, the flat band, to $\omega_{\bf k}=0$. The crucial step then is to construct the supercharge matrix ${\bf R}$ of \eqref{eq:ComplexSUSYCharge} by factorizing $H_B({\bf k})$ in \eqref{eq:magnonham1} as $H_B({\bf k})={\bf R}^\dagger({\bf k}) {\bf R}({\bf k})$. 
To keep in mind, such a decomposition should preserve the locality of $H_B({\bf k})$, namely, if $H_B({\bf k})$ consists only of nearest-neighbor hoppings, so should be reflected in the connectivities of ${\bf R}({\bf k})$. This would then yield a local fermion model with preserved topological signatures such as localized edge modes if the bosonic side has any.  

In essence, the factorization is tantamount to the square-rooting of $H_B({\bf k})$. We could produce this factorization using the graph square-rooting algorithm of Appendix~\ref{app:square_rooting_algorithm}. But instead, we can also draw insights from our graph correspondence --- one may opt for a decomposition such that ${\bf R}({\bf k})$ is a rectangular matrix.
Specifically, for the kagome-honeycomb case, it is a $2\times 3$ matrix as the partner lattice of kagome is the honeycomb lattice (Fig.~\ref{fig:susy-kagome-honeycomb}). 
This implies the Witten index is $\nu=1$ and there is a flat band on the kagome side.  

Empowered with the knowledge of this graph correspondence, we first identify the supercharge matrix ${\bf R}({\bf k})$. Introducing lower case indexing for the bosonic lattice and upper case indexing for the fermionic lattice, we can express it as
\begin{multline}\label{eq:susychargekagome}
    {\bf R}_{Ij}({\bf k}) = \\\sqrt{S}\sum_{\bf L}J_{\text{honeycomb-X}}({\bf n}_I-{\bf d}_j-{\bf L})e^{i{\bf k}\cdot({\bf n}_I-{\bf d}_j-{\bf L})} \,,
\end{multline}
where $J_{\text{honeycomb-X}}(\delta{\bf r}) = \sqrt{J}$ if $\delta{\bf r}$ is the nearest neighbor separation on the honeycomb-X lattice with honeycomb basis vectors ${\bf n}_I$ and zero otherwise. Together ${\bf d}_i$ and ${\bf n}_I$ form the basis of the bipartite honeycomb-X lattice.

Please note the simplicity of the step from \eqref{eq:magnonham1} to \eqref{eq:susychargekagome} arises from the choice of the ``canonical gauge'' where the Fourier transform is taken with the site locations $e^{i{\bf k}\cdot{\bf r}}$ and not the ``periodic gauge'' where it is taken with the unit cell location $e^{i{\bf k}\cdot{\bf R}}$ \cite{Cayssol2021}. If preferred, one can always switch to the periodic gauge after the Hamiltonians are identified.

The superpartner of the magnon model, \eg the fermionic Hamiltonian then derives by noting its Bloch form $H_F({\bf k})={\bf R}({\bf k}){\bf R}^\dagger({\bf k})$ which reads 
\begin{multline}
 H^F_{IJ}({\bf k}) = 3JS + \\
 S\sum_{\bf L}J_{\text{honeycomb}}({\bf n}_I-{\bf n}_J-{\bf L})e^{i{\bf k}\cdot({\bf n}_I-{\bf n}_J-{\bf L})} \,,
\end{multline}
where $J_{\text{honeycomb}}(\delta{\bf r})=J$ if $\delta{\bf r}$ connect nearest neighbors on the honeycomb lattice and zero otherwise.
This Hamiltonian has an identical spectrum as the magnons but the flat band. We immediately recognize that this fermionic model represents the well-known Dirac semimetal on the honeycomb lattice. 

\begin{figure}[t]
	\centering
	\includegraphics[width=\columnwidth]{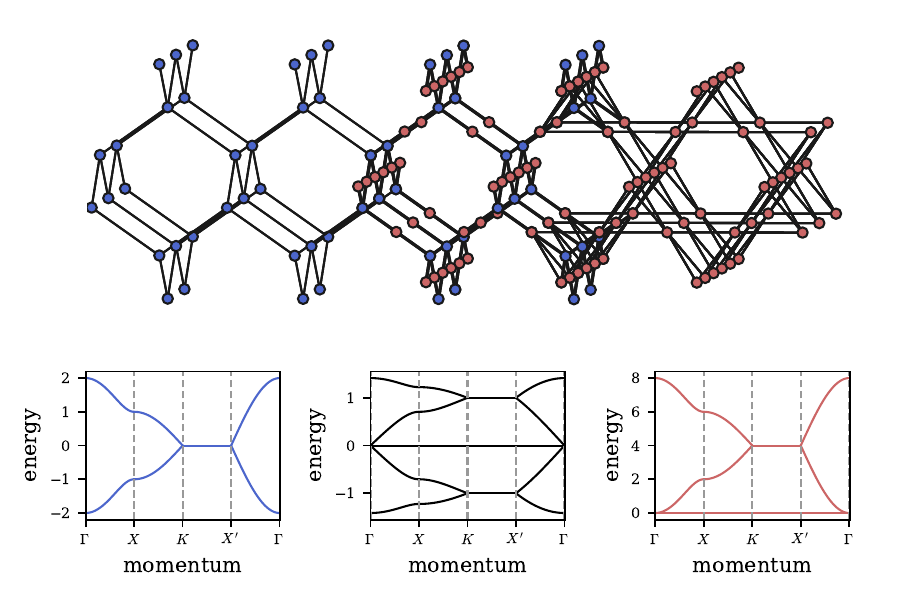}
	\caption{\textbf{Diamond-pyrochlore SUSY correspondence.} 
			Complex fermions (blue, left) on the diamond lattice 
			are supersymmetrically linked to complex bosons (red, right) on the pyrochlore lattice. 
			The middle panel shows the band structure of the supercharge lattice Hamiltonian on the diamond-X lattice.
			For the topological classification according to Table \ref{table:classificationBDI},
			we find, noting that the Witten index here is $\nu = 2$, that the nexus point in the supercharge spectrum 
			has a non-trivial topological invariant of $\pi_2 =-1$, see also the illustration in Fig.~\ref{fig:homotopy} of the Appendix.}
	\label{fig:susy-diamond-pyro}
\end{figure}

In summary, magnons described by a kagome Heisenberg model at saturation field have a fermionic partner with an index of $\nu=1$ demanding a flat band in their spectrum. This would seem like a complex procedure to discover a flat band, but it has an important benefit: it implies that a Heisenberg model on any example illustrated in the triptych-like figures of Section~\ref{sec:complex_SUSY} that exhibits a non-zero Witten index $\nu\neq 0$ leads to a magnon dispersion with a flat band in a saturation field -- and thereby constitutes a candidate system for a magnon crystal just below saturation. 
This allows us to go well beyond the known cases of kagome \cite{schnack2006exact} and pyrochlore \cite{hagymasi2021magnetization} antiferromagnets and postulate magnon crystals just below the saturation field, for instance, also for the squagome and the hyperkagome antiferromagnets along with a number of other two- and three-dimensional systems.

%%%%%%%%%%%%%%%%%%%%%%%%%%%%%%%%%%%%%%%%%%%%%%%%%%%%%%
% Ground-state manifolds
%%%%%%%%%%%%%%%%%%%%%%%%%%%%%%%%%%%%%%%%%%%%%%%%%%%%%%

\subsection{Parton dispersions}
\label{sec:parton_dispersions}

The phenomenon of displaying identical band structures by virtue of a graph correspondence finds realization in other frustrated magnets as well, specifically, in {\sl quantum} spin liquid (QSL) that have been discussed in certain spin-orbit coupled materials. Such QSL are unconventional phases of matter with one characteristic being the emergence of fractional excitations, called {\sl partons}, instead of more conventional magnons (as discussed above) or electronic quasiparticles. Depending on the underlying effective microscopic descriptions, the partons can range from being Abrikosov spinons \cite{abrikosov1965electron}, which are charge-neutral complex fermions carrying spin $S=1/2$, to Majorana fermions \cite{tsvelik1992new,Kitaev2006}, which are also charge neutral but do not carry any spin quantum number. While it remains an interesting open question if one can construct an exact SUSY mapping between different types of partons, we observe a curious similarity between the respective parton dispersions in two such distinct types of QSLs: One that features a {\sl spinon Fermi surface} of gapless Abrikosov spinons coupled to $U(1)$ gauge fields \cite{Lawlerhyperkagome2008}, and the other that features a {\sl Majorana Fermi surface} of Majorana fermions coupled to $\mathbb{Z}_2$ gauge fields \cite{Trebsthyperoctagon2014}. This identification turns out to ensue from the same type of graph correspondence that we have been discussing in our SUSY construction, but now between the underlying lattice geometries that harbor the QSLs. In other words, the graph correspondence enlightens the fact that the hopping Hamiltonians of partons on these two lattices must be isospectral (except for possible flat bands which do not play a substantial role here). 

\begin{figure}[t]
	\centering
	\includegraphics[width=\columnwidth]{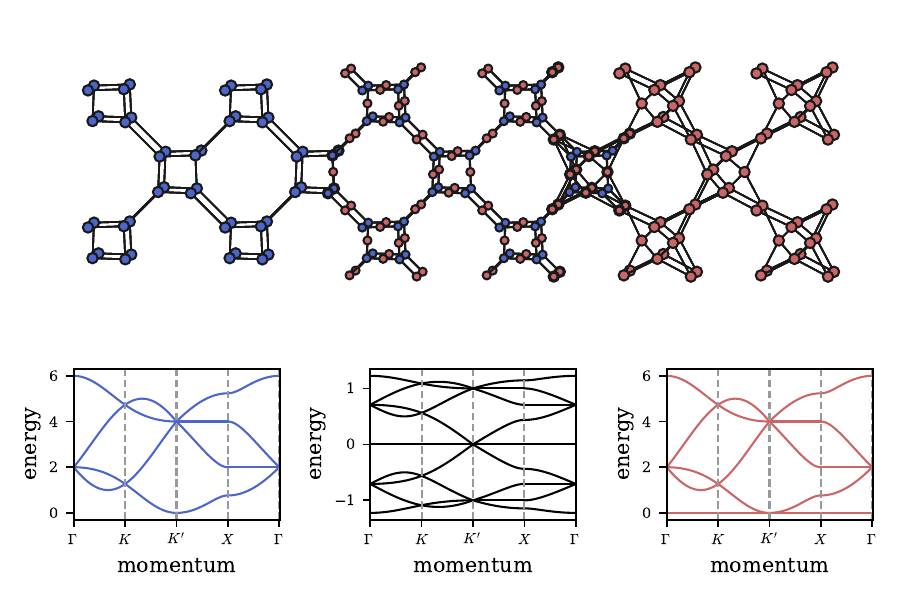}
	\caption{\textbf{Hyperoctagon-hyperkagome SUSY correspondence.}
			Complex fermions (blue, left) on the hyperoctagon lattice 
			are supersymmetrically linked to complex bosons (red, right) on the hyperkagome lattice. 
			The middle panel shows the band structure of the supercharge lattice Hamiltonian on the hyperoctagon-X lattice. 
			For the topological classification according to Table \ref{table:classificationBDI},
			we find, noting that the Witten index here is $\nu = 2$, that the nexus point in the supercharge spectrum 
			has a non-trivial topological invariant of $\pi_2 =+1$, see also the illustration in Fig.~\ref{fig:homotopy} of the Appendix.
	}
	\label{fig:susy-hyperoctagon-hyperkagome}
\end{figure}

The $U(1)$ QSL has been discussed in the context of a Heisenberg model on the hyperkagome lattice \cite{Lawlerhyperkagome2008}, after experimental indications of spin liquid behavior were reported for the hyperkagome iridate compound Na$_4$Ir$_3$O$_8$ \cite{Okamoto2007}.
Via our graph correspondence the spinon spectrum of this $U(1)$ QSL can be identified with the Majorana spectrum of a $\mathbb{Z}_2$ QSL emerging in a three-dimensional Kitaev model on the hyperoctagon lattice \cite{Trebsthyperoctagon2014} whose Hamiltonian consists of characteristic bond-directional Ising-like spin exchanges \cite{Trebst2022}.
Both of these lattices, the hyperkagome, and the hyperoctagon, can be obtained from our previous examples of the pyrochlore and the diamond lattice with a suitable depletion of tetrahedra or bonds, respectively, but more importantly in our context here, also from one another -- the hyperoctagaon lattice is the {\sl premedial} lattice of the hyperkagome lattice \cite{Trebsthyperoctagon2014}. That is, one obtains the hyperoctagon structure by shrinking each triangle of the hyperkagome lattice to a single vertex and respecting the connectivity of the original corner-sharing triangles -- but this is precisely the lattice square-rooting procedure of our SUSY lattice correspondence (Fig.~\ref{fig:lattice-construction-plaquettes}). Identifying these two lattice geometries as superpartners, we can also quickly construct the lattice of the supercharge that mediates the transformation between the two -- the hyperoctagon-X lattice illustrated in the middle of Fig.~\ref{fig:susy-hyperoctagon-hyperkagome}. 
The graph correspondence thus implies that the spinon excitation spectrum in the hyperkagome lattice has a bosonic partner on the hyperoctagon lattice. Surprisingly, however, we find it also coincides with the {\sl Majorana} excitation spectrum on the same hyperoctagon lattice. It turns out, that the real symmetric hopping matrix of the bosons is gauge equivalent via $b_j\to ib_j, j\in$ sublattice $A$ to the real antisymmetric hopping matrix of the Majorana fermions on the bipartite hyperoctagon lattice. Hence, both feature extended two-dimensional Fermi surfaces around the point of isotropic hoppings of the fermions on the individual lattices \footnote{Note that since the hyperkagome has a 6-site unit cell while the hyperoctagon has only 4 sites in its unit cell, this leads to a two-fold degenerate flat band in the spinon spectrum on the hyperkagome lattice (this is an example of the Witten index $\nu=2$).}. We have thus connected two enigmatic QSLs discussed in the literature via our SUSY framework.

We will return to Kitaev QSLs in Section~\ref{sec:topo_mechanics}, where we will exploit the fact that they can be cast in terms of free Majorana fermion models to formulate a SUSY connection to {\sl real} bosons and their classical analogs to discuss mechanical incarnations of Kitaev spin liquid physics.

%%%%%%%%%%%%%%%%%%%%%%%%%%%%%%%%%%%%%%%%%%%%%%%%%%%%%%
% Topological mechanics
%%%%%%%%%%%%%%%%%%%%%%%%%%%%%%%%%%%%%%%%%%%%%%%%%%%%%%

\section{Topological mechanics}
\label{sec:topo_mechanics}

We now turn to the second broader context in which we can apply our SUSY framework to establish connections between two seemingly distant fields -- topological mechanics and the physics of Majorana fermions. It rests on the principal observation that mechanical systems evolve around phase space coordinates $(q,p)$, \ie the classical limit of {\sl real} bosonic degrees of freedom whose most natural SUSY partners are real (Majorana) fermions. On a technical level, this will require us to expand our SUSY correspondence, which we had presented for the case of complex bosons and fermions in Section~\ref{sec:SUSY_lattice_models}, to the case of real bosons and fermions. We will see this also has implications on the accompanying SUSY lattice correspondence. 

Since the dynamics in mechanical setups is generally time-reversal symmetric and our SUSY correspondence %for topological mechanics 
respects this symmetry (as in the complex case), a natural starting point on the fermionic side is to consider time-reversal symmetric Majorana Hamiltonians. In the parlance of symmetry classes, the latter belong to symmetry class BDI 
in the ten-fold way \cite{Altland1997,Chiu2016classification} and admit a block-off-diagonal form as in \eqref{H1} when expressed in a suitable basis. 
As we will discuss in the following (Section~\ref{subsec:real_supersymmetry}), this matrix form implies that it is always possible to construct a {\sl local} supercharge that generates a representative fermion Hamiltonian from this class \cite{gong2022}. 
The locality is crucial in preserving the topology of the models identified by SUSY, \ie to carry over to the bosonic side. 
We demonstrate that this procedure is not only a natural connection between real fermions and bosons but offers another important reward -- the canonically conjugate positions and momenta get {\sl decoupled} in the resultant bosonic Hamiltonian, crucially allowing us to formulate classical analogs in terms of balls-and-springs networks. SUSY, thus, paves the way for constructing proper mechanical analogs of Majorana models that can be studied in table-top experiments. Notably, these mechanical systems will exhibit a topological response protected by SUSY, if the fermionic partner system has any. 

This section reviews previous work \cite{KaneLubensky2013, iadecola2016non, socolar2017mechanical, Attig2019} on this subject,  placing it within the larger SUSY framework we developed above for complex fermions. This includes identifying the class of the real Hamiltonian and how the real formalism fits into the topological classification presented above. It also includes a discussion of the Berry curvature, and how in our examples the real formalism has a finite curvature. Principal examples, which we will present in the following, will include mechanical analogs of topological superconductors (in Section~\ref{sec:KitaevChain}), Kitaev spin liquids (in Section~\ref{sec:MechanicalKSL}), and higher order topological insulators (in Section~\ref{sec:MechanicalHOTI}).

%%%%%%%%%%%%%%%%%%%%%%%%%%%%%%%%%%%%%%%%%%%%%%%%%%%%%%
\subsection{SUSY mapping for real fermions and bosons}
\label{subsec:real_supersymmetry}

\begin{figure}[t]
	\centering
	\includegraphics[width=\columnwidth]{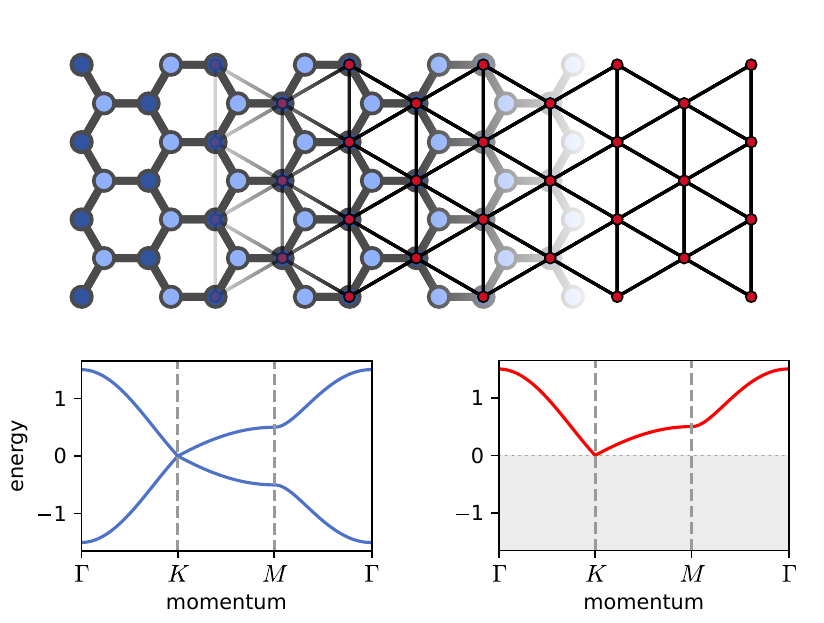}
	\caption{\textbf{SUSY correspondence for real fermions and bosons.} 
				Real (Majorana) fermions on a bipartite lattice (such as the honeycomb lattice on the left) 
				have an isospectral SUSY partner in the form of real bosons on one of its sublattices
				(such as the triangular lattice on the right). 
				Eigenmodes in these systems,  depicted in the lower panels, agree for all non-negative energies, 
				including potential zero modes.}
	\label{fig:MechanicalKitaevModel}
\end{figure}

We set out by recasting the SUSY formalism, outlined in Section~\ref{subsec:supersymmetry} for complex fermions and bosons, to their {\sl real} counterparts. Our starting point will be a model system on the fermionic side, \ie a Majorana fermion hopping model, that we define on a {\sl bipartite} lattice such as the honeycomb lattice depicted on the left in Fig.~\ref{fig:MechanicalKitaevModel}. Note that this choice of a bipartite lattice for the fermionic side is a first distinction to the complex SUSY scenario where we did not make such a restriction.
Adapting a suitable basis, one can cast such a Majorana Hamiltonian on any given bipartite lattice into a block off-diagonal form 
\begin{align}
	\label{Maj_ham1}
	{\cal H}_{F} & = {-\frac{i}{2}}\gamma^A_i {\bf A}_{ij} \gamma^B_j + {\rm h.c.} \,, \nonumber \\
			  & = {\frac{i}{2}}   
  \begin{pmatrix}
   \gamma^A & \gamma^B
  \end{pmatrix}
 \begin{pmatrix}
  & -{\bf A}\,\,\\
   {\bf A}^T & 
  \end{pmatrix}
  \begin{pmatrix}
   \gamma^A  \\
    \gamma^B
   \end{pmatrix}   \,,
\end{align}
where the matrix ${\bf A}$ represents the lattice adjacencies (or connections between the sublattices $A$ and $B$) weighted by appropriate hopping strengths and the $\gamma^A$ and $\gamma^B$ are the Majorana creation/annihilation operators that reside on the two sublattices. 
We now want this Hamiltonian to be the fermionic component of a SUSY Hamiltonian ${\cal H}_{\rm SUSY}= \frac{1}{2}\{\mathcal{Q}, \mathcal{Q}^\dagger \}$, generated by a supercharge ${\cal Q}$. To this end, let us consider the {\sl Hermitian} supercharge 
\begin{align}
\label{susychargereal}
 {\cal Q} & = 
  \gamma^A_i {\bf A}_{ij} {\hat q}_j + \gamma^B_i {\bf 1}_{ij} {\hat p}_j \,, \nonumber \\
  & = \begin{pmatrix}
   \gamma^A & \gamma^B
  \end{pmatrix}
  \begin{pmatrix}
   {\bf A} & \\
   & {\mathbf 1} 
  \end{pmatrix}  
  \begin{pmatrix}
   \hat{q} \\
   \hat{p}
  \end{pmatrix} \,,
\end{align}
which connects the Majorana fermion operators $\gamma^A$ and $\gamma^B$ (on the two sublattices of the fermionic lattice) to real bosonic operators $\hat{q}$ and $\hat{p}$, which are equal in number and conjugate to one another.
The SUSY Hamiltonian of this Hermitian supercharge ${\cal H}_{\rm SUSY}={\cal Q}^2\equiv {\cal H}_{F}+{\cal H}_{B}$ then block-decomposes into a fermionic and bosonic part.
The resultant bosonic Hamiltonian, in terms of the variables $(\hat{q},{\hat p})$ 
reads
\begin{align}\label{susyrealboson}
 {\cal H}_B & = {\frac{1}{2}}\sum_{ij} {\hat q}_i({\bf A}^T {\bf A})_{ij}{\hat q}_j +  \frac{1}{2} \sum_i \hat{p}_i \hat{p}_i \,, \nonumber \\
  & = {\frac{1}{2}}\begin{pmatrix}
   \hat{q} & \hat{p}
  \end{pmatrix}
  \begin{pmatrix}
    {\bf A}^T {\bf A} & \\
    & {\mathbf 1}
  \end{pmatrix}
  \begin{pmatrix}
   \hat{q} \\ \hat{p}
  \end{pmatrix}.
\end{align}

\begin{figure}[t]
	\centering
	\includegraphics[width=\columnwidth]{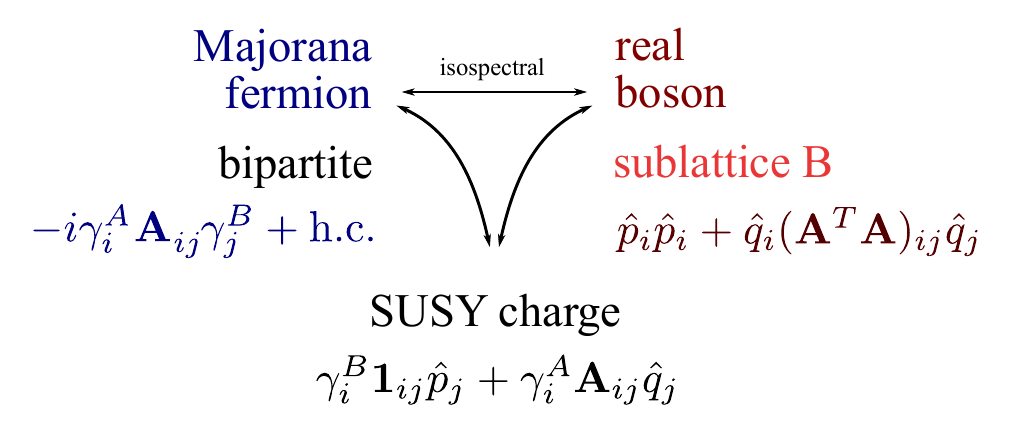}
	\caption{\textbf{SUSY correspondence of real bosons and Majorana fermions.}  Schematic relation between real bosons and Majorana fermions. Here, the real bosons reside on the sublattice of the bipartite Majorana fermion lattice. 
	}
	\label{fig:susy-techniques-real}
\end{figure}

Before we further inspect this bosonic Hamiltonian let us first point out a few noteworthy features that distinguish the construction so far from the complex case and, in particular, the lattice correspondence expounded before. 
First, let us introduce, in analogy to our discussion of complex fermion-boson SUSY in Section \ref{sec:SUSY_lattice_models}, the matrix
\begin{equation}
\label{eq:rigidity_matrix}
{\bf R}=
  \begin{pmatrix}
  {\bf A}   & \\
   & {\mathbf 1}
  \end{pmatrix} \,,
\end{equation}
which is representative of the supercharge ${\cal Q}$.
In the real case discussed here, it is block-diagonal.
Moreover, in the complex case, following the interpretation of the SUSY in terms of a graph correspondence, the fermionic and the bosonic models were defined on the two sublattices of a bipartite lattice (which we identified with the supercharge). The real case is strikingly different in this context: It is the Majorana fermions that now span the entire bipartite lattice under consideration, while their bosonic partners inhabit only {\sl one} of its sublattices. This is visualized in Fig.~\ref{fig:MechanicalKitaevModel} where we start with a Majorana fermion model on the honeycomb lattice (left) and end up with bosonic degrees of freedom that live on one of its two triangular sublattices (right). Technically, it is sublattice $B$ by virtue of the construction of ${\cal Q}$ in \eqref{susychargereal} 
\footnote{Of course, this has been a matter of choice. If one switches the tags $A$ and $B$ of the Majorana operators in \eqref{susychargereal}, then the bosonic modes will occupy the other sublattice.}. 
We have schematically summarized this SUSY correspondence between Majorana fermions and real bosons and their lattice correspondence in Fig.~\ref{fig:susy-techniques-real}. 

The spectral identification induced by the supercharge ${\cal Q}$ of \eqref{susychargereal} is subtle and distinct from the complex case. For the complex case, we had isospectrality between two Hermitian matrices, the two Bloch Hamiltonians ${\bf R}{\bf R}^\dagger$ and ${\bf R}^\dagger{\bf R}$. In the real case, the excitation spectra of ${\cal H}_F$ and ${\cal H}_B$ are obtained from the respective {\sl equations of motion} in the Fourier space
\begin{align}
\label{eq:eom1}
 \frac{\rm d}{{\rm d}t}\begin{pmatrix}
  {\gamma}_A \\ {\gamma}_B  
 \end{pmatrix} &= i
  \begin{pmatrix}
      & -{\bf A}({\bf k}) \\
      {\bf A}^\dagger({\bf k}) &
  \end{pmatrix}
  \begin{pmatrix}
   \gamma_A \\ \gamma_B  
  \end{pmatrix}  \,, \nonumber \\
  \frac{\rm d}{{\rm d}t}
  \begin{pmatrix}
  {\hat{p}} \\ {\hat{q}}  
 \end{pmatrix} &= i
  \begin{pmatrix}
     & -1 \\
     {\bf A}^\dagger({\bf k}){\bf A}({\bf k}) &
  \end{pmatrix}
  \begin{pmatrix}
   \hat{p} \\ \hat{q}  
  \end{pmatrix}  \,,
\end{align}
and diagonalizing the matrices on the right-hand side which are known as the Lie generators \cite{Roychowdhury2021}. In a compact form, the fermionic and the bosonic Lie generators read respectively as \[
	{\cal L}_F={\bf R}({\bf k}) \sigma_2 {\bf R}^\dagger({\bf k}) 
	\quad{\rm and}\quad 
	{\cal L}_B=\sigma_2 {\bf R}^\dagger({\bf k}) {\bf R}({\bf k}) \,,
\] 
where $\sigma_2$ is an extension of the Pauli matrix in the enlarged space of dimension $2N$ for $N$ number of operators of each flavor (both fermionic and bosonic) in the unit cell or equivalently, $2N$ sites in the unit cell. The forms of the two Lie generators suggest they are isospectral, again except for zero modes, with the excitation spectrum in the bosonic model being identified with the positive branch of the eigenvalues of ${\cal L}_B$, see the bottom panels of Fig.~\ref{fig:MechanicalKitaevModel} for our example system.

\subsubsection*{Classical limit}

We note that all the operators in our discussion above are quantum mechanical ones. In the appropriate classical limit, the boson Hamiltonian in \eqref{susyrealboson} represents a mechanical analog of the Majorana Hamiltonian in \eqref{Maj_ham1} in terms of a balls-and-springs model. Without loss of generality, we can assume the balls to be of unit masses with their dynamics described by the classical Hamiltonian 
\begin{align}\label{eq:classicalham1}
 {\cal H}_{\rm cl} &= \frac{1}{2}\sum_m e^2_m + \frac{1}{2}\sum_i p^2_i  \nonumber \\
 &= \frac{1}{2}\sum_{ij} q_i({\bf A}^T{\bf A})_{ij} q_j + \frac{1}{2}\sum_i p^2_i \,.
\end{align}
In the first line of the above equation, the spring extensions are denoted by $\{e_m\}$ which can be linearized in terms of the coordinates of the balls, $\{q^i\}$ via $e_m={\bf A}_{mi}q^i$ introducing a compatibility matrix ${\bf A}$. The matrix ${\bf R}$, introduced in \eqref{eq:rigidity_matrix}, extends this to accommodate all bosonic degrees of freedom $\{(q^i,p_i)\}$ and is referred to as the {\sl rigidity matrix} of the mechanical system. It thereby also renders a physical meaning to the supercharge \eqref{susychargereal} of our construction.  

Reexpressing the classical Hamiltonian in \eqref{eq:classicalham1} as 
\begin{align}
{\cal H}_{\rm cl} &= \sum_{ij} \frac{k_{ij}}{2}(q_i - q_j)^2 + \sum_i \frac{\kappa_i}{2} q^2_i + \frac{1}{2}\sum_i p^2_i\,,
\end{align}
we observe the mechanical model comprises both intersite and onsite springs of spring constants $k_{ij}$ and $\kappa_i$ respectively. Their parametric dependence can be read off as \cite{Attig2019}
\begin{align}
         k_{ij} &= -\sum_{a \in A} {\bf A}^T_{ia}{\bf A}_{aj}, \\
         \kappa_i &=  \sum_{a \in A} {\bf A}^2_{ia} - \sum_{b \in B} k_{ib}\,, 
\end{align}
where $k_{ij}$ are the off-diagonal elements of ${\bf A}^T{\bf A}$, that, by virtue of our SUSY construction, arise from the next-nearest-neighbor Majorana hopping (within the boson sublattice $B$) and $\kappa_i$ are the diagonal elements of ${\bf A}^T{\bf A}$ arising from the Majoranas hopping back and forth, modified by a contribution coming from the intersite springs. The matrix 
\[ 
	{\bf D} = {\bf A}^T{\bf A}
\] is known as the {\sl dynamical matrix} of the mechanical model.

The normal mode frequencies are obtained from the square root of the eigenvalues of ${\bf D}$, however, the normal modes are not the eigenstates of ${\bf D}$. As described in the previous section, they are instead obtained by diagonalizing the Lie generator ${\cal L}_B$ of the equations of motion \eqref{eq:eom1}. The positive branch of the eigenvalues of ${\cal L}_B$ coincides with the square-rooted eigenvalues of ${\bf D}$. In the following, we will unfold the topology associated with these normal modes in periodic systems. 

\subsubsection*{Topology of real bosons}

Previously in Section \ref{sec:SUSY_lattice_models}, we identified the topological invariants to classify the band topology in quadratic systems of complex fermions and bosons that are related by SUSY. We revisit this here for the {\sl real} case to illuminate the same for the normal modes in mechanical systems and unveil new topological invariants. 

We first need to settle the relationship between the real and complex formalisms. The hermitian supercharge discussed in the previous sections, such as in \eqref{susychargereal}, is focused on the BDI class. The general form is ${\cal Q} = \gamma {\bf R} {\bf x}$ where we combine the canonical coordinates into one vector ${\bf x} = (\hat q, \hat p)^T$ and have not placed any restrictions on ${\bf R}$ so that it describes a system with no symmetry. Written this way, the matrix ${\bf R}$ is a general real rectangular matrix but now appears double the size of the corresponding matrix in the complex formalism \eqref{eq:complexsusy_charge}. Switching from real to complex variables then produces a supercharge of the form
\begin{equation}
	\label{eq:susychargerealascomplex}
    {\cal Q}_{\rm SF} = \begin{pmatrix} c^\dagger &  c\end{pmatrix}
    \begin{pmatrix} {\bf R}_1 & {\bf R}_2\\ {\bf R}_2^* & {\bf R}_1^*\end{pmatrix}
    \begin{pmatrix} b \\ b^\dagger\end{pmatrix} \,.
\end{equation}
So this doubling turns out to exactly match the Hermitian part of Eq.~\eqref{eq:SUSYChargeR1R2} in our example from geometrically frustrated magnets in Section \ref{sec:GeoMagnets}. Hence, a change of variables from real to complex reveals indeed ${\cal Q} = {\cal Q}_{\rm SF}$. But we also see that at a matrix level, obtained by writing 
\begin{equation}
\label{eq:Qrealsymmetric}
{\cal Q} = \frac{1}{2}\begin{pmatrix} \boldsymbol{\gamma} & {\bf x}\end{pmatrix}
\begin{pmatrix} &{\bf R}\\ {\bf R}^T &\end{pmatrix}
\begin{pmatrix} \boldsymbol{\gamma} \\ {\bf x}\end{pmatrix},
\end{equation}
and writing ${\cal Q}_{\rm SF}$ in a similar way with spinor $(c, c^\dagger, b, b^\dagger)^T$, the topological class of ${\cal Q}$ and ${\cal Q}_{\rm SF}$ are the same, class BDI. Even though there is no physical symmetry in the real formalism beyond fermion parity, the matrix as viewed by the topological class has both ${\cal T}$ and ${\cal P}$ symmetry.  So the topological properties in the real formalism can readily be understood just by directly classifying the matrix defining the quadratic real supercharge. 

The use of this real formalism comes with a benefit: their Berry curvature signals a topological non-trivial system. To understand this, we need to step back and consider the subtlety of the real-formalism.

The supercharge for the complex case identifies individual fermionic eigenstates with their bosonic partners at equal (nonzero) energies as in \eqref{eq:Berry3}, thereby connecting the two Hilbert spaces. What demarcates the {\sl real} case is that such a spectral identification, instead of two Hamiltonians, applies to the two Lie generators ${\cal L}_F$ and ${\cal L}_B$ in \eqref{eq:eom1}. While ${\cal L}_F$ turns out to be the Majorana Hamiltonian itself and therefore a Hermitian operator, ${\cal L}_B$ is a {\em Krein-Hermitian} operator satisfying $\sigma_2{\cal L}_B={\cal L}_B^\dagger\sigma_2$ \cite{mostafazadeh2006krein, massarelli2022krein}. Accordingly, the bosonic normal modes obtained by diagonalizing ${\cal L}_B$ obey $\langle v_m|\sigma_2|v_n\rangle=[\sigma_3]_{m,n}$ while the fermionic ones, obtained by diagonalizing ${\cal L}_F$, obey $\langle u_m|u_n\rangle=\delta_{m,n}$. In other words, for the real case, the supercharge relates the members (the fermionic states) of a Hilbert space to those (the bosonic states) of a Krein space.

With the classification and Lie generator eigenvalue problems in mind, the SUSY map, discussed in Section~\ref{sec:classification} in \eqref{eq:normpreservingmap}, implies a constraint on the Berry phases associated with the same band in each of the three SUSY eigenproblems
\begin{align}\label{eq:berryphases}
 \theta_F^{(m)} = \theta_B^{(m)} + \theta_{\cal Q}^{(m)}\,.
\end{align}
This constraint is readily determined by expressing, for example, the fermionic states $|u_m({\bf k})\rangle$ that are eigenvectors of ${\cal L}_F$ in the Berry phase $\theta^{(m)}_F$ in terms of the bosonic eigenstates $|v_m({\bf k})\rangle$ that are Krein space eigenvectors of ${\cal L}_B$. Doing so leads to the relation
\begin{align}\label{eq:susyberryrealbosons2}
 &i\oint \langle u_m({\bf k}) | \partial_{\bf k} u_m({\bf k})\rangle \cdot {\rm d}{\bf k} = i\oint \langle v_m({\bf k}) |\sigma_2| \partial_{\bf k} v_m({\bf k})\rangle \cdot {\rm d}{\bf k} \nonumber \\
 &~~~~~~~~~~~~~~~~~~~~~~~~+ \oint \frac{{\rm Im} \left( \langle u_m({\bf k})|(\partial_{\bf k}{\bf R})|v_m({\bf k})\rangle \right)}{\sqrt{\omega_m({\bf k})}}\cdot {\rm d}{\bf k}\,,
\end{align}
where we identify the third term as the Berry phase of an eigenvalue problem associated with the supercharge ${\bf Q}$ (not discussed). 

This constraint allows for an alternative interpretation of the protection of zero modes in a SUSY system, as the constraint itself implies the existence of an alternative Berry potential. For the bosonic system, this SUSY Berry potential can be written as
\begin{align}
	\label{eq:susyberryrealbosons1}
	{\cal A}^{(m)}_{\rm SUSY} = 
		\langle v_m({\bf k})|i\sigma_2(\nabla_{\bf k} + \sigma_2\tilde{\bf R}^\dagger \nabla_{\bf k}\tilde{\bf R})|v_m({\bf k})\rangle,
\end{align}
where the single-particle bosonic state $|v_m({\bf k})\rangle$ is an eigenstate of ${\cal L}_B$ at energy $\omega_m({\bf k})$ (a similar statement is readily made for the fermionic system). ${\cal A}^{(m)}_{\rm SUSY}$ is mathematically identical to the fermionic Berry potential that gives rise to $\theta^{(m)}_F$ but defined {\em entirely} by the bosonic eigenvalue problem and the ${\bf R}$ matrix. Therefore, SUSY reveals an additional covariant derivative term on top of the conventional bosonic Berry connection $\langle v_m({\bf k})|i\sigma_2|\nabla_{\bf k}v_m({\bf k})\rangle$. The constraint of \eqref{eq:berryphases} allows a bosonic system to inherit the Berry phase of the SUSY-related fermionic system (and vice versa). 

The above perspective allows us to define the bosonic Berry phase in supersymmetric systems in an alternate way
\begin{align}
	\label{eq:susyberryrealbosons3}
 	\theta_{B, {\rm SUSY}}^{(m)} = \oint {\cal A}_{\rm SUSY}^{(m)}\cdot {\rm d}{\bf k}\,,
\end{align}
which can indeed reveal nontrivial windings $w\equiv\theta_B^{(m)}/\pi$ for the nodal points or lines in the spectrum of the normal modes in a topological mechanical system---examples of which will follow in the next Sections. Similar to the case of complex fermions and bosons presented in Appendix \ref{app:topology_complex}, the supersymmetric version of the Berry curvature for a mechanical system will follow from ${\cal F}^{(m)}_{\rm SUSY}=\nabla\times {\cal A}^{(m)}_{\rm SUSY}$ and will be useful in exploring the topology of Chern bands in apposite setups.

This discussion brings the SUSY formulation of topological mechanics to a special footing compared to the complex fermion-boson correspondence, where all our examples featured, in contrast to what has been discussed here, {\em identical} Berry phases and curvatures rendering a vanishing SUSY contribution to the underlying Berry connection (see again Appendix \ref{app:topology_complex}). In the mechanical systems, the topology in the bosonic models is revealed by the supersymmetric Berry potential in \eqref{eq:susyberryrealbosons1} while the conventional definition of the Berry connection signals merely a trivial phase.       

In the following, we will discuss one- and two-dimensional examples of mechanical systems which are the supersymmetric partner of the Kitaev chain, the Kitaev honeycomb spin liquid, and a higher-order topological insulator of Majorana fermions. 

%%%%%%%%%%%%%%%%%%%%%%%%%%%%%%%%%%%%%%%%%%%%%%%%%%%%%%
\subsection{Mechanical analog of the Kitaev chain:\\ the Kane-Lubensky chain}
\label{sec:KitaevChain}

\begin{figure}[t]
    \centering
    \includegraphics[width=\columnwidth]{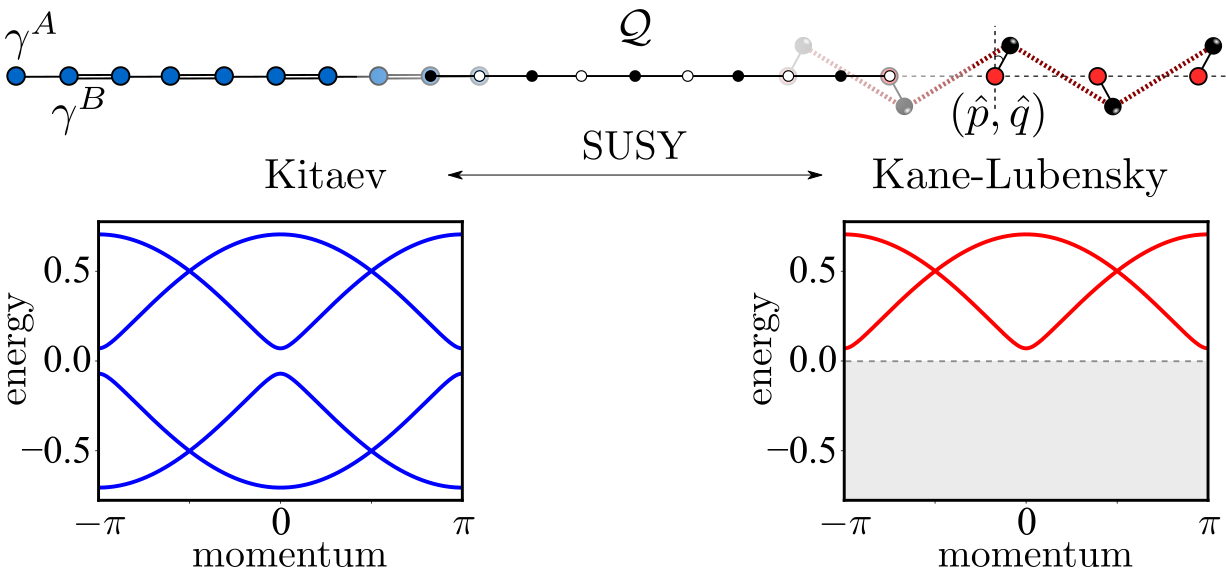}
    \caption{{\bf SUSY between the Kane-Lubensky chain and the Kitaev chain.} The superpartner of the Kitaev Majorana chain (left) is the Kane-Lubensky chain (right) of mechanical rotors. The dimerization of the Kane-Lubensky chain requires doubling the unit cell of the Kitaev chain. SUSY identifies the positive branch of the spectra on both sides.}
    \label{fig:Kane-Lubensky-Kitaev}
\end{figure}

In their groundbreaking paper on topological mechanics \cite{KaneLubensky2013}, Kane and Lubensky contemplated a mechanical analog of the Su--Schrieffer--Heeger (SSH) model by constructing a chain of rotors connected by springs with a dimerized unit cell, \ie two rotors per unit cell, as illustrated on the right of Fig.~\ref{fig:Kane-Lubensky-Kitaev}. This is an example of an isostatic mechanical frame where the number of constraints matches the number of degrees of freedom in the system. The SSH model  \cite{su1979solitons} itself constitutes what in today's terms is the first example of fermionic topological bands with gapless edge modes arising from a {\sl bulk-edge correspondence} \cite{bhbook}. Analogously, the mechanical model supports zero modes at the boundary for specific configurations in the parameter space when small angular displacements around a dimerized configuration of the rotors are connected to the spring lengths $l_n={\bf A}_{ni}\theta_i$ ($\bf A$ being the compatibility matrix). For a periodic chain with uniform rotors (of radius $r$, separated by a distance $a$), one can work in Fourier space to obtain \cite{chen2014nonlinear}
\begin{align}
 {\bf A}(k)=\begin{pmatrix}
             q_+ & q_- \\
             q_- & q_+e^{ik(2a)}
            \end{pmatrix},
\end{align}
where $q_{\pm}=r\cos{\bar{\theta}}(2r\sin{\bar{\theta}}\pm a)/\sqrt{a^2+4r^2\cos^2{\bar{\theta}}}$, $\bar{\theta}$ (the angular deviation from a vertical axis) specifying the dimerized configuration as in Fig.~\ref{fig:Kane-Lubensky-Kitaev}. 

The normal mode frequencies are obtained from the square root of the eigenvalues of the dynamical matrix ${\bf D}(k)={\bf A}^\dagger{\bf A}$ (a representative spectrum shown in Fig.~\ref{fig:Kane-Lubensky-Kitaev} with red solid lines for $r/a=0.5, \bar{\theta}=0.1$). In real space, for a chain of $N$ rotors, the dynamical matrix ${\bf D}={\bf A}^T{\bf A}$ is an $N\times N$ matrix of the form 
\begin{align}
     {\bf D} = 
     \begin{pmatrix}
      q_+^2 & q_-^2 & 0 & 0 & \dots & q_-^2 \\
      q_-^2 & q_+^2 & q_-^2 & 0 & \dots & 0 \\
      0 & q_-^2 & q_+^2 & q_-^2 & \dots & 0 \\
      \vdots & \vdots & \vdots & \vdots & \vdots & \vdots \\
      0 & 0 & \dots & q_-^2 & q_+^2 & q_-^2 \\
      q_-^2 & 0 & 0 & \dots & q_-^2 & q_+^2
     \end{pmatrix}
     \label{HKane-Lubenskyp1}.
\end{align}
Applying our SUSY correspondence, one can recognize the superpartner of this mechanical model is the {\sl Kitaev chain} \cite{kitaev2001unpaired} with the two hoppings specified by $q_{\pm}$. As on the mechanical/bosonic side, the chain is dimerized, we obtain a 4-site unit cell for the fermion model. The spectrum is shown in Fig.~\ref{fig:Kane-Lubensky-Kitaev} with blue solid lines (for $r/a=0.5, \bar{\theta}=0.1$) which is periodic over $k\in[-\pi/2,\pi/2]$ due to the unit-cell doubling.  

As an important extension, next, we demonstrate how one can construct a random Kitaev chain from a random one-dimensional phonon problem at the isostatic point identifying the two as superpartners of each other. For this, we consider a random $N\times N$ dynamical matrix which is structurally similar to the Kane-Lubensky dynamical matrix in \eqref{HKane-Lubenskyp1} in terms of connectivities and can be expressed as
\begin{align}
     \tilde{\bf D} = 
     \begin{pmatrix}
      a & b & 0 & 0 & \dots & b \\
      b & a & b & 0 & \dots & 0 \\
      0 & b & a & b & \dots & 0 \\
      \vdots & \vdots & \vdots & \vdots & \vdots & \vdots \\
      0 & 0 & \dots & b & a & b \\
      b & 0 & 0 & \dots & b & a
     \end{pmatrix}.
     \label{HKane-Lubenskyp22}
\end{align}
Here $a, b$ are coupling parameters of the phonon model assuming random values, unlike the Kane-Lubensky chain where they were specific functions of $r, a, \bar{\theta}$ \footnote{As an interesting side note, the matrix in \eqref{HKane-Lubenskyp22} can be recognized as a random chiral one-dimensional fermionic Hamiltonian when the diagonal elements are zero \cite{gurarie2003bosonic}}. 

To construct the fermionic model, one requires to identify a supercharge or equivalently, to derive a random compatibility matrix $\tilde{\bf A}$ from the dynamical matrix $\tilde{\bf D}$ such that $\tilde{\bf D}=\tilde{\bf A}^T\tilde{\bf A}$. It turns out that, for this one-dimensional problem, such a decomposition is indeed possible where the matrix $\tilde{\bf A}$ maintains the same degree of locality as $\tilde{\bf D}$ in terms of the connectivities and assumes an $N\times N$ form
\begin{align} 
     \tilde{\bf A} = 
        \begin{pmatrix}
            b_1 & b_2 & 0 & 0 & \dots & 0 \\
            0 & b_1 & b_2 & 0 & \dots & 0 \\
            0 & 0 & b_1 & b_2 & \dots & 0 \\
            \vdots & \vdots & \vdots & \vdots & \vdots & \vdots \\
            0 & 0 & \dots & 0 & b_1 & b_2 \\
            b_2 & 0 & 0 & \dots & 0 & b_1 
        \end{pmatrix},
        \label{HKane-Lubenskyp5}
\end{align}
where
\begin{align}
      b_1=x_i\sqrt{b}~;~b_2=\sqrt{b}/x_i~;~x_i=\pm \sqrt{\frac{a}{2b} \pm \sqrt{\frac{a}{2b}-1}},
\end{align}
(see Appendix~\ref{sec:AppKaneLubensky} for a detailed derivation). This compatibility matrix leads us to a Majorana Hamiltonian ${\cal H}_F$ identical to \eqref{Maj_ham1} but ${\bf A}$ replaced with $\tilde{\bf A}$. In other words, given a set of random parameters modeling a generic one-dimensional phonon problem restricted to nearest-neighbor connections (like the Kane-Lubensky chain), we can always derive a random Kitaev chain as its superpartner (Fig.~\ref{fig:Kane-Lubensky-Kitaev}) that shares the same excitation spectrum and the topology.

%%%%%%%%%%%%%%%%%%%%%%%%%%%%%%%%%%%%%%%%%%%%%%%%%%%%%%

\begin{figure*}[t]
    \centering
    \includegraphics[width=.9\linewidth]{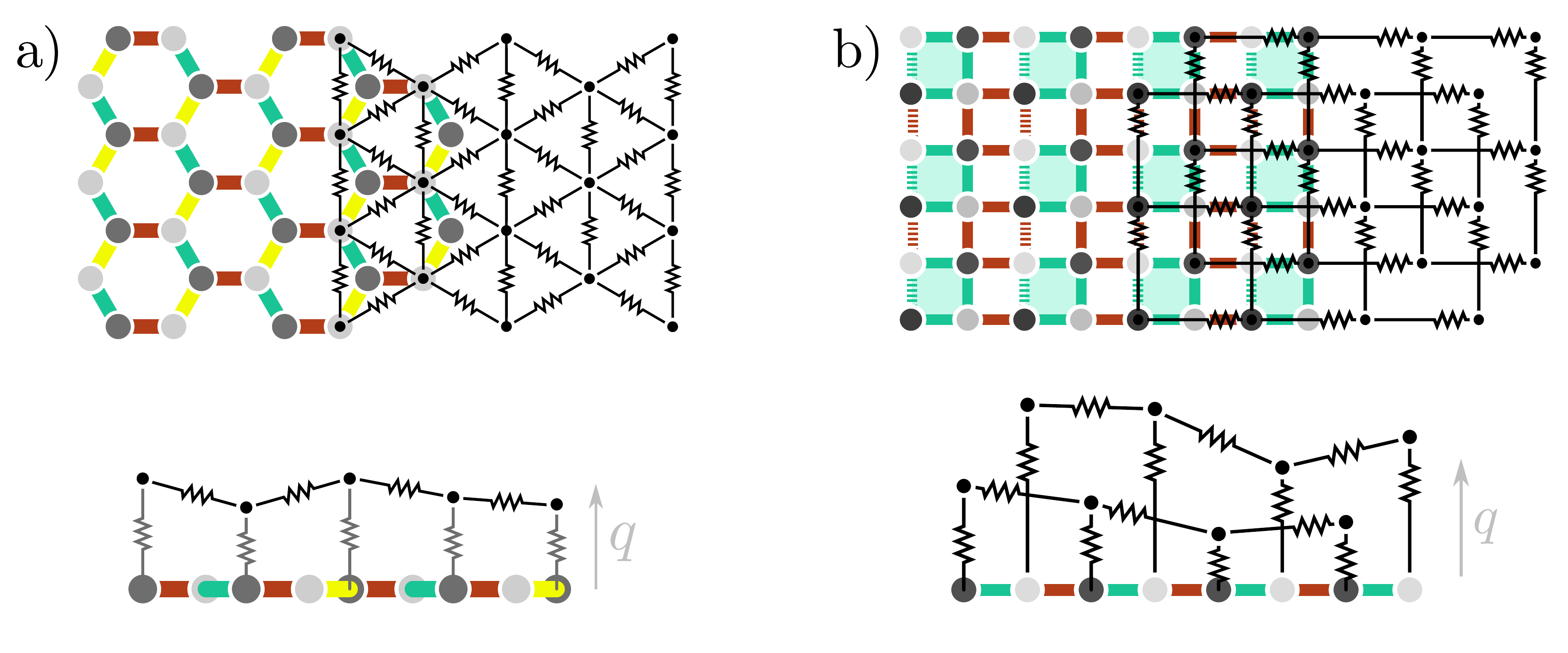}
    \caption{\textbf{Mechanical analogs of quantum spin liquids.} Left: (a) The mechanical Kitaev model is a balls-and-springs model on the triangular lattice, where the masses at each site are restricted to a movement along the axis perpendicular to the lattice plane (see the side view in the lower row). Right: (b) Mechanical analog of a second-order spin liquid on the square lattice. }
    \label{fig:MechanicalModels}
\end{figure*}

\subsection{Mechanical analog of the Kitaev spin liquid}
\label{sec:MechanicalKSL}

The Kitaev spin liquid \cite{Kitaev2006} is a paradigmatic state of matter where spin degrees of freedom {\sl fractionalize} into itinerant Majorana fermions and a static $\mathbb{Z}_2$ gauge structure. On the honeycomb lattice, the Majorana fermions exhibit a gapless Dirac spectrum in the absence of any time-reversal symmetry breaking terms and (roughly) isotropic couplings along the three distinct lattice directions. 
For strongly anisotropic couplings the Majorana band structure gaps out and one transitions into a topological phase that is primarily defined via its gauge structure as a $\mathbb{Z}_2$ toric code.
The magic of the Kitaev model is that all these statements have a rigorous analytical foundation. The original spin model with its bond-directional Ising-like exchange terms can be recast in terms of auxiliary Majorana fermion degrees of freedom as a quadratic Majorana Hamiltonian 
\begin{align}
	\label{eq:KitaevModel}
	{\cal H}_{\rm Kitaev} = \sum_{\langle i,j \rangle_\alpha} S_i^\alpha S_j^\alpha
					=  -i\gamma^A_i {\bf A}_{ij} \gamma^B_j + {\rm h.c.} \,,
\end{align}
that allows for an exact solution \cite{Kitaev2006}. 

But with an eye on our SUSY correspondence, we note that the right-hand-side of \eqref{eq:KitaevModel} is precisely of the form of the fermionic Hamiltonian \eqref{Maj_ham1} used in our SUSY correspondence for real fermions and bosons discussed above. The matrix ${\bf A}$ encodes the hopping of the Majorana fermions described by operators $\gamma^A$ and $\gamma^B$ on the underlying honeycomb lattice (with the two triangular sublattices $A$ and $B$ as shown in Fig.~\ref{fig:MechanicalModels}). Following the steps outlined in Section~\ref{subsec:real_supersymmetry}, we can write down its SUSY partner in terms of real bosons, take its classical limit, and arrive at a balls and springs model on the triangular lattice \footnote{Our model can also be considered to be a particularly simple balls and springs incarnation of `mechanical graphene', which has been first put forward in \cite{socolar2017mechanical}.} as illustrated in Fig.~\ref{fig:MechanicalModels}~a) and b). Each mass, located at a site of this triangular lattice, is restricted to a movement along an axis {\sl perpendicular} to the lattice plane, and is connected via two types of springs to both the plane and its neighboring masses.

This mechanical Kitaev model retains many interesting features of its quantum mechanical counterpart \cite{Attig2019}.
For instance, its mechanical response exhibits the Dirac spectrum familiar from the quantum model near isotropic coupling -- a notable feature for the mechanical system in the sense that on the level of individual springs, one always has quadratic excitation spectra and the formation of a linear dispersion constituting the Dirac cone is as such direct evidence of many-body physics. We find that, also in this mechanical model, the Dirac cones have nonzero windings when the supersymmetric Berry phase \eqref{eq:susyberryrealbosons3} is computed, with its quantized value $\pm \pi$ solely attributed to the closed-path integral of the SUSY-induced additional term in \eqref{eq:susyberryrealbosons1}. When moving away from isotropic couplings into the strongly anisotropic toric-code limit of the quantum model, the mechanical analog will again go along and now exhibit a {\sl gapped} excitation spectrum. This is again a somewhat unusual situation for a mechanical system as  
the opening of a gap and the absence of any well-defined low-energy modes implies that the mechanical system remains rigid for low-frequency drives -- again a many-body phenomenon since individual springs defy any small-frequency rigidity.

Notably, such a mechanical Kitaev model not only retains classical analogs of the itinerant Majorana fermions in its propagating phonon modes but also exhibits its underlying $\mathbb{Z}_2$ gauge structure with classical analogs of the static vison excitations of the quantum model. For instance, one can excite a pair of such gauge excitations by switching the sign of an intersite interaction, just like in the quantum case, only that it is a spring constant in the classical context. Individual visons can then be moved around by flipping additional spring constants, \eg to study braiding in the classical model. In a similar spirit, a Majorana version of the Kekule tight-binding model studied in Ref.~\onlinecite{iadecola2016non} would lead to a mechanical analog through our (real) SUSY prescription; adopting a suitably designed dynamical protocol can enable braiding of the topological zero modes therein. One could also set up a scattering experiment where propagating phonon modes and spatially arranged gauge excitations give rise to interference effects, akin to studying the scattering of itinerant Majorana fermions in the presence of massive vison excitations in the quantum model. 

%%%%%%%%%%%%%%%%%%%%%%%%%%%%%%%%%%%%%%%%%%%%%%%%%%%%%%
\subsection{Mechanical analog of higher-order topological insulator}
\label{sec:MechanicalHOTI}     

Being restricted to symmetry class BDI for the Majorana fermion models in our SUSY correspondence of real bosons and fermions, we, unfortunately, miss one of the more interesting features of the Kitaev honeycomb model -- the formation of a gapped topological phase of the Majorana fermions in the presence of time-reversal symmetry breaking (which brings the system into symmetry class D). But despite the absence of such a phase in the classification of topological insulators in symmetry class BDI in two spatial dimensions \cite{Chiu2016classification}, there is still another possibility for the formation of highly non-trivial band structure -- the formation of higher-order topological insulator (HOTI) \cite{hughes-bernevig_quadrupole_insulators}.

\begin{figure}[t]
    \centering
    \includegraphics[width=\columnwidth]{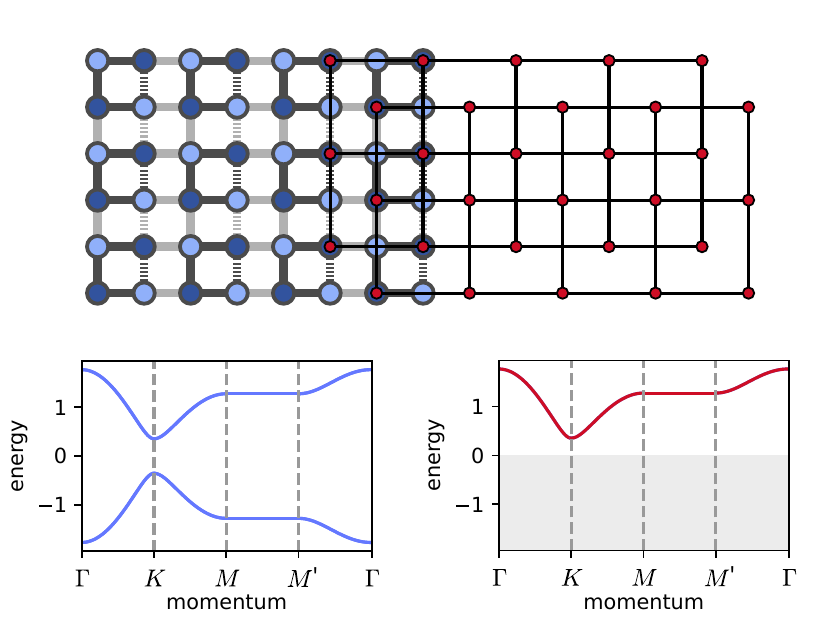}
    \caption{\textbf{SUSY for a Majorana fermion higher-order topological insulator.} Majorana fermions on a decorated square lattice have an isospectral SUSY partner in the form of real bosons on two disjoint copies of the square lattice. Eigenmodes in these systems agree for positive energies.}
    \label{fig:MechanicalSOTI}
\end{figure}

Such a HOTI for itinerant Majorana fermions  
has been discussed in the context of a second-order Kitaev spin liquid \cite{vatsal_HOTI_magnet} where the non-trivial topology manifests itself in gapless Majorana {\sl corner modes}. While the original spin model, in that case, has been formulated on the five-coordinated Shastry-Sutherland lattice (in order to retain the exact solvability of the spin model), we can here proceed with a somewhat simpler starting point and follow the work of Benalcazar {\it et al}. \cite{hughes-bernevig_quadrupole_insulators} in defining a free Majorana fermion model on the square lattice with staggered couplings among the elementary plaquettes, as illustrated in Fig.~\ref{fig:MechanicalSOTI}.
Every such elementary plaquette shall be pierced by a $\pi$-flux so that the Bloch Hamiltonian takes the form
\begin{align}
      {\bf H}_{\rm HOTI}({\bf k}) = i
      \begin{pmatrix}
         & {\bf A}^\dagger({\bf k}) \\ 
         -{\bf A}({\bf k}) &  
       \end{pmatrix},
\end{align}
where 
\begin{align}
        {\bf A}({\bf k}) =
        \begin{pmatrix}
         t + \lambda e^{-ik_x} & -t - \lambda e^{ik_y} \\
         t + \lambda e^{-ik_y} & \phantom{-}t + \lambda e^{ik_x}
        \end{pmatrix},
\end{align}
with the coupling parameters $t$ and $\lambda$ denoting the interplaquette and intraplaquette hopping strengths, respectively. 
Feeding this Majorana Hamiltonian into our SUSY correspondence, deriving its real bosonic SUSY partner, and taking its classical
limit, we arrive at a balls-and-springs model (Fig.~\ref{fig:MechanicalModels}) whose onsite and intersite springs have spring constants $t^2+\lambda^2$ and $-t\lambda$, respectively, and the masses are restricted to move along an out-of-plane axis only.  

This mechanical analog of a Majorana HOTI (or the second-order Kitaev spin liquid of Ref.~\onlinecite{vatsal_HOTI_magnet}) retains, by construction, the gapped spectrum (Fig.~\ref{fig:MechanicalSOTI}) and the topological response of its quantum mechanical ancestor. Most strikingly, this includes the formation of floppy (gapless) corner modes, which on the mechanical level are tightly interwoven with the onsite spring couplings. The latter are sensitive to the number of neighbors, which, in a setup with open boundary conditions (\ie a system with actual edges and corners), gives rise to a spatial variation of these couplings along the boundary of the mechanical system. Notably, a non-Abelian extension of the supersymmetric Berry connection in \eqref{eq:susyberryrealbosons1} also enables constructing {\em bosonic} Wilson loop operators to decode the topology in this mechanical analog of a HOTI. Such a bosonic topological phase is characterized, like their fermionic counterparts \cite{hughes-bernevig_quadrupole_insulators}, by a {\em quantized quadruple moment} that, in this case, follows from the eigenvalues of the bosonic Wilson loops as discussed in detail in Ref.~\onlinecite{vatsal_HOTI_magnet}. 
When the coupling parameters of the mechanical model are steered away from the topological trivial regime -- where the corner masses experience a restoring force -- into the topological regime, these corner masses become essentially free, \ie they can be moved arbitrarily far from their equilibrium positions without any restoring force. This is the mechanical equivalent of a floppy corner mode.

%%%%%%%%%%%%%%%%%%%%%%%%%%%%%%%%%%%%%%%%%%%%%%%%%%%%%%
% Outlook
%%%%%%%%%%%%%%%%%%%%%%%%%%%%%%%%%%%%%%%%%%%%%%%%%%%%%%

\section{Outlook}
Our work introduces an application of SUSY to identify free theories of bosons and fermions as superpartners on an extensive variety of lattice geometries that are of high relevance to several condensed matter setups from frustrated magnets to superconductors. At its core, the formulation entails a prescription of squaring and square-rooting Hamiltonians, a mapping that can be elegantly understood in terms of lattice adjacencies in arbitrary spatial dimensions. This mapping then gives rise to topological invariants associated with the single-particle Witten index $\nu$ of the SUSY theory and a five-fold way classification of the supercharge. 

The mapping from bosons to fermions is a specific highlight for it opens the door to the discovery and/or control of unexpectedly low energy modes in bosonic systems. One could, for example, use this mapping to scan through materials databases seeking topologically protected phonon/magnon zero modes in associated nearest neighbor truncated phonon/magnon models. Next neighbor perturbations would then lift these zero modes to finite energy but leave them at an unexpectedly low energy due to locality. In addition to scanning through a materials database (such as Ref.~\cite{li2021computation} for topological phonons), one could also use the mapping to design materials with these unexpected low energy modes. In the end, such a program, with an unwanted finite locality replacing the unwanted occupation of bulk modes, is similar to the current search and design of electronic \cite{slager2013space,kruthoff2017topological,Bradlyn2017,Regnault2022,Elcoro2021,Xu2020b} as well as, more recently, phononic topological materials \cite{peng2022phonons}. 

A conceptually interesting aspect of the SUSY is that it also allows to establish a connection between the Riemannian structures associated with the manifolds of quantum states (Bloch states) of different lattice models that share identical band structures. Such information is encoded in the {\sl quantum geometric tensor} \cite{provost1980riemannian} 
\[
\chi_{ij}^m=\langle \partial_{k_i}u_m|(1-|u_m\rangle\langle u_m|)|\partial_{k_j}u_m\rangle \,,
\] 
whose real part yields the quantum metric $g_{ij}$ measuring the distance between nearby quantum states and the imaginary part, the Berry curvature \cite{cheng2010quantum}. We observe that, for a number of lattice models connected by SUSY, the imaginary part, i.e. the Berry curvature, appears to be identical for the fermionic and bosonic counterparts, but there is a small, but nonvanishing difference in the associated metric elements $g_{ij}$. This should have an immediate impact on physical observables that connect to the quantum metric, which include the anomalous Hall conductivity in a nonuniform electric field \cite{kozii2021intrinsic} or the diamagnetic response of flat band superconductors \cite{torma2018quantum}, and should warrant future exploration, particularly exploring these phenomena in SUSY-related bosonic systems.

Another interesting aspect of our SUSY framework is that it also encapsulates theoretical recipes to realize `square-root topological phases' \cite{Schomerus2017,Mizoguchi2020,Ezawa_2020,Mizoguchi2021} including the $2^n$-th root series \cite{Marques_2021, Marques20212n, dias2021matryoshka}. In addition, it gives way to envision topological states in bosonic systems by squaring, an approach which, so far, has received limited attention \cite{Xu2020}, despite the recent proposal \cite{bomantara2022square} that periodically driven systems could unveil new possibilities to realize such exotic phases in the lab. For generic quadratic bosonic systems generated by our SUSY correspondence, a local topology-preserving map can be constructed to reduce them to number-conserving Hamiltonians \cite{Chaudhary_2021}. Studying how SUSY influences such mappings in the space of their fermionic counterparts is an interesting future direction. 

While the present work pertains to the non-interacting limit of bosonic and fermionic systems, extending it to interacting systems, such as in the model of \eqref{eq:InteractingSUSY}, could solve a number of interesting puzzles in the field of strongly correlated electrons. It could be that Fendley's discovery \cite{Fendley_2016} of a SUSY-like mapping between states in the transverse field Ising model/Kitaev chain system may point the way. It is `SUSY-like' because the mapping preserves the states' energy up to small corrections in the system size. So there is reason to believe that an interacting version of the phenomena discussed in this paper, especially of the Kitaev chain example we discuss, indeed exists. Taking a cue from a recently identified strategy for square-rooting the functions of bosonic number operators \cite{Fiete2020}, Heisenberg spin models may also serve as a fertile playground for extending our work to interacting systems.  Lastly, if such an interacting version of SUSY exists, it may also provide insights into $\mathbb{Z}_2$ quantum spin liquids whose fusion rule $e\times m=\varepsilon$ identifies bosonic spinons paired with $\mathbb{Z}_2$ visons as fermionic spinons \cite{Essin2013} but are difficult to study beyond exactly solvable models. Perhaps the right step towards solving these puzzles is to connect the SUSY discussed in this manuscript, which requires locality and is revealed by lattice geometry, to those discovered to describe certain interacting phase transitions on the surface of certain symmetry-protected topological insulators \cite{Grover2014,Yao2017,Li2017}.

The most pressing need for such an interacting extension could be to explain the origin of intertwined orders \cite{Fradkin2015} in the cuprates and other high-temperature superconductors. Intertwined orders arise due to a seemingly fine-tuned competition between several ordering tendencies. Physicists working on these systems did not expect accidental degeneracies. They proposed them only after decades of work and surprising discoveries. However such accidental low-energy phenomena are a natural consequence of the topology discussed in this paper. While our models exhibit strictly zero-energy modes, actual materials would have soft modes at finite but low energy, parametrically small in the strength of further neighbor interactions. If someone constructs such a theory, then perhaps it would indeed solidify SUSY as a pillar of condensed matter physics.

%%%%%%%%%%%%%%%%%%%%%%%%%%%%%%%%%%%%%%%%%%%%%%%%%%%%%%
% Acknowledgments
%%%%%%%%%%%%%%%%%%%%%%%%%%%%%%%%%%%%%%%%%%%%%%%%%%%%%%

\begin{acknowledgments}
We thank V. Dwivedi, A. Rosch, and in particular M. Zirnbauer for enlightening discussions. The Cologne group acknowledges partial funding from the DFG within Project-ID 277146847, SFB 1238 (project C02), and within Project-ID 277101999, CRC 183 (project A04). KR thanks the sponsorship, in part, by the Swedish Research Council during the initial stage of this work. ST thanks the Center for Computational Quantum Physics at the Flatiron Institute, New York, for hospitality in the final stages of writing this manuscript. The numerical simulations were performed on the CHEOPS cluster at RRZK Cologne. This material is based upon a collaboration started at the Kavli Institute for Theoretical Physics, performed in part at the Aspen Center for Physics, and supported by the National Science Foundation under Grants No. NSF OAC-1940243, NSF PHY-1748958, and NSF PHY-1607611. 
\end{acknowledgments}

%%%%%%%%%%%%%%%%%%%%%%%%%%%%%%%%%%%%%%%%%%%%%%%%%%%%%%
% Bibliography
%%%%%%%%%%%%%%%%%%%%%%%%%%%%%%%%%%%%%%%%%%%%%%%%%%%%%%

\bibliography{susy}

%%%%%%%%%%%%%%%%%%%%%%%%%%%%%%%%%%%%%%%%%%%%%%%%%%%%%%
% Appendices
%%%%%%%%%%%%%%%%%%%%%%%%%%%%%%%%%%%%%%%%%%%%%%%%%%%%%%

\newpage
\appendix

%%%%%%%%%%%%%%%%%%%%%%%%%%%%%%%%%%%%%%%%%%%%%%%%%%%%%%
% Square-rooting algorithm
%%%%%%%%%%%%%%%%%%%%%%%%%%%%%%%%%%%%%%%%%%%%%%%%%%%%%%

\section{Graph square-rooting algorithm}
\label{app:square_rooting_algorithm}

The algorithm for taking graph square roots discussed in this manuscript was first introduced as a ``lattice construction" algorithm in Ref.~\cite{Attig2017}. Here we derive it in a more conceptual manner, starting solely from squaring an adjacency matrix, 
\begin{equation}
    (\mathbf{A}^2)_{ik} = \sum_j \mathbf{A}_{ij} \mathbf{A}_{jk} \,.
    \label{eq:graph_edge_weights_squaring_Appendix}
\end{equation}
It is apparent that an edge between vertices $i$ and $k$ in the \emph{squared} graph corresponds to (the sum of all) combinations of edges going from vertex $i$ to $k$ via intermediate vertices $j$. In more practical terms, this constitutes the \emph{next-nearest neighbors} of the graph.

It is well known that in a bipartite graph, the next-nearest neighbors of one of the two sublattices only reside within that sublattice. Therefore, if one would construct a bipartite graph for a given lattice in such a way that the given lattice arises from taking next-nearest neighbors in the bipartite graph, one would naturally define a \emph{second sublattice} which can be viewed as the superpartner to the given lattice.

The last piece of the algorithm is to show how exactly one can define a bipartite graph in such a way for a given lattice. This construction is at the heart of the lattice construction algorithm and relies on the observation that locally, a $z$-coordinated site of one sublattice in the bipartite graph results in a fully connected plaquette (or clique in graph language) with $z$ vertices in the other sublattice upon squaring. c.f. Fig.~\ref{fig:lattice-construction-plaquettes}. Therefore, to go the other way, we propose to replace all $z$ cliques with $z$-coordinated sites, constituting the new sublattice and thus producing the bipartite graph.

For practical applications, note that every $z$ clique contains cliques of order $z-1$. One therefore needs to mandate that the algorithm should start by replacing maximal plaquettes in descending order of their size. This will guarantee that one replaces a minimal amount of plaquettes. It is further known from computer science that finding all maximal cliques in a graph is generally an NP-hard problem. However, in a physics situation, the graphs one usually has to deal with are lattices whose unit cells have a rather finite number of elements which in turn puts an upper limit on how large cliques can become and thereby greatly alleviates the problem of clique identification.

Note also, that the replacement of plaquettes can only guarantee the connectivity of the graph, but not its labels. In a physical situation, hopping (or interaction) strengths would appear as labels in the adjacency matrix, which further complicates the calculation of the labels within the bipartite graph. 

In total, the entire lattice construction algorithm can be summarized as follows, assuming one is given a lattice with connectivities labeled by the set of numbers $\{{\bf t}_{ik}\}$:
\begin{enumerate}
    \item Identify all cliques within the lattice.
    \item In descending order let ${\mathcal X}$ be either the full set or a subset of cliques that includes all the vertices and do the following to every clique in ${\mathcal X}$:
    \begin{enumerate}
            \item Add a new vertex to the center.
            \item Remove the edges.
            \item Add new edges between the new vertex and the old vertices.
    \end{enumerate}
    {\bf Assert:} All the old edges of the initial lattice are removed, and only the newly added edges are retained.  
    \item Assign labels ${\bf A}_{ij}$ to the newly added edges; reduce the number of labels using graph symmetries. The result is a matrix of the form
    \begin{equation}
        {\bf A} = \begin{pmatrix} & {\bf A}_{\rm I-II}\\ {\bf A}_{\rm II-I}&\end{pmatrix}
    \end{equation}
    \item Use the following to produce a system of equations that link the newly created labels ${\bf A}_{ij}$ to the old labels
    \begin{align}
        {\bf t}_{ik} &= \sum_{j}{\bf A}_{ij}{\bf A}_{jk}\,,
    \end{align}
    where ${\bf t}_{ik}$ is non-zero only when $i$ and $k$ refer to the old vertices.
    \item Solve these equations to obtain fitting labels for the bipartite lattice. {\bf Assert:} a solution is found.
    \item Construct the partner of ${\bf t}_{ik}$ from \eqref{eq:graph_edge_weights_squaring_Appendix} with $i$ and $k$ now referring to the new vertices.
\end{enumerate}

This algorithm provides a solution, a partner lattice, to the square rooting problem, but other solutions are possible. The line graph algorithm is essentially the same algorithm as above, but step 2 is replaced by starting from the smallest cliques, i.e., the edges (see \ref{sec:otherlatticecorrespondences}). Once the algorithm is performed for these cliques, all higher cliques are automatically removed, for larger cliques always contain the smaller ones. Also, if step 2 is applied a second time to the partner lattice, it does not necessarily yield the original lattice. It may produce yet another partner. Hence, backtracing step 2, in addition to the line graph construction, is another replacement for this step.

The labels in step 3 are generally complex numbers. In this manuscript, we have limited ourselves to choosing real numbers and to time-reversal invariant problems.

Step 5 involves solving a non-linear system of equations that may not always yield a solution or may yield multiple solutions. A solution was found in all cases we studied in this manuscript. But for problems with multiple Wyckoff centers in the unit cell, a solution may not be possible. Still, it could be obtained by fine-tuning just one or two parameters, such as experimentally, by applying pressure or temperature. This case would identify SUSY regions in the phase diagram.

In summary, this algorithm provides a method to generate a partner lattice of a given lattice. It does not deliver a unique solution but only one compliant with graph rules and squaring. 

%%%%%%%%%%%%%%%%%%%%%%%%%%%%%%%%%%%%%%%%%%%%%%%%%%%%%%
% Five-fold way classification of SUSY models 
%%%%%%%%%%%%%%%%%%%%%%%%%%%%%%%%%%%%%%%%%%%%%%%%%%%%%%

\section{Five-fold way classification of SUSY models}
\label{app:classification}

In this Appendix, we will present a detailed discussion of the symmetry-based classification of Hermitian supercharges. A key result of this approach is that by imposing the conventional symmetries that classify the Hamiltonians of free fermionic systems \cite{Schnyder2008classification, Kitaev2009periodic, ryu2010topological, ludwig2015topological}, the supercharge can belong to one of the five distinct classifying spaces, which leads us to a {\em five-fold way classification} as will be elaborated in the following. 

A supercharge is fundamentally characterized by the Witten index $\nu$: its Bloch spectrum features precisely $\nu$ flat bands at zero energy. The topology of additional band-degeneracy points at zero energy coinciding with the flat bands, so-called {\em nexus points} \cite{Heikkil2015, Bradlyn2016, chang2017nexus, das2020topological}, has a remarkable dependence on this Witten index $\nu$ that can be understood in terms of homotopy maps \cite{roychowdhury2018classification}. 
Of specific interest in the parlance of condensed matter setups are: 
(i) the zeroth homotopy group $\pi_0({\cal M})$, which is the set of mappings of a single point into a classifying (or topological) space $\cal M$ and can lead to topological invariants like {\rm sign}[Pfaffian] (the emergence of such an invariant has been discussed in, \eg Refs.~\cite{bzduvsek2017robust} and \cite{Roychowdhury2018}), 
(ii) the first or the fundamental homotopy group $\pi_1({\cal M})$, which is the set of mappings of loops into $\cal M$ and gives rise to invariants such as the Berry phase, 
(iii) the second homotopy group $\pi_2({\cal M})$, the set of mappings of closed surfaces into $\cal M$ leading to invariants like the Chern number, and (iv) the third homotopy group $\pi_3({\cal M})$, the set of mappings of closed volumes into $\cal M$ characterizing, \eg Hopf insulators \cite{moore2008topological}. In what follows, we will elucidate the five-fold classification of supercharges for each value of $\nu$ specializing in the homotopy groups $\pi_n$ from $n=0$ to $3$. 
In a subsequent Appendix, Appendix~\ref{app:topologymodelexamples}, we will provide a number of examples by classifying the topology of some of the lattice models in two and three dimensions that are discussed in the main text.

Let us start our discussion by considering the matrix representing the Hermitian supercharge ${\cal Q}_{\rm H}$ in \eqref{eq:complexsusy_charge}, which, by the very nature of our SUSY construction, is chiral, \ie it takes the general form
\begin{align}
\label{susyH1}
{\bf H} = \begin{pmatrix}
      & {\bf R} \\
      {\bf R}^\dagger &
    \end{pmatrix} \,.
\end{align} 
Symmetry-wise it, therefore, belongs to one of the five chiral classes AIII, BDI, CII, CI, or DIII in the ten-fold way classification \cite{ryu2010topological}. Table~\ref{table:classification11} summarizes how the time-reversal ${\cal T}$ and the particle-hole ${\cal P}$ operator behave in each one of these classes (while all have the chiral symmetry, \ie ${\cal C}^2=1$) and the resultant classifying spaces ($\cal M$) of the (spectrally-flattened) Hamiltonians \cite{Kitaev2009periodic}.

\begin{table}[t]
\begin{tabular}{|c|c|c|c|c|}
\hline
Cartan label & ${\cal T}^2$ & ${\cal P}^2$ & ${\cal C}^2$ & Classifying space \\ 
\hline\hline
AIII & 0 & 0 & 1 & $U(N)$ \\
\hline
BDI & 1 & 1 & 1 & $O(N)$ \\
\hline
CII & -1 & -1 & 1 & $Sp(2N)$ \\
\hline
CI & 1 & -1 & 1 & $U(N)/O(N)$ \\
\hline
DIII & -1 & 1 & 1 & $U(2N)/Sp(2N)$ \\
\hline
\end{tabular}
\caption{{\bf Five chiral symmetry classes} which a chiral Hamiltonian can belong to. The last column is of the corresponding classifying spaces of the K-theory \cite{Kitaev2009periodic}. The notation  `$/$' here refers to the coset space.}
\label{table:classification11}
\end{table}

The antiunitary symmetry operators, ${\cal T}$ and ${\cal P}$, and the unitary symmetry operator ${\cal C}$ in each one of these classes impose specific conditions on the matrix ${\bf H}$ when we consider it as a first-quantized Hamiltonian, namely,
\begin{align}\label{symonH}
 {\cal T}{\bf H}{\cal T}^{-1}={\bf H}~;~{\cal P}{\bf H}{\cal P}^{-1}=-{\bf H}~;~{\cal C}{\bf H}{\cal C}^{\dagger}=-{\bf H}\,.
\end{align}
Adapting a canonical representation of these symmetries in each of the five classes of Table~\ref{table:classification11}, one can understand what these conditions have to imply for the matrix ${\bf R}$ which we list in Table~\ref{table:classification22}.
\begin{table}[b]
\begin{tabular}{|c|c|c|c|c|}
\hline
Cartan & \multirow{2}{1em}{${\cal T}$} & \multirow{2}{1em}{${\cal P}$} & \multirow{2}{1em}{${\cal C}$} & \multirow{2}{6.5em}{condition on $\bf R$} \\  
label & & & &  \\
 \hline\hline
AIII & --- & --- & $\sigma_3$ & --- \\
\hline
BDI & ${\cal K}$ & $\sigma_z{\cal K}$ & $\sigma_3$ & $\bf R^*={\bf R}$ \\
\hline
CII & $\sigma_3\otimes i\sigma_2{\cal K}$ & $-1\otimes i\sigma_2{\cal K}$ & $\sigma_3\otimes 1$ & $(i\sigma_2){\bf R}^*(-i\sigma_2)={\bf R}$ \\
\hline
CI & $\sigma_1{\cal K}$ & $-i\sigma_2{\cal K}$ & $\sigma_3$ & $\bf R^T={\bf R}$ \\
\hline
DIII & $i\sigma_2{\cal K}$ & $\sigma_1{\cal K}$ & $\sigma_3$ & $\bf R^T=-{\bf R}$ \\
\hline
\end{tabular}
\caption{{\bf Canonical representation of the symmetries in the five chiral classes.} $\cal K$ denotes complex conjugation.
The last column enlists the condition to be obeyed by the matrix ${\bf R}$ in each of these classes in compliance with \eqref{symonH}.}
\label{table:classification22}
\end{table}

Let us now remind us of the fact that SUSY renders a Witten index $\nu$ to each ${\bf R}$ in terms of its dimensions: for ${\bf R}$ being an arbitrary $M\times N$ matrix, $\nu=N-M$. Therefore, when ${\bf R}$ is a square matrix, which is referred to as the isostatic case, the Witten index is $\nu=0$ whereas for a rectangular $\bf R$ matrix, the nonisostatic cases, $\nu\neq 0$, the latter exclusively manifesting in flat bands in conjunction with (degenerate) point nodes in the band structures of $\bf H$. 

For the flat bands in a SUSY problem always appear at zero energy, their topology cannot be classified in terms of the ten-fold way. In fact, for the first three classes -- AIII, BDI, and CII -- the different values of $\nu$ (the isostatic and the nonisostatic cases) exhibit distinct topology as discussed by two of us \cite{roychowdhury2018classification} in terms of homotopy groups $\pi_n$ (mapping the $n$-dimensional sphere ${\cal S}_n$ to the classifying spaces ${\cal M}$ in Table~\ref{table:classification11}. For the last two classes -- CI and DIII -- ${\bf R}$ must be a square matrix and one can as well compute the stable homotopy groups of the respective classifying spaces (following Bott's original work~\cite{bott1959stable}), which will be {\em independent} of $\nu$. This leads us to the aforementioned five-fold way classification of the supersymmetric lattice models, \ie the supercharges, which is shown in Table~\ref{table:classification33}, where we list the possible topological invariants, for each of the five symmetry classes, as a function of the Witten index and the homotopy groups up to $\pi_3$ relevant for models in $d=1,2,3$ dimensions. For the reader's convenience, we also reorganize the results from this table for specific Witten indices in
Tables \ref{table:classification34} ($\nu = 0$) and \ref{table:classification35} ($\nu = 1$).

\begin{table}[t]
  \begin{tabular}{|c|c|cccc|}
  \hline
   Cartan label &  $\quad \nu \quad$ & $\quad \pi_0 \quad$ & $\quad \pi_1 \quad$ & $\quad \pi_2 \quad$ & $\quad \pi_3 \quad$
   \\
  \hline\hline
  \multirow{3}{2em}{AIII} & 0 &  0 & $\mathbb{Z}$ & 0 & $\mathbb{Z}$ \\
  & 1 & 0 & 0 & $\mathbb{Z}$ & 0 \\
  & $\geq$ 2 & 0 & 0 & 0 & 0 \\
  \hline\hline
  \multirow{5}{2em}{BDI} & 0 & $\mathbb{Z}_2$ & $\mathbb{Z}_2$ & 0 & $\mathbb{Z}$ \\
  & 1 & 0 & $\mathbb{Z}_2$ & 0 & $\mathbb{Z}$ \\
  & 2 & 0 & 0 & $\mathbb{Z}$ & $\mathbb{Z}$ \\
  & 3 & 0 & 0 & 0 & $\mathbb{Z}$ \\
  & $\geq$ 4 & 0 & 0 & 0 & 0 \\
  \hline\hline
  \multirow{2}{2em}{CII} & 0 & 0 & 0 & 0 & $\mathbb{Z}$ \\
  & $\geq$ 1 & 0 & 0 & 0 & 0 \\
  \hline\hline
  \multirow{2}{2em}{CI} & 0 & 0 & $\mathbb{Z}$ & $\mathbb{Z}_2$ & $\mathbb{Z}_2$ \\
  & $\geq$ 1 & 0 & $\mathbb{Z}$ & $\mathbb{Z}_2$ & $\mathbb{Z}_2$ \\
  \hline\hline
  \multirow{2}{2em}{DIII} & 0 & 0 & $\mathbb{Z}$ & 0 & 0 \\
  & $\geq$ 1 & 0 & $\mathbb{Z}$ & 0 & 0 \\
  \hline
  \end{tabular}
  \caption{{\bf Five-fold way classification table of SUSY lattice models.} 
   For each of the five chiral symmetry classes, the table indicates topological invariants ($\mathbb{Z}_2$, $\mathbb{Z}$) as a function of Witten index $\nu$ and homotopy groups $\pi_n$. It thereby generalizes Table \ref{table:classificationBDI} of the main text, which is specific to symmetry class BDI. For symmetry classes AIII, BDI, and CII, Witten indices $\nu$ greater than the listed values exhibit trivial topology, while for the class CI and DIII, the topology turns out to be the same for any $\nu\neq 0$ and $\nu=0$. For the reader's convenience, we have also resorted the results from this Table for fixed Witten indices in Tables \ref{table:classification34} ($\nu = 0$) and \ref{table:classification35} ($\nu = 1$).}
  \label{table:classification33}
\end{table}

Note in the above discussion, the Witten index $\nu$ is of fundamental importance and not the spatial dimensionality of a model unlike what is typically done in the ten-fold way classification table to reveal strong invariants \cite{Schnyder2008classification, Kitaev2009periodic, ryu2010topological, ludwig2015topological}.
When computing a topological invariant of a lattice model, we often resort to the band structures in $d$ dimensions, characterized by the pseudomomenta ${\bf k}$, for which a $d$-dimensional torus $\mathbb{T}_d$ should be considered as the base manifold instead of the sphere ${\cal S}_d$. But in doing so, it turns out we only miss out on the information about weak topological invariants (which can exist only in translation symmetric systems). To note, in presence of such a parameter $\bf k$, the conditions noted in Table~\ref{table:classification22} become 
\begin{align}\label{constraints}
 &{\rm BDI:}~~\bf R^*({\bf k})={\bf R}(-{\bf k}) \,, \nonumber \\
 &{\rm CII:}~~~(i\sigma_2){\bf R}^*({\bf k})(-i\sigma_2)={\bf R}(-{\bf k}) \,, \nonumber \\
 &{\rm CI:}~~~~\bf R^T({\bf k})={\bf R}(-{\bf k}) \,, \nonumber \\
 &{\rm DIII:}~~\bf R^T({\bf k})=-{\bf R}(-{\bf k}) \,.
\end{align}
Here we seek to explore the topology of the nodes in the band structures of $\bf H$ at zero energy that involve flat bands. The relevant parameter is what is known as the {\em node dimensionality}, which is distinct from the spatial dimensionality. The node dimensionality turns out to be dictated primarily by the Witten index. For characterization of such nodes in the ten Atland-Zirnbauer (AZ) classes {\em without} the flat bands (\ie the isostatic case, $\nu=0$) in terms of homotopy groups, we refer to Ref.~\cite{bzduvsek2017robust} where this has previously been established. The present work generalizes these ideas to the non-zero Witten index $\nu \neq 0$ for the five chiral AZ classes.

\begin{table}[t]
  \begin{tabular}{|c|cccc|}
  \hline
   \multirow{2}{5.5em}{  Cartan label} & \multicolumn{4}{c|}{$\nu = 0$} \\
    &  $\quad \pi_0 \quad$ & $\quad \pi_1 \quad$ & $\quad \pi_2 \quad$ & $\quad \pi_3 \quad$
    \\
  \hline\hline
  AIII  &  0 & $\mathbb{Z}$ & 0 & $\mathbb{Z}$ \\
  BDI & $\mathbb{Z}_2$ & $\mathbb{Z}_2$ & 0 & $\mathbb{Z}$ \\
  CII  & 0 & 0 & 0 & $\mathbb{Z}$ \\
  CI & 0 & $\mathbb{Z}$ & $\mathbb{Z}_2$ & $\mathbb{Z}_2$ \\
  DIII & 0 & $\mathbb{Z}$ & 0 & 0 \\
  \hline
  \end{tabular}
  \caption{{\bf Five-fold classification table of SUSY lattice models} (the supercharges) for fixed Witten index $\nu=0$
  	providing the topological invariants for each symmetry class and the first four homotopy groups $\pi_n$.}
  \label{table:classification34}
\end{table}

\begin{table}[t!]
  \begin{tabular}{|c|cccc|}
  \hline
   \multirow{2}{5.5em}{  Cartan label} & \multicolumn{4}{c|}{$\nu = 1$} \\
    &  $\quad \pi_0 \quad$ & $\quad \pi_1 \quad$ & $\quad \pi_2 \quad$ & $\quad \pi_3 \quad$
    \\
  \hline\hline
  AIII  &  0 & 0 & $\mathbb{Z}$ & 0 \\
  BDI & 0 & $\mathbb{Z}_2$ & 0 & $\mathbb{Z}$ \\
  CII  & 0 & 0 & 0 & 0 \\
  CI & 0 & $\mathbb{Z}$ & $\mathbb{Z}_2$ & $\mathbb{Z}_2$ \\
  DIII & 0 & $\mathbb{Z}$ & 0 & 0 \\
  \hline
  \end{tabular}
  \caption{{\bf Five-fold classification table of SUSY lattice models} (the supercharges) for fixed Witten index $\nu=1$
  	providing the topological invariants for each symmetry class and the first four homotopy groups $\pi_n$.}
  \label{table:classification35}
\end{table}

The following Appendix will illuminate the topology of several lattice models all of which have a chiral Bloch Hamiltonian as in \eqref{susyH1} representing their Hermitian supercharge ${\cal Q}_{\rm H}$. All these examples belong to symmetry class BDI. We seek to find appropriate topological invariants that fit the BDI class in Table~\ref{table:classification33} by computing the relevant $\pi_n$ for a given Witten index $\nu$.  

%%%%%%%%%%%%%%%%%%%%%%%%%%%%%%%%%%%%%%%%%%%%%%%%%%%%%%
% Explicit examples
%%%%%%%%%%%%%%%%%%%%%%%%%%%%%%%%%%%%%%%%%%%%%%%%%%%%%%

\section{Topology of the SUSY lattice models in BDI class}
\label{app:topologymodelexamples}

\begin{figure*}
    \centering
    \includegraphics[width=2\columnwidth]{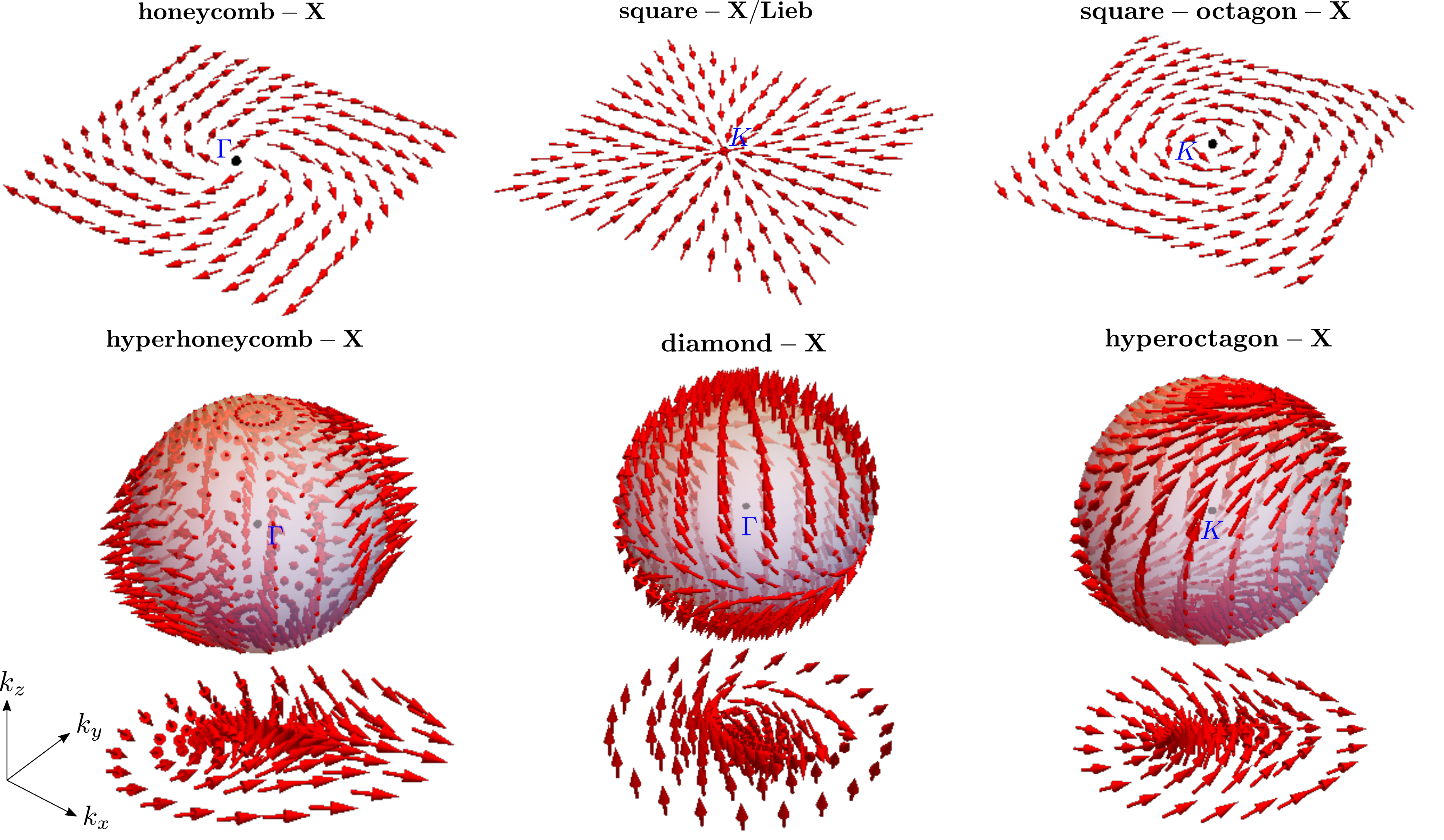}
    \caption{\textbf{Topological invariants from homotopy.} The three examples for two-dimensional lattice geometries in the top row exhibit a non-trivial homotopy associated with $\pi_1$, \ie a non-trivial Berry phase around the nexus points located at high-symmetry points in the Brillouin zone (indicated in blue). 
    Shown are the supercharge lattices honeycomb-X (Fig.~\ref{fig:susy-kagome-honeycomb}), square-X/Lieb lattice (Fig.~\ref{fig:susy_trio_square_checkerboard}), and square-octagon-X (Fig.~\ref{fig:susy-squareoctagon-squagome}). The three examples for three-dimensional lattice geometries in the bottom row, exhibit a non-trivial homotopy for $\pi_2$ (and are hence plotted over a unit sphere ${\cal S}_2$ and its projection onto $\mathbb{R}^2$), \ie they exhibit a non-trivial skyrmion number when projecting onto the two-dimensional plane (bottom row). Shown are the supercharge lattices hyperhoneycomb-X (Fig.~\ref{fig:susy-hyperhoneycomb}), diamond-X (Fig.~\ref{fig:susy-diamond-pyro}), and hyperoctagon-X (Fig.~\ref{fig:susy-hyperoctagon-hyperkagome}). The topology in these examples follows from the homotopy classification of the BDI class against the Witten index $\nu$ as shown in Table~\ref{table:classificationBDI}. For all examples plotted are the different components of the vector ${\bf d}({\bf k})$ that represents an effective Bloch Hamiltonian linearized around the band degeneracy points at zero energy [see \eqref{linearized}].}
    \label{fig:homotopy}
\end{figure*}

In order to appreciate the topology of SUSY models with non-zero Witten index $\nu\neq 0$, which we have classified in the previous Appendix, we return to some of the principal examples from the main text (illustrated in the triptych-like figures of Section \ref{sec:complex_SUSY}) and elucidate the non-trivial topological invariants with the nexus points of their respective supercharge models. The latter arise from nontrivial homotopy groups $\pi_1$ and $\pi_2$ in two- and three-dimensional lattice geometries, respectively. Note that all examples are in symmetry class BDI.

We will focus on the part of the band structures near the band degeneracy points at zero energy that include the flat bands and might exhibit additional band degeneracies, \eg in the form of Dirac crossings, at the so-called nexus points. The basic idea now is to project onto the relevant bands of the original chiral Bloch Hamiltonian ${\bf H}(\bf k)$ near such a band-degeneracy point and derive an effective Hamiltonian $\tilde{\bf H}(\bf k)$ involving only those bands. This effective Hamiltonian $\tilde{\bf H}(\bf k)$ is not generally in a chiral form but we can adopt a basis in which the chiral symmetry operator is diagonal and re-express $\tilde{\bf H}(\bf k)$ in that basis to restore its chiral form. Depending on its dimension $N$, $\tilde{\bf H}(\bf k)$ can be written in terms of the suitable SU$(N)$ generators ${\bf \Lambda}$ as 
\begin{align}\label{linearized}
    \tilde{\bf H}(\bf k) = {\tilde{\bf H}}_0 + {\bf d}({\bf k})\cdot \Lambda \,,
\end{align}
where ${\tilde{\bf H}}_0$ is the effective Hamiltonian at the degeneracy point. We will then inspect the profile of the vector ${\bf d}({\bf k})$ in the Fourier space (as illustrated in Fig.~\ref{fig:homotopy}) over loops or spheres encompassing the degeneracy point depending on the Witten index $\nu$ of the supercharge to begin with. 

Let us illustrate this procedure with examples for some of the lattice models with $\nu\neq 0$ studied in this manuscript in the following:\\

\noindent
{\em Honeycomb-X lattice} (Fig.~\ref{fig:susy-kagome-honeycomb}): For the honeycomb-kagome SUSY correspondence, the Hermitian supercharge yields a Bloch Hamiltonian on the intermediate honeycomb-X lattice that has a flat band and a three-fold degeneracy at the $\Gamma$-point of the Brillouin zone. With a Witten index of  $\nu=1$ in this case ${\bf R}({\bf k})$ is a $2\times 3$ matrix. Around the $\Gamma$-point, we obtain a $3\times 3$ effective Hamiltonian $\tilde{\bf H}(\bf k)$ which we linearize to obtain the vector ${\bf d}(\bf k)$. This vector turns out to have two components that, when plotted over the Brillouin zone, reveal a full winding over any closed path around the $\Gamma$-point (see Fig.~\ref{fig:homotopy}). 
We are therefore able to conclude that this model exhibits nontrivial topology, just as  Table~\ref{table:classification33} suggests for $\pi_1$ with $\nu=1$. As long as the degeneracy involving the flat band is indeed present, we find perturbations cannot destroy such a winding pattern rendering it a topological nature. \\

\noindent
{\em Square-X~/~Lieb lattice} (Fig.~\ref{fig:susy_trio_square_checkerboard}): For the square-X (Lieb) lattice, we have a Witten index of $\nu=1$ and ${\bf R}({\bf k})$ is a $1\times 2$ matrix. The original Block Hamiltonian ${\bf H}({\bf k})$ is $3\times 3$ matrix, and no further projection is required. The band structure has a three-fold degeneracy at $K=(\pm\pi,\pm\pi)$. The winding of the resultant two-component ${\bf d}({\bf k})$ vector, obtained by linearizing ${\bf H}({\bf k})$ around one of the $K$-points, is shown in Fig.~\ref{fig:homotopy} revealing a nontrivial topology due to $\pi_1$. \\

\noindent
{\em Square-octagon-X lattice} (Fig.~\ref{fig:susy-squareoctagon-squagome}): The square-octagon-X lattice mediates the SUSY correspondence between the square-octagon and the squagome lattice. This is an example with Witten index $\nu=2$ and an ${\bf R(k)}$ of dimensions $4\times 6$. A four-fold degenerate point is present along with two flat bands at the four corners of the Brillouin zone with $K=(\pm \pi/(2+\sqrt{2}),0)$ and $K'=(0,\pm \pi/(2+\sqrt{2}))$. Adapting a basis in which the chiral symmetry is diagonal, we find the effective $4\times 4$ Hamiltonian to take a chiral form with an effective $1\times 3$ $\tilde{\bf R}$-matrix. However, after linearizing around one of the $K$-points at play, we find the vector ${\bf d}({\bf k})$ is a two-component only. The winding of this vector around the $K$-point is shown in Fig.~\ref{fig:homotopy}, for which we need to immediately express a word of caution: while the plot might seem to suggest that there is a nontrivial topology associated with $\pi_1$ for this $\nu=2$ model (and a trivial topology for $\pi_2$), 
this is unexpected for BDI class at $\nu=2$, see Table \ref{table:classificationBDI}. To test if this is an accidental degeneracy, we added third-neighbor connections but that did not gap out the nexus point. One possible explanation is the existence of an additional symmetry, such as a lattice symmetry, that changes the class, but we have not investigated this issue any further.

In a similar fashion to the two-dimensional examples discussed above, we can now move on to analyze some three-dimensional SUSY models:\\  

\noindent
{\em Diamond-X lattice} (Fig.~\ref{fig:susy-diamond-pyro}): 
The Hermitian supercharge responsible for the SUSY correspondence between the diamond and pyrochlore lattices can be envisaged as a chiral Bloch Hamiltonian on the intermediate diamond-X lattice. This is another example of Witten index $\nu=2$ and an ${\bf R}({\bf k})$ of dimensions $2\times 4$ (akin to the square-octagon-X lattice discussed above). A four-fold degenerate point is present along with two flat bands at the $\Gamma$-point of the Brillouin zone.
After linearizing the effective Hamiltonian, which is a $4\times 4$ chiral matrix with an off-diagonal block $\tilde{\bf R}({\bf k})$ of dimensions $1\times 3$, around the $\Gamma$-point, we find the vector ${\bf d}({\bf k})$ is of three components. Plotted over the Brillouin zone (see Fig.~\ref{fig:homotopy}), it reveals a hedgehog texture surrounding the $\Gamma$-point arising from a nontrivial topology due to $\pi_2$ as Table~\ref{table:classification33} suggests. Projecting this texture onto two spatial dimensions reveals a skyrmion, a whirl in the texture which is another manifestation of the non-trivial topology at play here. \\

\noindent
{\em Hyperhoneycomb-X lattice} (Fig.~\ref{fig:susy-hyperhoneycomb}): The Hermitian charge mediating the SUSY correspondence between the hyperhoneycomb lattice and its SUSY partner is a Bloch Hamiltonian on the hyperhoneycomb-X lattice. This is a $\nu=2$ system with a fourfold degeneracy at the $\Gamma$-point of the Brillouin zone and two flat bands. After linearizing, the effective Hamiltonian takes the form a $4\times 4$ chiral matrix with an off-diagonal block $\tilde{\bf R}({\bf k})$ of dimensions $1\times 3$, around the $\Gamma$-point. We find the vector ${\bf d}({\bf k})$ is of three components. When plotted over the Brillouin zone (see Fig.~\ref{fig:homotopy}), it reveals a hedgehog texture surrounding the $\Gamma$-point (and a two-dimensional skyrmionic projection) arising from a nontrivial topology due to $\pi_2$ as Table~\ref{table:classification33} suggests. \\  

\noindent
{\em Hyperoctagon-X lattice} (Fig.~\ref{fig:susy-hyperoctagon-hyperkagome}): The Hermitian charge mediating the SUSY correspondence between the hyperoctagon lattice and its SUSY partner, the hyperkagome lattice, manifests as a Bloch Hamiltonian on the hyperoctagon-X lattice. This is yet another $\nu=2$ system with fourfold degeneracy at the $K$-point of the Brillouin zone and two flat bands. After linearizing around the $K$-point, the effective Hamiltonian takes the form of a $4\times 4$ chiral matrix with an off-diagonal block $\tilde{\bf R}({\bf k})$ of dimensions $1\times 3$. We find the vector ${\bf d}({\bf k})$ is of three components. Plotted over the Brillouin zone (see Fig.~\ref{fig:homotopy}), it reveals a hedgehog texture surrounding the $K$-point (and a two-dimensional skyrmionic projection) arising from a nontrivial topology due to $\pi_2$ as indicated by Table~\ref{table:classification33}.

\section{Additional examples of SUSY lattice correspondences and beyond line graph construction}
\label{app:MoreSUSYExamples}

\begin{figure}[t]
    \centering
    \includegraphics[width=\columnwidth]{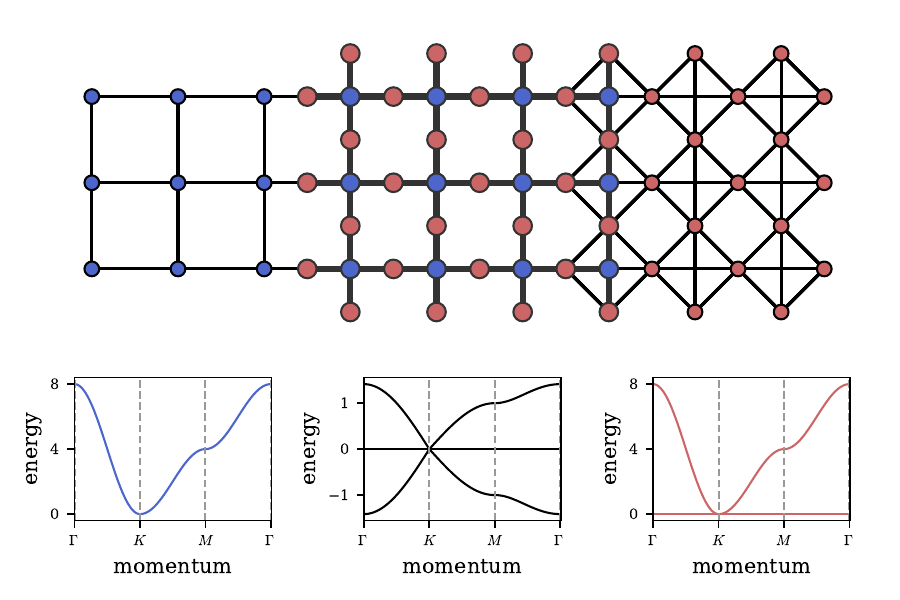}
    \caption{\textbf{SUSY corresponding square and checkerboard lattices.} Complex fermions (blue, left) on the square lattice are supersymmetrically linked to complex bosons (red, right) on the checkerboard lattice. The mapping can be established with a supercharge which can be interpreted as the adjacency matrix of a square-X lattice (center plot), \ie a square lattice with additional sites on every bond whose two sublattices are the square and checkerboard lattices respectively. For the topological classification according to Table \ref{table:classificationBDI}, we find, noting that the Witten index here is $\nu = 1$, that the nexus point in the supercharge spectrum has a non-trivial topological invariant of $\pi_1 = +1$, see also the illustration in Fig.~\ref{fig:homotopy} of the Appendix.}
    \label{fig:susy_trio_square_checkerboard}
\end{figure}

In addition to the SUSY lattice correspondences for tight-binding models of complex fermions and bosons presented in the main manuscript, we provide a few more instances to widen its applicability. The first one coincides with the line graph construction and the second one reproduces an identical partner with no additional flat bands. In distinction, the third and the fourth examples take us beyond the line graph construction in which our SUSY links two lattices as the superpartners that are not the line graph of one another, yet, (one of them) hosting flat bands. A characteristic of the supercharge lattice in these two cases is that in the associated bipartite graph, the connections between the sites of one sublattice are mediated by more than one site from the other sublattice, unlike the examples shown so far. It turns out such a bipartite model can host multiple flat bands (given by the Witten index) at Dirac-type band crossing(s) in its tight-binding spectrum. In the latter two examples, the multiplicity of the connectivity in the supercharge lattice provides a new control for the number of flat bands in addition to the geometry of the graph. 

The first example is a SUSY lattice correspondence between the square lattice and the checkerboard lattice (which is sometimes also referred to as planar pyrochlore lattice) via the square-X lattice (Fig.~\ref{fig:susy_trio_square_checkerboard}) that is commonly referred to as the Lieb lattice \cite{Lieb1989}. The additional site in the checkerboard unit cell with regard to the square lattice manifests itself in a single flat band in the checkerboard band structure, pointing to an extensive ground-state manifold of the Heisenberg antiferromagnet on this lattice \cite{Moessner1998b}. 

\begin{figure}[t]
    \centering
    \includegraphics[width=\columnwidth]{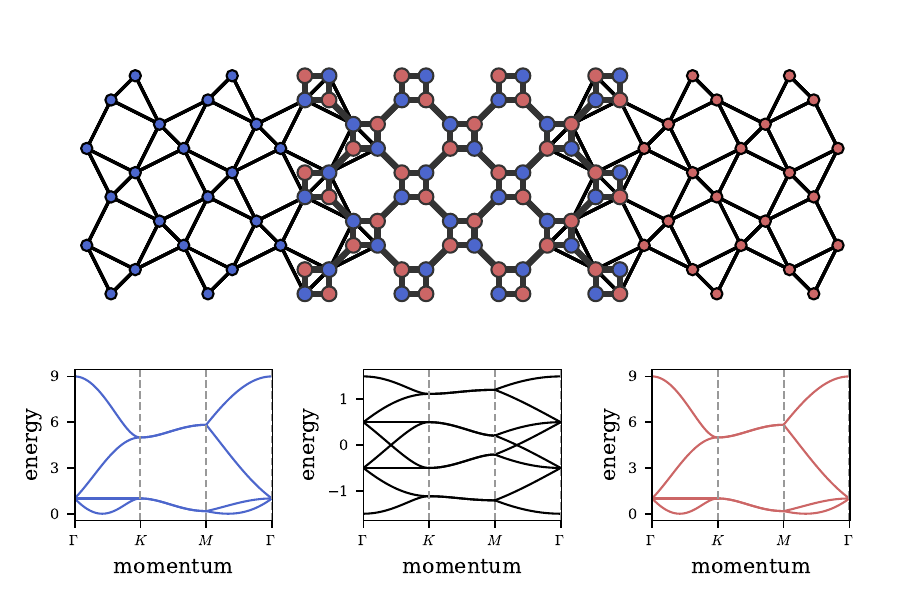}
    \caption{\textbf{SUSY corresponding Shastry-Sutherland lattices.} Complex fermions (blue, left) on the Shastry-Sutherland lattice are supersymmetrically linked to complex bosons (red, right) on the same lattice. The mapping can be established with a supercharge which can be interpreted as the adjacency matrix of a square-octagon lattice (center plot) whose two sublattices are both Shastry-Sutherland lattices. For the topological classification according to Table \ref{table:classification34}, we find, noting that the Witten index here is $\nu = 0$, that the Fermi surface in the supercharge spectrum has a trivial topological invariant of $\pi_1 = 0$.}
    \label{fig:susy_trio_shastry_sutherland}
\end{figure}

The second example (Fig.~\ref{fig:susy_trio_shastry_sutherland}) pertains to the square-octagon lattice whose two sublattices each constitute a so-called Shastry-Sutherland lattice \cite{ShastrySutherland1981}, which has been extensively discussed in the context of SrCu$_2$(BO$_3$)$_2$ \cite{Kageyama1999,Miyahara1999,Koga2000}. 
This SUSY lattice correspondence allows us to formulate a spin-fermion SUSY correspondence between the spin physics on the Shastry-Sutherland lattice and fermions on the square-octagon lattice, the planar variant of the three-dimensional scenario that we discussed in the main text when connecting the hyperoctagon lattice and a 3D generalization of the Shastry-Sutherland lattice (Fig.~\ref{fig:3DShastrySutherland-hyperoctagon-correspondence}). 

\begin{figure}[t]
    \centering
    \includegraphics[width=\columnwidth]{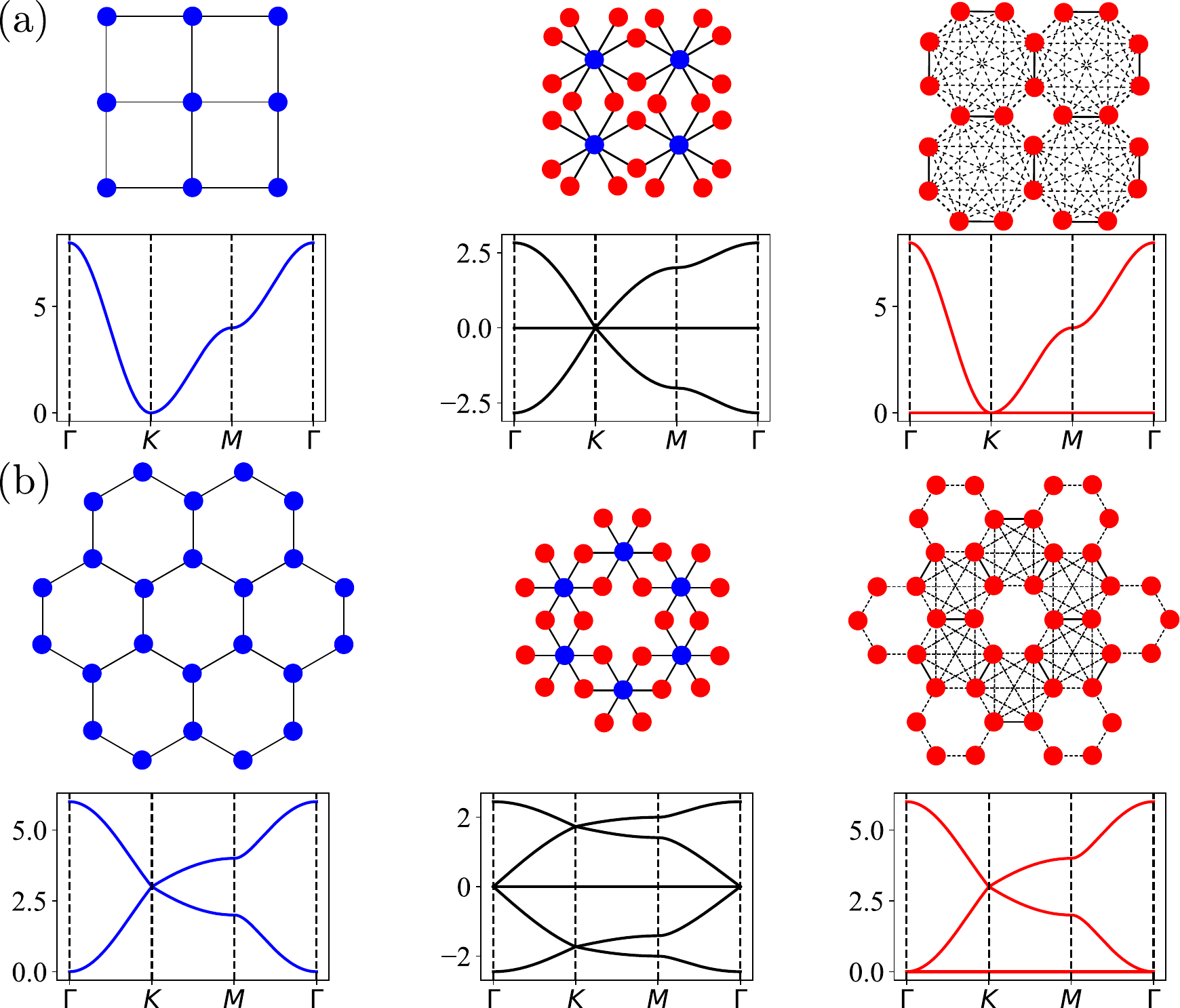}
    \caption{\textbf{More examples on the (SUSY-enabled) beyond line graph construction.} (a) Complex fermions (blue, left) on the square lattice are supersymmetrically linked to complex bosons (red, right) on the square-octagon lattice with modulated and all-to-all connections within the octagonal plaquettes. The mapping is established with a supercharge interpreted as the adjacency matrix of the lattice (center plot) where the connections between one type of sublattice are mediated by one/multiple sites from the other sublattice placing such a bipartite network beyond the split graph line graph construction. (b) Another similar construction with the complex fermions (blue, left) on the honeycomb lattice with its partner bosonic model on another decorated honeycomb lattice with two of the three adjacent hexagons around each honeycomb vertex allowing all-to-all connections within themselves. The Witten index is $\nu=3$ in (a) and $\nu=4$ in (b) rendering the Fermi surface in both supercharge spectra a trivial topology with $\pi_1 = 0$ (according to Table \ref{table:classification33}).}
    \label{fig:beyond_line_graph_more}
\end{figure}    

The third example [Fig.~\ref{fig:beyond_line_graph_more} (a)] demonstrates how to design a supersymmetric partner of the square lattice (blue, left) which is not its line graph (unlike the first example), yet hosts flat bands. The partner turns out to be a square-octagon lattice (red, right). The supercharge lattice, in this case, admits multiple connectivities between the sublattices \ie the blue sites are doubly connected with each other via the red sites, because of which the square-octagon lattice formed by the red sites develops modulated and all-to-all hoppings in the octagonal plaquettes. The supercharge has Witten index $\nu=3$ exacting three flat bands in the tight-binding spectrum of the square-octagon side.

Constituting our fourth example Fig.~\ref{fig:beyond_line_graph_more} (b) illuminates another instance of a multiply-connected supercharge lattice which is of relevance in the context of flat band quantum materials. The supercharge lattice is realized in the family of compounds CrX$_3$ ($X=$ Cl, Br, I) where the Cr cations inhabit one sublattice, and the halogen (X) anions the other \cite{wu2022optical} and establishes a supersymmetric link between the nearest-neighbor honeycomb network (blue, left) and its non-line graph partner (red, right) -- another honeycomb network but with modulated and all-to-all hoppings within $2/3$ of the honeycomb plaquettes. The charge has Witten index $\nu=4$, hosting four flat bands at $E=0$ that also appear on the red honeycomb network which comprises the halogen sites in CrX$_3$.  

Our final example in this appendix is to illustrate the real fermion-boson SUSY applying to the square-octagon lattice that hosts a Kitaev spin liquid \cite{yang2007mosaic}. Following the prescription of the real fermion-boson SUSY in Section~\ref{subsec:real_supersymmetry}, we connect the Majorana fermions on this lattice to a spring-mass network forming the Shastry-Sutherland lattice. This is in a similar spirit to what is done in Ref.~\cite{Attig2019} where a mechanical analog of the Kitaev honeycomb model has been devised. However, this new mechanical model importantly allows us to also vary the flux sectors of this model thereby probing both gapped and gapless Majorana spectra in the $\pi$-flux and zero-flux phases, respectively.  

In detail, the Majorana Hamiltonian on the square-octagon lattice is defined as  
\begin{align}
      {\bf H}_F({\bf k}) = i
      \begin{pmatrix}
         & -{\bf A}^\dagger({\bf k}) \\ 
         {\bf A}({\bf k}) &  
       \end{pmatrix},
\end{align}
with
\begin{align}
        &{\bf A}({\bf k}) =  \nonumber \\
        &t\begin{bmatrix}
         1/z_x & 1/z_y & (z_xz_y)^{1/\sqrt{2}} & 0 \\
         z_y & z_x & 0 & (z_xz_y)^{-1/\sqrt{2}} \\
         0 & (z_y/z_x)^{1/\sqrt{2}} & 1/z_y & z_x \\
         (z_x/z_y)^{1/\sqrt{2}} & 0 & 1/z_x & z_y 
        \end{bmatrix},
\end{align}
where $z_{x,y}\equiv e^{ik_{x,y}}$ and $t$ denotes the hopping strengths. The unit cell here comprises eight sites included in a basis
$\begin{pmatrix}
       \gamma^A_{1{\bf k}} & \gamma^A_{2{\bf k}} & \gamma^A_{3{\bf k}} & \gamma^A_{4{\bf k}} &
       \gamma^B_{1{\bf k}} & \gamma^B_{2{\bf k}} & \gamma^B_{3{\bf k}} & \gamma^B_{4{\bf k}}
\end{pmatrix}^T$.

This fermionic Hamiltonian can be generated by a supercharge whose matrix form is
\begin{align}
      {\bf R}({\bf k}) = \begin{pmatrix}
         {\bf 1} &  \\ 
         & {\bf A}({\bf k})  
       \end{pmatrix}.
\end{align}
The bosonic model it yields is defined on the Shastry-Sutherland lattice with the sites hosting unit masses and the bonds, the springs including intrasite contributions as the other examples. This is an intriguing case of a mechanical setup where the phonon spectrum exhibits a manifold of Goldstone modes that form a {\em line} in momentum space (akin to a Fermi surface). The spectrum also exhibits a dispersionless section along a path between certain high-symmetry points ($\Gamma$ to $K$) in the Brillouin zone. This example illuminates novel topological properties of phonon bands beyond those studied in Ref.~\cite{Attig2019} and opens up a new avenue for realizing exotic phonon dispersions in mosaic structure lattices.

%%%%%%%%%%%%%%%%%%%%%%%%%%%%%%%%%%%%%%%%%%%%%%%%%%%%%%
% Topological invariants
%%%%%%%%%%%%%%%%%%%%%%%%%%%%%%%%%%%%%%%%%%%%%%%%%%%%%%

\section{Topology for complex fermion-boson SUSY correspondence}
\label{app:topology_complex}

Topological twists (or defects) in fermionic bands can be characterized by quantized phases or integer-valued invariants originating from the homotopy classification of tight-binding Hamiltonians \cite{TKNN1982, Kohmoto1985, ryu2010topological}. For instance, the Berry phase has been an incredibly useful tool to characterize the topology of (nodal) band structures in electronic systems \cite{Zak1989, Niureview2010, quantizedBerry_hughes2011, Chan2016}. 

\begin{figure}
    \centering
    \includegraphics[width=\columnwidth]{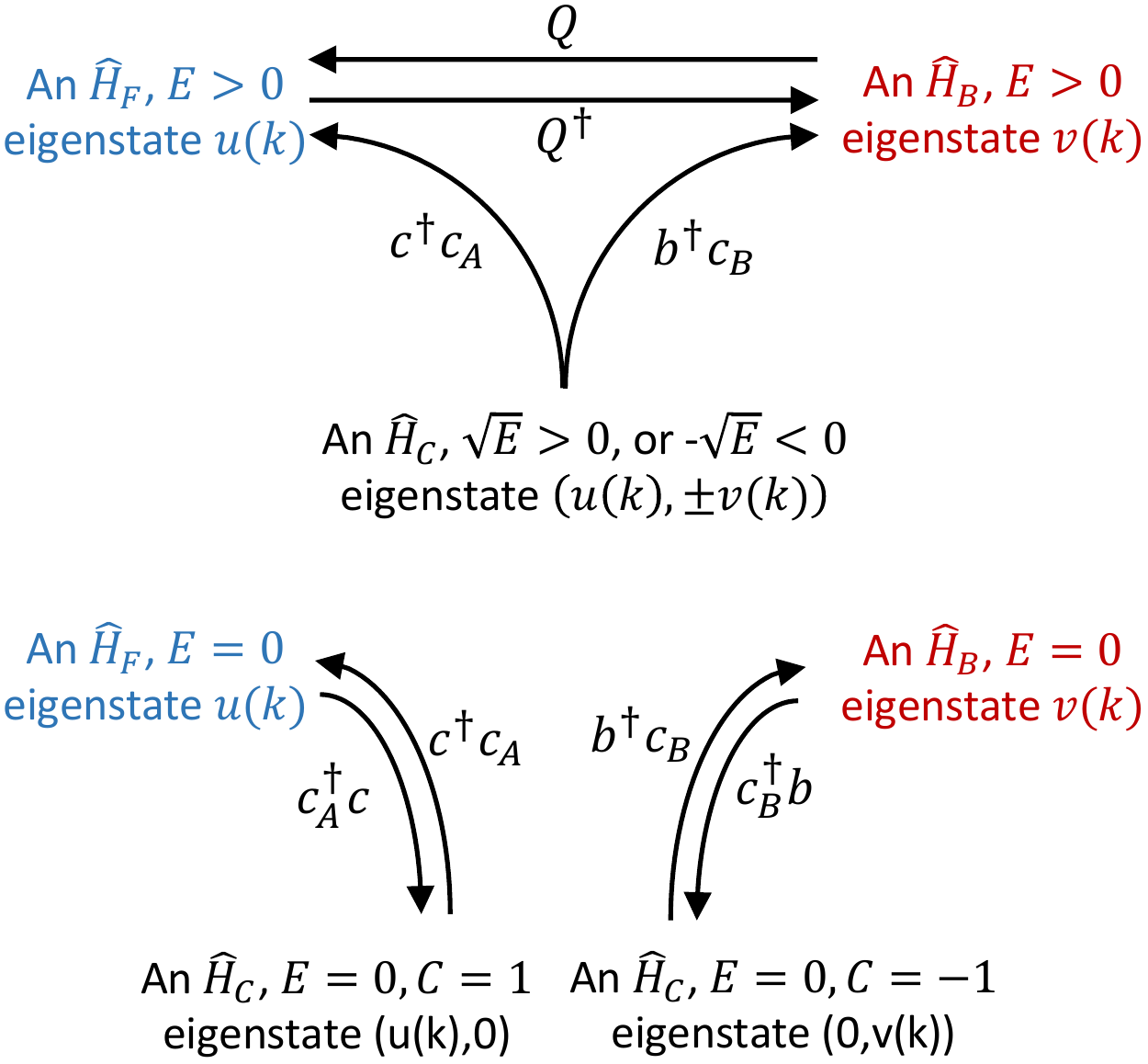}
    \caption{{\bf An eigenstate ``olog’’ of SUSY states} \cite{spivak2014category} showing the mappings of single particle eigenstates between the chiral fermion model ${\cal H}_C$ associated with the supercharge [see Eq.~\eqref{eq:complexsusy_charge}], the boson model ${\cal H}_B$ and the fermion model ${\cal H}_F$ [see Eq.~\eqref{eq:complexsusy_H_SUSY}]. Notice the zero energy space of ${\cal H}_C$ splits into two sectors, one on the $A$ sublattice that maps to eigenstates of ${\cal H}_F$ and one on the $B$ sublattice that maps to eigenstates of ${\cal H}_B$. Following a given eigenstate of ${\cal H}_C$ with finite energy leads to the same state independent of which path is taken. Note: the mapping under $Q$, $Q^\dagger$ of the many-body ${\cal H}_{\rm SUSY}$ states, follows a similar pattern, with $Q$, preserving $R = N_F + N_B$, the so-called ``$R$" symmetry \cite{hall1991u}, mapping the eigenstates in Fock sector $(N_F,N_B)$ to those in Fock sector $(N_F+1,N_B-1)$ and $Q^\dagger$ mapping $(N_F,N_B)$ to $(N_F-1,N_B+1)$.}
    \label{fig:susymapping}
\end{figure}

{\em Berry Phases.} For the SUSY lattice models at hand, specified via the Hamiltonians ${\bf RR^\dagger}$ and ${\bf R^\dagger R}$, the Berry phases associated with the corresponding Bloch wavefunctions turn out to be intimately related. Using the wavefunction relations of Eq.~\eqref{eq:Berry3}, illustrated in Fig.~\ref{fig:susymapping}, one can show that if the closed-orbit Berry phase for a (fermionic) eigenfunction of ${\bf RR^\dagger}$ is given by
\begin{align}\label{eq:Berry1}
 \theta_F^{{\bf RR^\dagger}} &= i\oint \langle u({\bf k}) | \partial_{\bf k} u({\bf k})\rangle \cdot {\rm d}{\bf k} \,,
\end{align}
then the one for the (bosonic) eigenfunction of ${\bf R^\dagger R}$ at the same energy will be 
\begin{align}
	\label{eq:Berry2}
  	\theta_B^{{\bf R^\dagger R}} 
		&= i\oint \langle v({\bf k}) | \partial_{\bf k} v({\bf k})\rangle \cdot {\rm d}{\bf k} \\ \nonumber 
		&=\theta_F^{{\bf RR^\dagger}}
			-\oint \frac{{\Im}(\langle u({\bf k})|(\partial_{\bf k}{\bf R})|v({\bf k})\rangle)}{\sqrt{\omega_n({\bf k})}}\cdot {\rm d}{\bf k} \,.
\end{align}
Of particular interest here is the additional term in the second line, which points to a potential difference in the bosonic and fermionic Berry phases. It is a direct consequence of the wavefunction correspondence in \eqref{eq:Berry3} and notably involves both the fermionic {\sl and} bosonic states. If we reexpress it entirely in terms of the bosonic states by further utilizing the mapping in \eqref{eq:Berry3}, we arrive at a SUSY version of the bosonic Berry phase
\begin{align}
 \label{eq:Berry4}
  	\theta_B^{\rm SUSY} 
		&= i\oint \langle v({\bf k}) | \partial_{\bf k} v({\bf k})\rangle \cdot {\rm d}{\bf k} \nonumber \\
		&~~~~~~+\oint \frac{{\Im}(\langle v({\bf k})|{\bf R}^\dagger(\partial_{\bf k}{\bf R})|v({\bf k})\rangle)}{\omega_n({\bf k})}\cdot {\rm d}{\bf k}\,.
\end{align}
Only if the integrand in the second augmenting term can be written as $\partial_{\bf k}{\cal F}({\bf k})$ for some function ${\cal F}({\bf k})$, then the conventional Berry phases $\theta_F$ and $\theta_B$, acquired under adiabatic evolution along any closed orbit, in the two supersymmetrically related systems will be equal. Such a vanishing of the augmenting term would also be ensured by 
imposing that the matrix ${\bf R}^\dagger(\partial_{\bf k}{\bf R})$ at its heart is Hermitian and $[{\bf R}^\dagger,\partial_{\bf k}{\bf R}]=0$.  (For a Hermitian ${\bf R}$ these two conditions are in fact equivalent \footnote{Such a condition is noted in Ref.~\cite{Flynn2020} that studied Berry phase in quadratic bosonic systems}). For the honeycomb-kagome correspondence described above, where both band structures (of ${\bf RR^\dagger}$ and ${\bf R^\dagger R}$) feature Dirac points, the second term in \eqref{eq:Berry4} indeed vanishes. As a result, the conventional Berry phases $\theta_F$ and  $\theta_B$ in the two systems around these point nodes are identical and admit quantized values $\pm\pi$.

\begin{figure}[t]
    \centering
    \includegraphics[width=1.0\columnwidth]{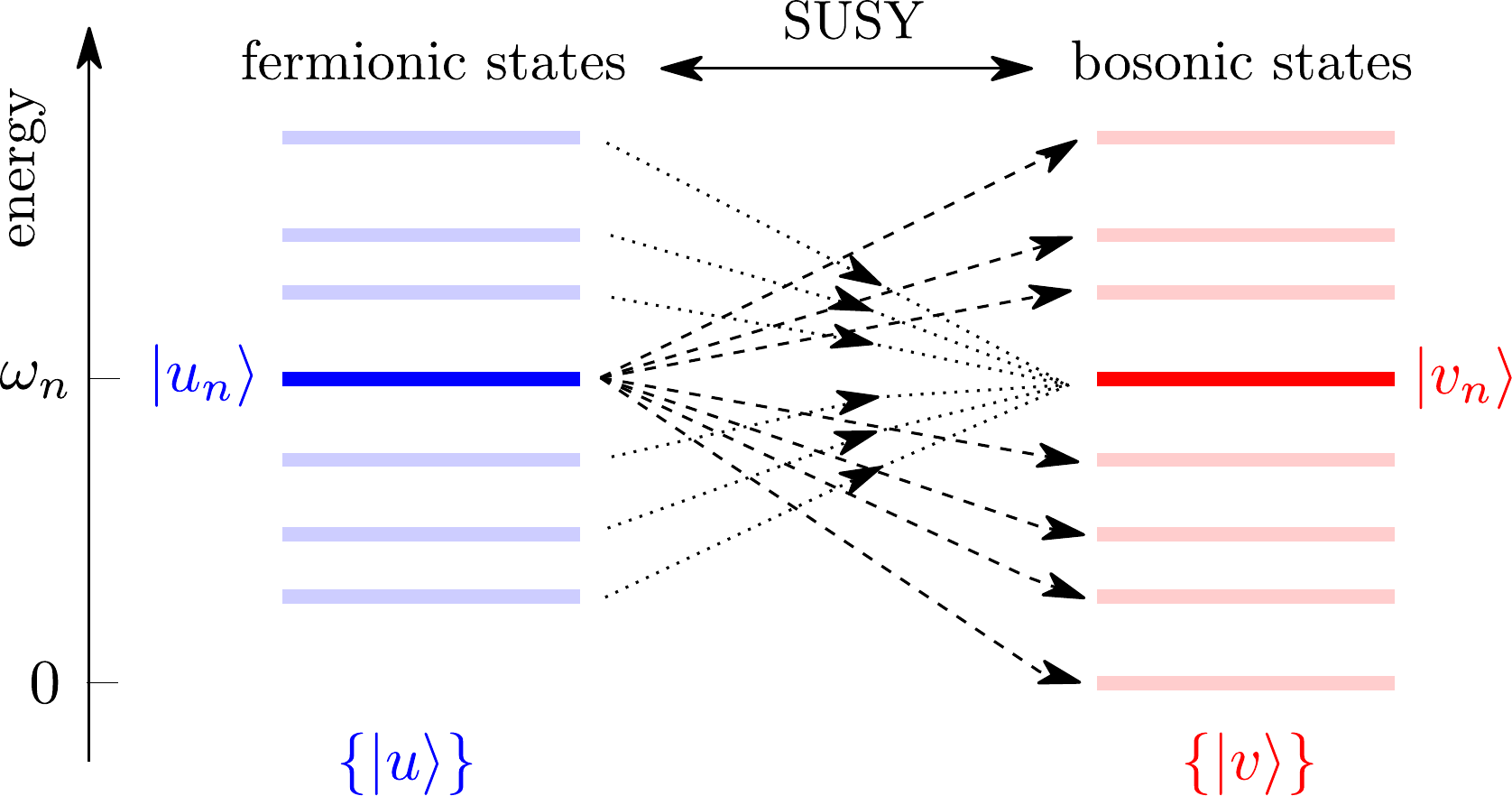}
    \caption{{\bf SUSY contribution to the bosonic Berry curvature.} The SUSY contribution to the bosonic Berry curvature consists of two distinct transitions between the fermionic vector space $\{|u\rangle\}$ (blue) and the bosonic vector space $\{|v\rangle\}$ (red): (i) $|u_n\rangle \rightarrow |v_{m\neq n}\rangle$ shown in dashed arrows and (ii) $|u_{m\neq n}\rangle \rightarrow |v_{n}\rangle$ shown in dotted arrows (detailed in the text).}
    \label{fig:susyBerrycurv}
\end{figure}

{\em Berry Curvature.} Similarly, one can study how the Berry {\sl curvature} is related between the bosonic and the fermionic tight-binding models that are connected by such a supersymmetric identification. For a given (isolated) fermionic band $|u_n\rangle$ in the spectrum of ${\cal H}_F\equiv {\bf RR^\dagger}$
the Berry curvature can be cast as
\begin{align}\label{berrycurv1}
 {\cal F}_n^{(u)}({\bf k})= -2~{\rm Im}\Bigg[\sum_{m\neq n} \frac{\langle u_n|\partial_x{\cal H}_F |u_m\rangle\langle u_m|\partial_y{\cal H}_F |u_n\rangle}{(\omega_n-\omega_m)^2}\Bigg],
\end{align}
where $\partial_i\equiv \partial/\partial k_i$. 
Then the corresponding supersymmetric bosonic partner state in the spectrum of ${\cal H}_B\equiv {\bf R^\dagger R}$ given by $|v_n\rangle$ will have a Berry curvature of 
\begin{align}\label{berrycurv2}
 {\cal F}_n^{(v)}({\bf k}) &= {\cal F}_n^{(u)}({\bf k}) \nonumber \\
 &+ 2~{\rm Im} \Bigg[ \sum_{m\neq n} \frac{\langle v_n|\partial_x{\bf R}^\dagger|u_m\rangle\langle u_m|\partial_y{\bf R}|v_n\rangle}{(\omega_n-\omega_m)} \nonumber \\
 &~~~- \sum_{m\neq n} \frac{\langle u_n|\partial_x{\bf R}|v_m\rangle\langle v_m|\partial_y{\bf R}^\dagger|u_n\rangle}{(\omega_n-\omega_m)} \Bigg] \,,
\end{align}
which like the Berry phase above is found to be augmented by an additional term.

The expression in \eqref{berrycurv1}  formally resembles the result of a second-order perturbation theory calculation. It is a gauge-invariant form of the Berry curvature that admits contributions owing to the transitions from all other bands to the given $n$-th band considering $\partial_i{\cal H}$ as a perturbation. In a similar spirit, one can inspect the supersymmetric contribution in \eqref{berrycurv2} which reveals two noteworthy features. Firstly, the supersymmetric contribution recognizes $\partial_i{\bf R}$ and $\partial_i{\bf R}^\dagger$, \ie {\sl derivatives} of the supercharges, as first-order perturbations. The conventional term instead concerns the derivative of the fermionic Hamiltonian $\partial_i{\cal H}_F$ as second-order perturbations instead. Secondly, the supersymmetric contribution comprises two complementary components signifying two distinct types of transitions (admixing the bosonic and fermionic Hilbert spaces) -- the first term in the square bracket in \eqref{berrycurv2} is due to the transitions from the bosonic Bloch state $|v_n\rangle$ to all eigenvectors in the vector space of its fermionic superpartner indexed with $m\neq n$, as schematically illustrated in Fig.~\ref{fig:susyBerrycurv}. In contrast, the second term is due to the transitions from the fermionic eigenvector $|u_n\rangle$ (the superpartner of $|v_n\rangle$) at the same energy $\omega_n$, to all the eigenvectors in the bosonic vector space $\{|v\rangle\}$ indexed with $m\neq n$. 
Similar to what we have seen for the Berry phase above, it turns out that in the honeycomb-kagome correspondence, when we gap out the Dirac cones with a small inversion symmetry breaking parameter to isolate the uppermost band in both cases, these two terms nullify each other rendering the fermionic and bosonic Berry curvatures identical ${\cal F}^{(v)}({\bf k}) = {\cal F}^{(u)}({\bf k})$. \

De facto, it seems that for the SUSY connection of {\sl complex} fermions and bosons discussed here, there is no distinction of the SUSY version of the Berry phase and its original variant, and similarly for the Berry curvature. At least this is what we find for a number of examples discussed in Section~\ref{sec:complex_SUSY}. A completely different picture awaits us when we turn to the case of {\sl real} fermions and bosons, which we discussed in Section~\ref{sec:topo_mechanics}, in the context of topological mechanics. 

%%%%%%%%%%%%%%%%%%%%%%%%%%%%%%%%%%%%%%%%%%%%%%%%%%%%%%
% Topological Mechanics
%%%%%%%%%%%%%%%%%%%%%%%%%%%%%%%%%%%%%%%%%%%%%%%%%%%%%%

\section{Kane-Lubensky chain to Kitaev chain}
\label{sec:AppKaneLubensky}

Let us complement the discussion of the SUSY correspondence of one of our principal examples in topological mechanics, between the (bosonic) Kane-Lubensky chain and the (fermionic) Kitaev chain, in the main text with a more detailed derivation here.
In particular, let us demonstrate how to derive a {\em local} compatibility matrix $\tilde{\bf A}$ of the form given in \eqref{HKane-Lubenskyp5} (of the main text) from the Kane-Lubensky dynamical matrix 
\begin{align}
     \tilde{\bf D} = 
     \begin{pmatrix}
      a & b & 0 & 0 & \dots & b \\
      b & a & b & 0 & \dots & 0 \\
      0 & b & a & b & \dots & 0 \\
      \vdots & \vdots & \vdots & \vdots & \vdots & \vdots \\
      0 & 0 & \dots & b & a & b \\
      b & 0 & 0 & \dots & b & a
     \end{pmatrix}
     \equiv
     \tilde{\bf A}^T\tilde{\bf A} \,,
     \label{HKane-Lubenskyp56}
\end{align} 
that enables constructing a Kitaev chain as a superpartner of the phonon problem. The strategy is to first construct an incidence matrix of the graph represented by the phonon chain which entails interactions up to the next nearest neighbors as reflected in $\tilde{\bf D}$, thereby retaining the features of locality same as $\tilde{\bf D}$. This is important to construct a local fermionic Hamiltonian via our SUSY approach. It provides a simple mathematical demonstration of the graph square-rooting algorithm presented earlier applied to formulate supersymmetric partners of random nearest-neighbor one-dimensional bosonic models.    

To proceed, the first step is to project out the diagonal terms of $\tilde{\bf D}$ to construct a weighted adjacency matrix that contains the connectivities (including the bond strengths) between distinct sites of the phonon chain. Note the parameters $a, b$ in \eqref{HKane-Lubenskyp56} are arbitrary positive numbers representing a generic one-dimensional phonon model with an adjacency matrix
\begin{align}
 \tilde{\cal A} = 
  \begin{pmatrix}
    0 & b & 0 & 0 & \dots & b \\
    b & 0 & b & 0 & \dots & 0 \\
    0 & b & 0 & b & \dots & 0 \\
    \vdots & \vdots & \vdots & \vdots & \vdots & \vdots \\
    0 & 0 & \dots & b & 0 & b \\
    b & 0 & 0 & \dots & b & 0
  \end{pmatrix}.
\label{HKane-Lubenskypp2}
\end{align}
Regarding the underlying phonon lattice as a one-dimensional graph with connectivities between the next nearest neighboring sites only, the incidence matrix induces a map from the sites, denoted by $\{i\}$, to the bonds of this graph, denoted by $\{\alpha\}$, which we can write as \begin{align}
      \tilde{\cal I} = 
       \begin{pmatrix}
         x_1\sqrt{b} & \sqrt{b}/x_1 & 0 & 0 & \dots & 0 \\
         0 & x_2\sqrt{b} & \sqrt{b}/x_2 & 0 & \dots & 0 \\
         0 & 0 & x_3\sqrt{b} & \sqrt{b}/x_3) & \dots & 0 \\
         \vdots & \vdots & \vdots & \vdots & \vdots & \vdots \\
         \sqrt{b}/x_N & 0 & 0 & \dots & 0 & x_N\sqrt{b}
      \end{pmatrix}
      \label{HKane-Lubenskypp3}
     \end{align}
for a periodic $N$-site (and $N$-bond) isostatic chain. We then determine the parameters $\{x_\alpha\}$, which is a set of $N$ real-valued numbers on the bonds, by imposing the condition 
\begin{align}
 \tilde{\cal I}^T\tilde{\cal I} = b
   \begin{pmatrix}
    f_1 & 1 & 0 & 0 & \dots & 1 \\
    1 & f_2 & 1 & 0 & \dots & 0 \\
    0 & 1 & f_3 & 1 & \dots & 0 \\
    \vdots & \vdots & \vdots & \vdots & \vdots & \vdots \\
    0 & 0 & \dots & 1 & f_{N-1} & 1 \\
    1 & 0 & 0 & \dots & 1 & f_N
   \end{pmatrix}
   = \tilde{\bf D}
\label{HKane-Lubenskyp4}
\end{align}
with $f_\alpha = x_\alpha^2+1/x_\alpha^2$ or, equivalently,
\begin{align}
 x_\alpha^2+1/x_\alpha^2 = a/b,
\end{align}
that has four solutions
\begin{align}
 x_i = \pm \sqrt{\frac{a}{2b} \pm \sqrt{\frac{a}{2b}-1}},
\end{align}
independent of $\alpha$, leading to the $N\times N$ compatibility matrix $\tilde{\bf A}$ noted in \eqref{HKane-Lubenskyp5} of the main text. 

%%%%%%%%%%%%%%%%%%%%%%%%%%%%%%%%%%%%%%%%%%%%%%%%%%%%%%
\end{document}